\documentclass[twocolumn]{aastex631}

\usepackage{xspace}
\usepackage[nointegrals]{wasysym}
\usepackage{mathrsfs}
\usepackage{enumitem}
\usepackage{mathtools}

\newcommand{\link}[1]{{\small{\texttt{\underline{#1}}}}}

\def\eg{{\it e.g.,}\xspace}

\def\ie{{\it i.e.,}\xspace}

\def\gsim{~\rlap{$>$}{\lower 1.0ex\hbox{$\sim$}}}
\def\lsim{~\rlap{$<$}{\lower 1.0ex\hbox{$\sim$}}}
\def\wpmsq{{W/m$^{-2}$}}
\def\tidalq{{$Q$}}

\newcommand{\be}{\begin{equation}}
\newcommand{\ee}{\end{equation}}

\def\tess{{\it TESS}\xspace}
\def\kepler{{\it Kepler}\xspace}

\def\msun{{M_\odot}}

\def\mearth{{M_\oplus}}
\def\rearth{{R_\oplus}}

\def\al26{{$^{26}$Al}\xspace}
\def\k40{{$^{40}$K}\xspace}
\def\th232{{$^{232}$Th}\xspace}
\def\U{{$^{235}$U}\xspace}
\def\UU{{$^{238}$U}\xspace}
\def\co2{{CO$_2$}\xspace}

\def\vplanet{\texttt{\footnotesize{VPLanet}}\xspace}
\def\atmesc{\texttt{\footnotesize{AtmEsc}}\xspace}
\def\binary{\texttt{\footnotesize{BINARY}}\xspace}
\def\distorb{\texttt{\footnotesize{DistOrb}}\xspace}

\def\distrot{\texttt{\footnotesize{DistRot}}\xspace}
\def\eqtide{\texttt{\footnotesize{EqTide}}\xspace}
\def\galhabit{\texttt{\footnotesize{GalHabit}}\xspace}

\def\poise{\texttt{\footnotesize{POISE}}\xspace}
\def\radheat{\texttt{\footnotesize{RadHeat}}\xspace}
\def\spinbody{\texttt{\footnotesize{SpiNBody}}\xspace}
\def\stellar{\texttt{\footnotesize{STELLAR}}\xspace}
\def\thermint{\texttt{\footnotesize{ThermInt}}\xspace}

\def\vsp{\texttt{\footnotesize{VSPACE}}\xspace}
\def\bigpl{\texttt{\footnotesize{BigPlanet}}\xspace}
\def\vplot{\texttt{\footnotesize{VPLot}}\xspace}

\def\nbody6{\texttt{\footnotesize{NBODY6++}}\xspace}
\def\mercury{\texttt{\footnotesize{MERCURY}}\xspace}

\def\hnbody{\texttt{\footnotesize{HNBODY}}\xspace}

\def\apjs{{\it Astrophys.~J.~Supp.}}

\def\apjl{{\it Astrophys.~J.~Lett.}}

\def\aap{{\it Astron.~\&~Astrophys.}}

\begin{document}

\title{VPLanet: The Virtual Planet Simulator}

\author{Rory Barnes}
\affiliation{Astronomy Department, University of Washington, Box 951580, Seattle, WA 98195, USA}
\affiliation{NASA Virtual Planetary Laboratory Lead Team, USA}

\author{Rodrigo Luger}
\affiliation{Center for Computational Astrophysics, Flatiron Institute, New York, NY 10010, USA}
\affiliation{NASA Virtual Planetary Laboratory Lead Team, USA}

\author{Russell Deitrick}
\affiliation{Center for Space and Habitability, University of Bern, Gesellschaftsstrasse 6, CH-3012, Bern, Switzerland}
\affiliation{NASA Virtual Planetary Laboratory Lead Team, USA}

\author{Peter Driscoll}
\affiliation{Department of Terrestrial Magnetism, Carnegie Institution for Science, 5241 Broad Branch Rd, Washington, DC, 20015, USA}
\affiliation{NASA Virtual Planetary Laboratory Lead Team, USA}

\author{Thomas R. Quinn}
\affiliation{Astronomy Department, University of Washington, Box 951580, Seattle, WA 98195, USA}
\affiliation{NASA Virtual Planetary Laboratory Lead Team, USA}

\author{David P. Fleming}
\affiliation{Astronomy Department, University of Washington, Box 951580, Seattle, WA 98195, USA}
\affiliation{NASA Virtual Planetary Laboratory Lead Team, USA}

\author{Hayden Smotherman}
\affiliation{Astronomy Department, University of Washington, Box 951580, Seattle, WA 98195, USA}

\author{Diego V. McDonald}
\affiliation{Astronomy Department, University of Washington, Box 951580, Seattle, WA 98195, USA}

\author{Caitlyn Wilhelm}
\affiliation{Astronomy Department, University of Washington, Box 951580, Seattle, WA 98195, USA}
\affiliation{NASA Virtual Planetary Laboratory Lead Team, USA}

\author{Rodolfo Garcia}
\affiliation{Astronomy Department, University of Washington, Box 951580, Seattle, WA 98195, USA}
\affiliation{NASA Virtual Planetary Laboratory Lead Team, USA}

\author{Patrick Barth}
\affiliation{Max Planck Institute for Astronomy, K{\" o}nigstuhl 17, 69117 Heidelberg, Germany}

\author{Benjamin Guyer}
\affiliation{Astronomy Department, University of Washington, Box 951580, Seattle, WA 98195, USA}

\author{Victoria S. Meadows}
\affiliation{Astronomy Department, University of Washington, Box 951580, Seattle, WA 98195, USA}
\affiliation{NASA Virtual Planetary Laboratory Lead Team, USA}

\author{Cecilia M. Bitz}
\affiliation{Atmospheric Sciences Department, University of Washington, Box 951640, Seattle, WA 98195, USA}
\affiliation{NASA Virtual Planetary Laboratory Lead Team, USA}

\author{Pramod Gupta}
\affiliation{Astronomy Department, University of Washington, Box 951580, Seattle, WA 98195, USA}
\affiliation{NASA Virtual Planetary Laboratory Lead Team, USA}

\author{Shawn D. Domagal-Goldman}
\affiliation{Planetary Environments Laboratory, NASA Goddard Space Flight Center, 8800 Greenbelt Road, Greenbelt, MD 20771, USA}
\affiliation{NASA Virtual Planetary Laboratory Lead Team, USA}

\author{John Armstrong}
\affiliation{Department of Physics, Weber State University, Ogden, UT 84408-2508, USA}
\affiliation{NASA Virtual Planetary Laboratory Lead Team, USA}

\begin{abstract}

We describe a software package called \vplanet that simulates fundamental aspects of planetary system evolution
over Gyr timescales, with a focus on investigating habitable worlds. In this initial release, eleven physics modules are included that model internal, atmospheric,
rotational, orbital, stellar, and galactic processes.  Many of these modules can be coupled
simultaneously to simulate the evolution of terrestrial planets, gaseous planets, and stars. The code is validated by reproducing a selection of observations and past results. \vplanet is written in C and designed so that the user can choose the physics modules to apply to an individual object at runtime without recompiling, \ie a single executable can simulate the diverse phenomena that are relevant to a wide range of planetary and stellar systems. This feature is enabled by matrices and vectors of function pointers that are dynamically allocated and populated based on user input.  The speed and modularity of \vplanet enables large parameter sweeps and the versatility to add/remove physical phenomena to assess their importance. \vplanet is publicly available from a repository that contains extensive
documentation, numerous examples, Python scripts for plotting and data management, and infrastructure for community input and future development.

\end{abstract}

\section{Introduction\label{sec:intro}}

Exoplanetary systems display a diversity of morphologies, including a wide range of orbital architectures and planetary densities. This heterogeneity likely leads to a wide range of evolutionary trajectories that result in disparate planetary properties. As astronomers and astrobiologists probe these worlds to determine their atmospheric and surface properties, a comprehensive model of the physical effects that sculpt a planet can help prioritize targets and interpret observations. Here we describe a software package called \vplanet that self-consistently simulates many processes that influence the evolution of gaseous and terrestrial planets in a range of stellar systems. Our approach allows for coupling of simple models by simultaneously solving ordinary and partial differential equations (ODEs and PDEs) and explicit functions of time to track the evolution of, and feedbacks among, interior, atmospheric, stellar, orbital, and galactic processes.
Below we describe a set of physical models (called modules) that simulates these phenomena, as well as their assumptions and limitations. We validate the code, including both individual modules and a subset of module combinations, by reproducing key observations of the Earth, Solar System bodies, and known stellar systems. Where observations are lacking or unavailable, we reproduce a selection of previously published results. The software is open source and includes documentation, examples, and the opportunity for community involvement.\footnote{\url{https://github.com/VirtualPlanetaryLaboratory/vplanet}}

While \vplanet is designed to model an arbitrary planetary system, the primary motivation for creating \vplanet is to investigate the potential habitability of exoplanets with a single code that can capture feedbacks across the range of physical processes that affect a planet's evolution. TRAPPIST-1 \citep{Gillon16,Gillon17} is a good example of why coupled processes are needed to understand planetary evolution. In that system, the host star dimmed during the pre-main sequence, orbital interactions between planets are strong, tidal heating and rotational braking are significant, and stellar activity can drive atmospheric mass loss. As discussed below, \vplanet combines a set of theoretical models to provide a first order approach to investigate these processes and the feedbacks among them.

We define {\it habitable} to mean a planet that supports liquid surface water. Since astrochemical and planet formation studies find that most small planets form with significant inventories of water and bioessential elements \citep[\eg][]{vanDishoeck14,Morbidelli18}, the most pertinent question for potentially habitable planets may be ``do they still have water?'' The presence of water is controlled by the planet's interior, atmosphere, orbit, host star, and even the galaxy. Our approach does not presume {\it a priori} that any one process dominates the evolution because numerous processes can severely impact a planet's potential to support liquid water, even in the habitable zone \citep[HZ;][]{Kasting93,Kopparapu13}. For example, tidal heating may be strong enough to drive a runaway greenhouse \citep{Barnes13}, the pre-main sequence evolution of low-mass stars may desiccate (remove water from) a planet \citep{LugerBarnes15}, and stable resonant orbital oscillations can drive extreme eccentricity cycles \citep{Barnes15_res}. Furthermore, all these processes can operate in a given system, and hence a rigorous model of planetary system evolution, including habitability, should include as broad a range of physics as possible \citep{MeadowsBarnes18}.

Coupled evolutionary models could be valuable tools for interpreting the environments of newly discovered Earth-sized planets with nearly Earth-like levels of incident stellar radiation (``instellation'') orbiting other stars. These planets may be capable of supporting liquid water, but their habitability is currently an open question given the plethora of physical processes at play, \ie a planet's presence in the star's habitable zone does not imply the planet supports liquid water. Moreover, worlds like Proxima Centauri b \citep{AngladaEscude16} and TRAPPIST-1 c--g \citep{Gillon17} orbit bright enough host stars that their atmospheres will be probed with future ground- and space-based facilities \citep{Meadows18,Lincowski18}, which will provide constraints on otherwise unobservable surface habitability. This paper describes a benchmarked model of planetary system evolution that can be used to simulate the evolution of $\sim 1\mearth$, $\sim 1\rearth$ terrestrial planets with approximately Earth-like material properties and structure, with the goal of understanding their potential for habitability.

Though we are interested in Earth-like planets, the model is not restricted to Earth-radius and Earth-mass planets; rather, the model is intended to  investigate a wide range of planetary systems.
The phenomena described above are also important for uninhabitable worlds such as GJ 1132 b \citep{BertaThompson15} and 40 Eri b \citep{Ma18}. \vplanet can be used to simulate many such planets (and moons) to infer their histories.

The individual modules of \vplanet can simulate many aspects of planetary evolution, but the ability to couple modules together facilitates novel investigations, as demonstrated in several previously-published studies. \cite{Deitrick18a,Deitrick18b} combined the orbital, rotational, and climate modules to show that potentially habitable exoplanets can become globally glaciated if their orbital and rotational properties evolve rapidly and with large amplitude. \cite{Fleming18} showed that the coupled stellar-tidal evolution of tight binary stars can lead to orbital evolution that ejects circumbinary planets. \cite{Lincowski18} combined stellar evolution and atmospheric escape to track water loss and oxygen build-up on the TRAPPIST-1 planets. Finally, \cite{Fleming19} simulated the coupled stellar-tidal evolution of binary stars to show how unresolved binaries impact gyrochronology age estimates of stellar populations.

\vplanet is a fast and flexible code that combines a host of semi-analytic models, which are all written in C. It can provide quick insight into an individual planetary system with a single simulation (such as calculating tidal heating or stellar evolution), or can perform parameter sweeps and generate ensembles of evolutionary tracks that can be compared to observations. Alternatively, \vplanet can be combined with  machine learning algorithms to identify key parameters \citep{Deitrick18b}. These capabilities can provide direct insight or can complement research with more sophisticated tools, such as 3-D global circulation models, which are too computationally expensive to explore vast parameter spaces. For example, \vplanet can be used to isolate the most important phenomena (see $\S$~\ref{sec:vplanet}) and explore parameter space, and then more complicated models can target interesting regions of that parameter space to provide further insight and observational predictions.

The modularity and flexibility of \vplanet is enabled by matrices and vectors of function pointers, in which individual elements represent addresses of functions. In this framework, a user can specify a range of physics to be simulated, \ie select modules at runtime, and the software dynamically allocates the memory and collates the appropriate derivatives for integration. This approach allows users to trivially add or remove physics, \eg tides or stellar evolution, and assess their relative importance in the system's evolution. Moreover, this approach allows a ``plug and play''  development scheme in which new physics can be added and coupled with minimal effort.

The \vplanet code and its ecosystem have been designed to ensure  transparency, accessibility, and reproducibility. For example, figures presented below that are derived from \vplanet output include a link (in the electronic version of the paper) to the location in the \vplanet repository that contains the input files and scripts that generate the figure, \eg \href{https://github.com/VirtualPlanetaryLaboratory/vplanet/tree/master/examples}{\link{examples}}. The approximate run time for the simulation(s) necessary to generate the figure is also listed. The software is open source and freely available for all to use.

The objectives of \vplanet are to 1) simulate newly discovered exoplanets to assess the probability that they possess surface liquid water, and hence are viable targets for biosignature surveys, 2) model diverse planetary and stellar systems, regardless of potential habitability, to gain insights into their properties and history,  and 3) enable transparent and open science that contributes to the search for life in the universe. This paper is organized as follows. In $\S$~\ref{sec:vplanet} we describe the \vplanet algorithm and underlying software that enables flexibility in how the modules are connected. Readers interested only in the validation of the physics can skip to $\S$~\ref{sec:atmesc} in which we begin the
qualitative description of the 11 fundamental modules and demonstrate reproducibility of previously published results.  Details describing the quantitative results and implementation are relegated to Appendices \ref{app:atmesc}--\ref{app:thermint}. Briefly, the 11 modules, in alphabetical order, are as follows: simple thermal atmospheric escape models with \atmesc ($\S$~\ref{sec:atmesc}, App.~\ref{app:atmesc}), an analytic model of circumbinary planet orbital evolution with \binary ($\S$~\ref{sec:binary}, App.~\ref{app:binary}), 2nd and 4th
order secular models of orbital evolution with \distorb($\S$~\ref{sec:distorb}, App.~\ref{app:distorb}), a
semi-analytic model for rotational axis evolution due to
orbital evolution and the stellar torque with \distrot ($\S$~\ref{sec:distrot}, App.~\ref{app:distrot}), an approximate model for tidal effects with \eqtide
($\S$~\ref{sec:eqtide}, App.~\ref{app:eqtide}), a model of Oort Cloud object orbits adopted to capture wide binary orbits that includes galactic migration with \galhabit, ($\S$~\ref{sec:galhabit}, App.~\ref{app:galhabit}), an energy balance climate model with an
explicit treatment of ice sheet growth and retreat with \poise
($\S$~\ref{sec:poise}, App.~\ref{app:poise}),
radiogenic heating throughout planetary interiors with \radheat ($\S$~\ref{sec:radheat}, App.~\ref{app:radheat}), an $N$-body orbital model with \spinbody ($\S$~\ref{sec:spinbody}, App.~\ref{app:spinbody}), stellar
evolution, including the pre-main sequence, with \stellar ($\S$~\ref{sec:stellar}, App.~\ref{app:stellar}),  and an internal thermal and magnetic evolution model that is calibrated to Earth and Venus with \thermint ($\S$~\ref{sec:thermint}, App.~\ref{app:thermint}). In $\S$~\ref{sec:multi} we reproduce previous results that couple multiple modules. In $\S$~\ref{sec:conclusions} we discuss
the value of the  coupled model and how it may be used to prioritize targets for life-detection observations. Appendix \ref{app:symbols} is a list of all symbols arranged by module, and Appendix~\ref{app:support} describes \vplanet's accessibility as well as customized tools that streamline its usage.

\section{The \texttt{VPLanet} Algorithm\label{sec:vplanet}}

Models of planetary system evolution must be both comprehensive enough to simulate
a planetary system with an arbitrary architecture, as well as flexible enough to ensure that only appropriate physics are applied to individual system members. The \vplanet approach is to engineer a software framework in which the user selects established models, and the executable assembles the appropriate subroutines to calculate the evolution.  After checking for consistent input, the code simulates the entire system by solving the relevant equations simultaneously. This approach provides a simple interface and, more importantly, the opportunity to isolate important processes by easily turning on/off certain physics without needing to recompile. For example, the role of radiogenic heating in a planet's evolution can be toggled by simply removing the module name (\radheat) from the appropriate line in the input file.

\begin{figure}[ht]
\begin{center}
\includegraphics[width=0.45\textwidth,trim={1cm 7cm 7cm 1cm}]{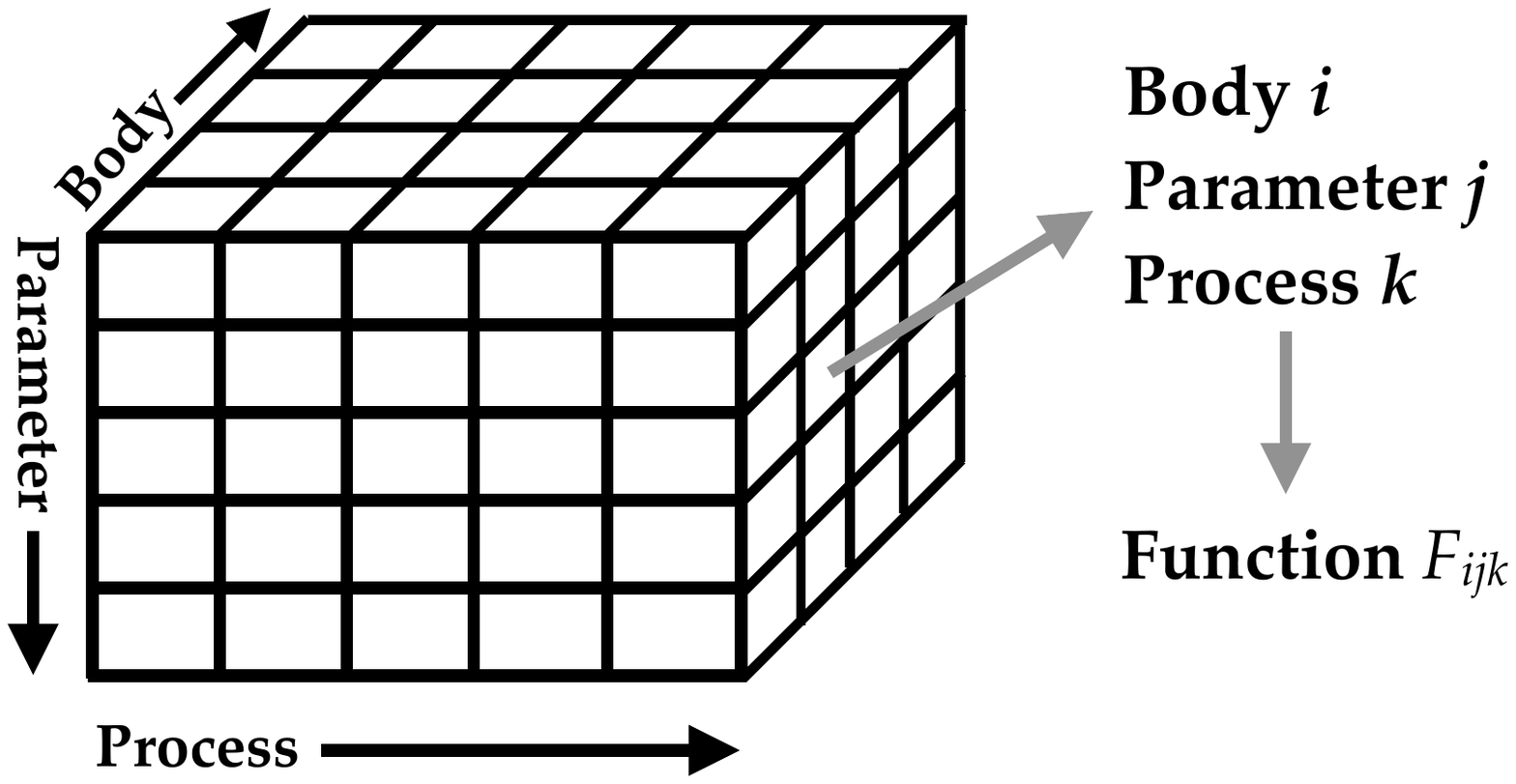}
\end{center}
\caption{\label{fig:matrix} Schematic of the matrix of function pointers used by \vplanet to integrate a system forward. Each element points to a function that typically contains an ODE. During each timestep, \vplanet loops through each body, parameter, and process to solve the entire system simultaneously.}
\end{figure}

The key software design that permits this flexibility is the use of arrays and
matrices of function pointers, \ie the elements contain memory addresses of functions. With this approach \vplanet reads in the modules requested by
the user and assembles the appropriate governing equations into a multi-dimensional
matrix in which one dimension corresponds to the bodies, the second to variables, and
the third to processes that modify the variables, see Fig.~\ref{fig:matrix}. Parameters
that are directly calculated are called ``primary variables'' and all other parameters are derived from them.

\vplanet is written in C, whose standard version includes the function pointer matrices described above. Moreover, C generates an executable that is usually as fast or faster than any other computer language, and speed of calculation is critical for the high-dimensional problem of planetary system evolution. Note that all modules are in C, but the support scripts, described in more detail in Appendix~\ref{app:support}, are written in python.

The modules described in the next 11 sections all solve ODEs or are explicit functions of time (with the exception of \poise, which solves PDEs in latitude and time). Thus, to couple the modules together
and simulate diverse phenomena, \vplanet loops over the dimensions of the function
pointer matrix and solves the equations simultaneously. For parameters influenced by
multiple processes, \vplanet calculates the derivatives and then sums them to obtain the
total derivative.

As an
example, consider a system consisting of a star, a close-in planet (\ie with an orbital period less than 10 days), and a more distant planetary companion. The star undergoes stellar evolution, the inner planet and the star experience mutual tidal effects, and the two planets gravitationally perturb each other. If the user applies the \stellar and \eqtide modules to the star, the \eqtide, \distorb, and \distrot modules to the inner planet, and the \distorb and \distrot modules for the outer planet, then \vplanet will simulate the coupled stellar-orbital-tidal-rotational evolution of the 3 objects. In this case, the eccentricity of the inner planet is modified by tidal effects and gravitational
perturbations from the other planet, so in the matrix the inner planet's dimension for eccentricity processes would have 2 members for the two derivatives.

\vplanet integrates the system of equations using a fourth order Runge-Kutta
method with dynamical timestepping. At each timestep, the values of all the derivatives
are used to calculate a timescale for each process on each primary variable:
\begin{equation}
T_j = \eta \times \min\Big|\frac{x_j}{dx_j/dt}\Big|,
\label{eq:timestep}
\end{equation}
where $T_j$ is the the timescale of the $j$th process, $\eta$ is a number less than
1 that is tuned to provide the desired accuracy, and $x_j$ is a primary variable, \eg the
star's luminosity or a planet's obliquity.  All primary variables are updated over the same timescale $T_j$, \ie the timestep is set by the fastest-changing variable. In many cases some variables are updated at a faster cadence than necessary for convergence, but this methodology ensures that all effects are
properly modeled. Note that the timestep is computed from individual derivatives, not the summed derivatives, to ensure that all phenomena are accurately modeled. In the two-planet system described above, the derivatives for eccentricity from both tides and perturbations are calculated and included in Eq.~\ref{eq:timestep}. This approach is similar to the stand-alone code \eqtide\footnote{https://github.com/RoryBarnes/EqTide} \citep{Barnes17}, which served as the foundation for \vplanet.

\begin{figure}[ht]
\begin{center}
\includegraphics[width=0.45\textwidth,trim={8cm 0cm 8cm 0cm}]{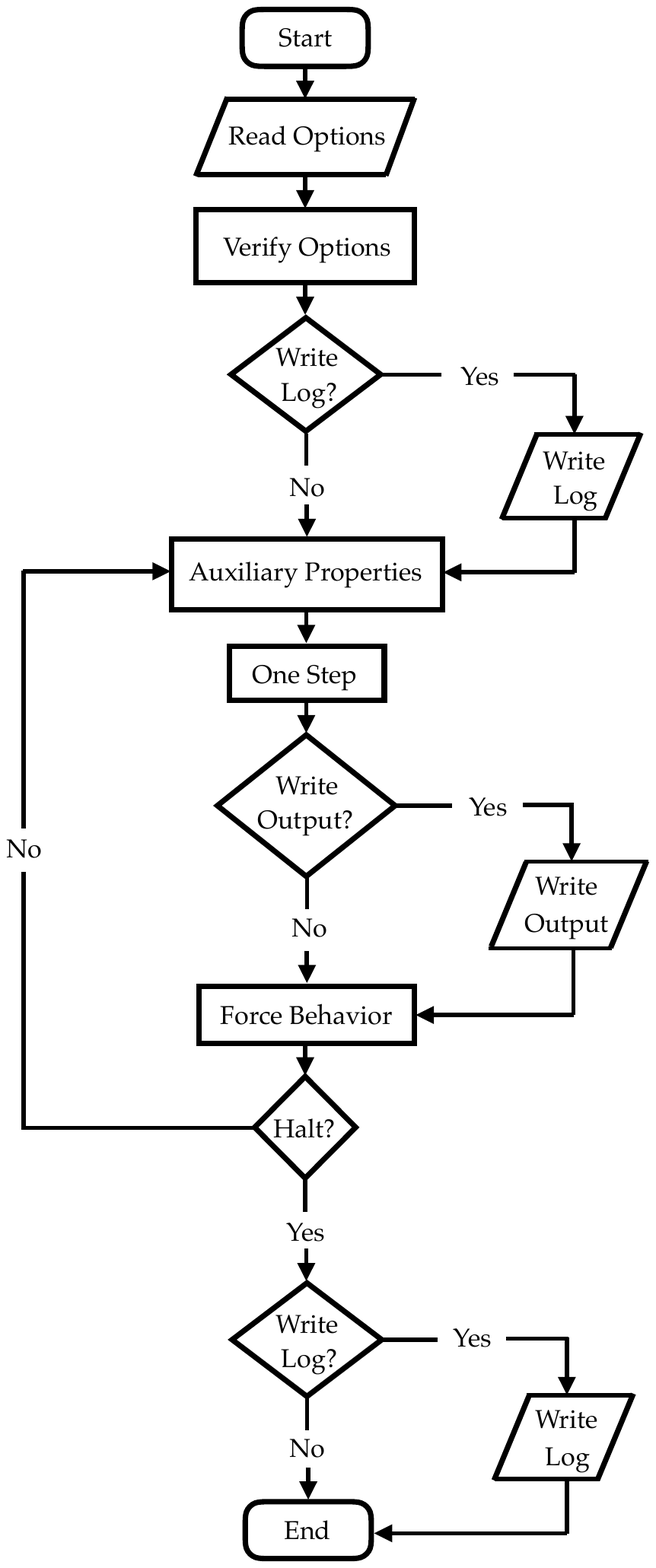}
\end{center}
\caption{\label{fig:flowchart} The \vplanet flow chart. See text for more details.}
\end{figure}

The \vplanet operational flow chart is shown in Fig.~\ref{fig:flowchart}. \vplanet first reads
in the options set by the user in $\ge2$ input files. The primary input file contains
the top-level instructions, such as integration parameters (if any), units, and
the list of the members of the system. Each system member has a ``body file'' which contains all the initial conditions, module-specific parameters, and output option selections for that object. After reading in all the options, they are vetted
for completeness and inconsistencies in a process called ``verify.''  After passing this step,
the physical state and evolution may be self-consistently calculated.

After verification, a log file can be written in which the initial state of the system
is recorded in SI units, which are the units used in all internal \vplanet calculations,
although we note that the user can specify both the input and output units in the primary input file. At this point,
if the user requested an integration, it begins. The evolution is broken up into
four parts: 1) calculation of ``auxiliary properties,'' \ie parameters needed to calculate the derivatives of the primary variables, 2) one step
(forward or backward) is taken, 3) the state of the system is written to a file if requested, 4) if necessary, changes to the matrix are implemented in ``force behavior'' (\eg if a planet becomes tidally locked, its rotational frequency derivative due to tides is removed), and 5) checks for any threshold the user set to
halt the integration are performed (\eg the user may specify that the execution should be terminated once all water is lost from a planet).  Once the integration
is complete, the final conditions of the system can be recorded in the log file.

Note that in addition to coupling modules through simultaneous solutions of ODEs and PDEs,
coupling also can occur in planetary interiors. For example, in a tidally heated terrestrial
planet, the tidal power is a function of temperature \citep[\eg][]{DriscollBarnes15},
but the temperature is also a function of the tidal power. \vplanet couples these connections during the Auxiliary Properties step in Fig.~\ref{fig:flowchart}.

In the following sections, we briefly describe and validate each of the 11 modules that control the evolution of planetary systems in \vplanet. More detailed information about each module can be found in the Appendices.

\section{Atmospheric Escape: \texttt{AtmEsc}\label{sec:atmesc}}

\begin{figure}[h!]
\begin{center}
\includegraphics[width=0.35\textwidth]{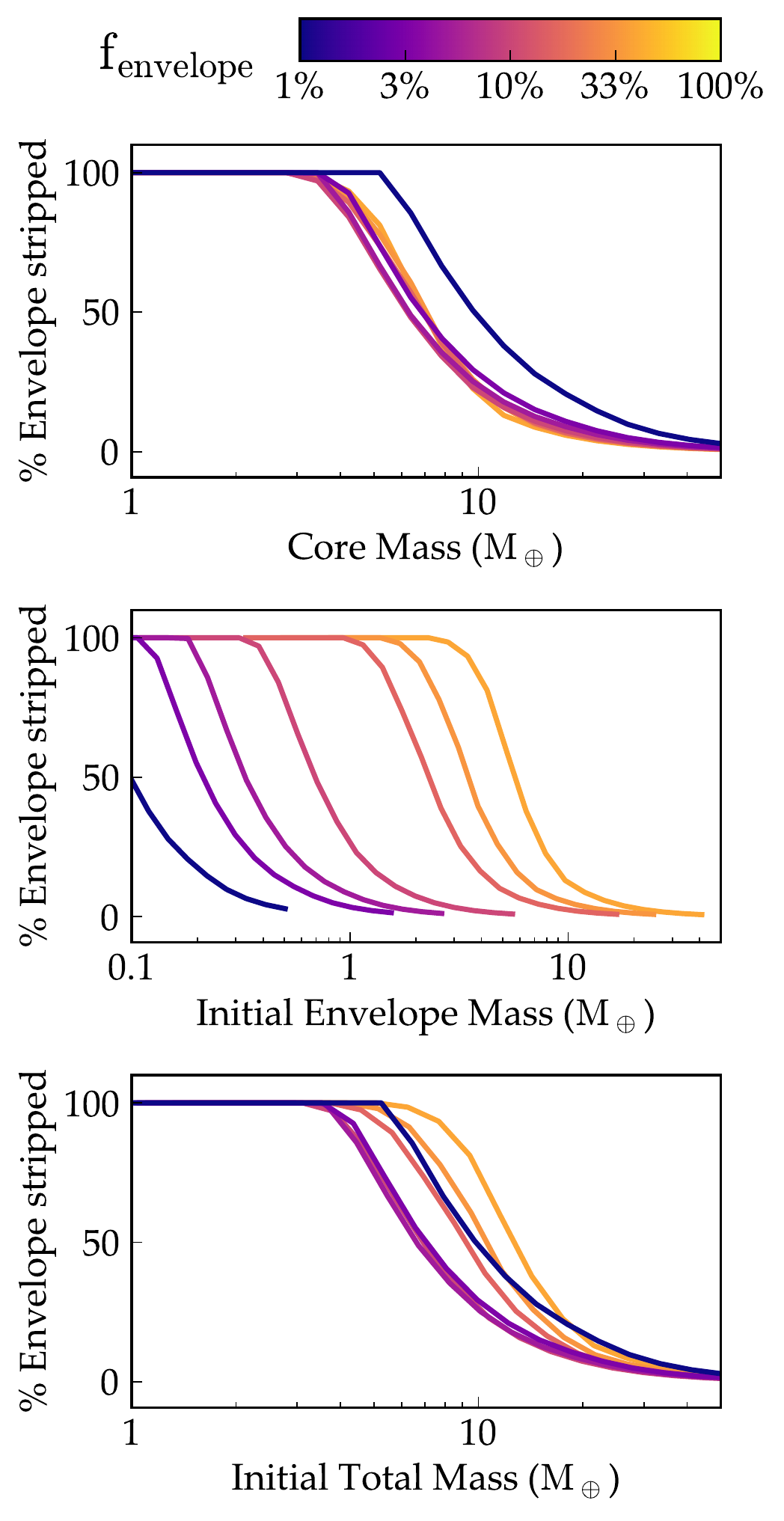}
\end{center}
\caption{\label{fig:kepler36} Reproduction of Figure 3 in \citet{Lopez2013} using \vplanet. The various curves correspond to the fraction of a planet's gaseous envelope lost to photoevaporation as a function of the planet core mass (top), the initial envelope mass (center), and the initial total planet mass (bottom). Colors correspond to different initial envelope mass fractions, ranging from 1\% (dark blue) to 45\% (dark orange). We recover the result of \citet{Lopez2013} that, at fixed instellation, the core mass shows the tightest correlation with the fraction of the envelope that is lost for a mini-Neptune. Approximate runtime: 30 seconds. \href{https://github.com/VirtualPlanetaryLaboratory/vplanet/tree/master/examples/AtmEscKepler-36}{\link{examples/AtmEscKepler-36}}}
\end{figure}

The erosion of a planet's atmosphere due to extreme stellar
radiation is among the biggest challenges to its habitability,
particularly around low mass stars \citep[\eg][]{Lissauer07,Scalo07,LugerBarnes15}.
The \atmesc module models the escape of planetary atmospheres and their surface volatiles following
simple parametric energy-limited and diffusion-limited prescriptions,
which are discussed in detail in Appendix~\ref{app:atmesc}. In order to validate our approach, in Figure~\ref{fig:kepler36} we present a reproduction of Figure~3 in \citet{Lopez2013} using \atmesc. In that study, the authors used a coupled thermal evolution~/~photoevaporation model to explain the density dichotomy in the Kepler-36 system, which hosts two highly irradiated planets: a low-density mini-Neptune and a high-density super-Earth. The authors showed how, at fixed instellation, the initial core mass dictates the evolution of the gaseous envelope of a sub-giant planet: planets with high-mass cores hold on to their envelopes, while those with low-mass cores are more easily stripped by photoevaporation. As in \citet{Lopez2013}, we plot the percentage of  b's gaseous envelope that is lost as a function of the core mass (top panel), initial envelope mass (center panel), and initial total mass (bottom panel) for different initial envelope mass fractions, assuming an escape efficiency $\epsilon_{XUV} = 0.1$, a planet age of 5 Gyr, instellation one hundred times that received by Earth, and the XUV evolution model of \citet{Ribas05} for a solar-mass star. We model the planet's radius with the evolutionary tracks for super-Earths of \cite{Lopez12} and \cite{LopezFortney14}.

We find, as those authors did, that the core mass displays the tightest correlation with the fraction of the envelope that is lost. Our values generally agree, although \atmesc predicts slightly higher escape rates at lower core mass. For instance, in the top panel, \atmesc finds that planets with core masses up to $\sim 3-4 \mathrm{M}_\oplus$ are completely stripped for all values of the envelope fraction; in the top panel of Figure 3 in \citet{Lopez2013}, only planets with masses less than $\sim 2-3 \mathrm{M}_\oplus$ are completely evaporated. This result is likely due to the fact that the study of \citet{Lopez2013} employs a fully coupled thermal evolution/atmospheric escape model, in which the simulation tracks the evolution of the internal entropy and luminosity of the planet and any feedbacks with the escape process. In contrast, \atmesc makes use of pre-computed radius grids as a function of mass and irradiation to calculate the atmospheric escape rate, and so is unable to capture feedback effects from the escape on the thermal evolution. However, at higher core mass the difference between the two studies is on the order of tens of percent, well within the uncertainties in parameters like $\epsilon_{XUV}$ and other observational constraints.

\section{Circumbinary Planet Orbits: \texttt{BINARY}\label{sec:binary}}

The recent discovery of transiting circumbinary planets (CBPs) by \textit{Kepler} provides intriguing laboratories to probe the orbital dynamics of such systems.  Non-axisymmetric gravitational perturbations from the central binary force CBPs into oscillating orbits that display short and long-term non-Keplerian behavior. The \binary module computes the orbit of a massless test particle on a circumbinary orbit using the analytic theory derived by \citet{Leung2013}. By generalizing the work of \citet{LeePeale2006} to the case of an eccentric central binary orbit, \citet{Leung2013} modeled the orbit of a massless CBP as a combination of the circular motion of the CBP's guiding center with radial and vertical epicylic oscillations induced by non-axisymmetric components of the central binary's gravitational potential. We discuss the full model and implementation details of the \citet{Leung2013} theory in App.~\ref{app:binary}.

\begin{figure*}[t!]
\includegraphics[width=\textwidth]{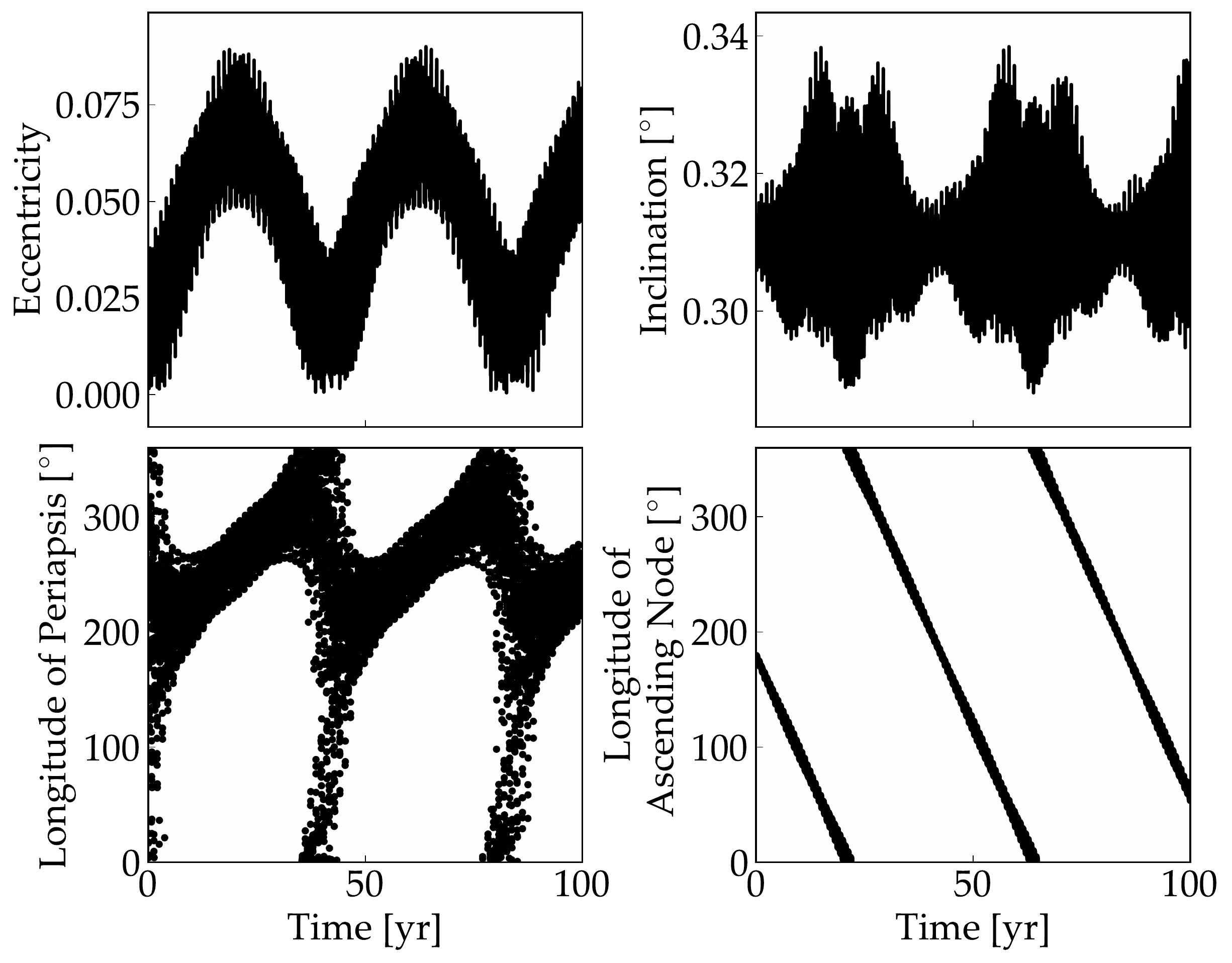}
\label{fig:binaryComp}
\caption{\vplanet simulation of the orbital evolution of Kepler-16b using the \binary module, which may be compared to Figure 4 of \citet{Leung2013}.  All CBP orbital elements are measured in Jacobi coordinates. \textit{Top left:} CBP orbital eccentricity versus time. \textit{Top right:} CBP orbital inclination relative to the plane of the binary versus time. \textit{Bottom left:} CBP longitude of periapse versus time. \textit{Bottom right:} CBP longitude of ascending node versus time.  Approximate runtime: 25 seconds. \href{https://github.com/VirtualPlanetaryLaboratory/vplanet/tree/master/examples/CircumbinaryOrbit}{\link{examples/CircumbinaryOrbit}}}
\end{figure*}

\citet{Leung2013} validated their analytic formalism against direct N-body simulations, and here we validate our implementation of their analytic theory by reproducing their Figure 4 that depicts the orbital evolution of the CBP Kepler-16 b \citep{Doyle2011}.  For the initial conditions, we used the orbital parameters for both the binary and CBP given in Table 1 in \citet{Leung2013} and set the CBP's $e_{\text{free}} = 0.03$ following \citet{Leung2013}. The results of our validation simulation are shown in Fig.~\ref{fig:binaryComp} and are in excellent agreement with Figure 4 from \citet{Leung2013}. We further validate \binary by comparing the intermediate quantities used in the \citet{Leung2013} theory (see Appendix~\ref{app:binary}) for the Kepler-16 system in Table~\ref{tab:binary}. We find that \binary perfectly reproduces nearly all of the intermediate quantities from \citet{Leung2013}. The maximum error between the \binary values and \citet{Leung2013} is $0.02\%$ on $n_k$, a negligible difference.

\begin{deluxetable}{lcc}
\tabletypesize{\small}
\tablecaption{\binary Intermediate Values for Kepler-16} \label{tab:binary}
\tablewidth{0pt}
\tablehead{
\colhead{Parameter [units]} & \colhead{\citet{Leung2013}} & \colhead{\binary}
}
\startdata
$R_0$ [AU] & $0.7016$ & $0.7016$ \\
$n_k$ [yr$^{-1}$] & $10.0823$ & $10.0839$ \\
$n_0/n_k$ & $1.00702$ & $1.00702$ \\
$\kappa_0/n_k$ & $0.99224$ & $0.99224$ \\
$\nu_0/n_k$ & $1.02158$ & $1.02158$ \\
$C_0$ & $0.000159$ & $0.000160$ \\
$C^0_1$ & $-0.000282$ & $-0.000282$ \\
$C^0_2$ & $-0.000589$ & $-0.000589$ \\
$C^0_3$ & $-0.000049$ & $-0.000049$ \\
$C^+_1$ & $0.000005$ & $0.000005$ \\
$C^+_2$ & $-0.000033$ & $-0.000033$ \\
$C^+_3$ & $-0.000006$ & $-0.000006$ \\
$C^-_1$ & $0.035772$ & $0.035776$ \\
$C^-_2$ & $0.002438$ & $0.002440$ \\
$C^-_3$ & $0.000110$ & $0.000110$ \\
\enddata \vspace*{0.1in}
\tablecomments{Validation values taken from Table 2 in \citet{Leung2013}.}
\end{deluxetable}

\section{Approximate Orbital Evolution: \texttt{DistOrb}\label{sec:distorb}}
\begin{figure*}
    \centering
    \includegraphics[width=\textwidth]{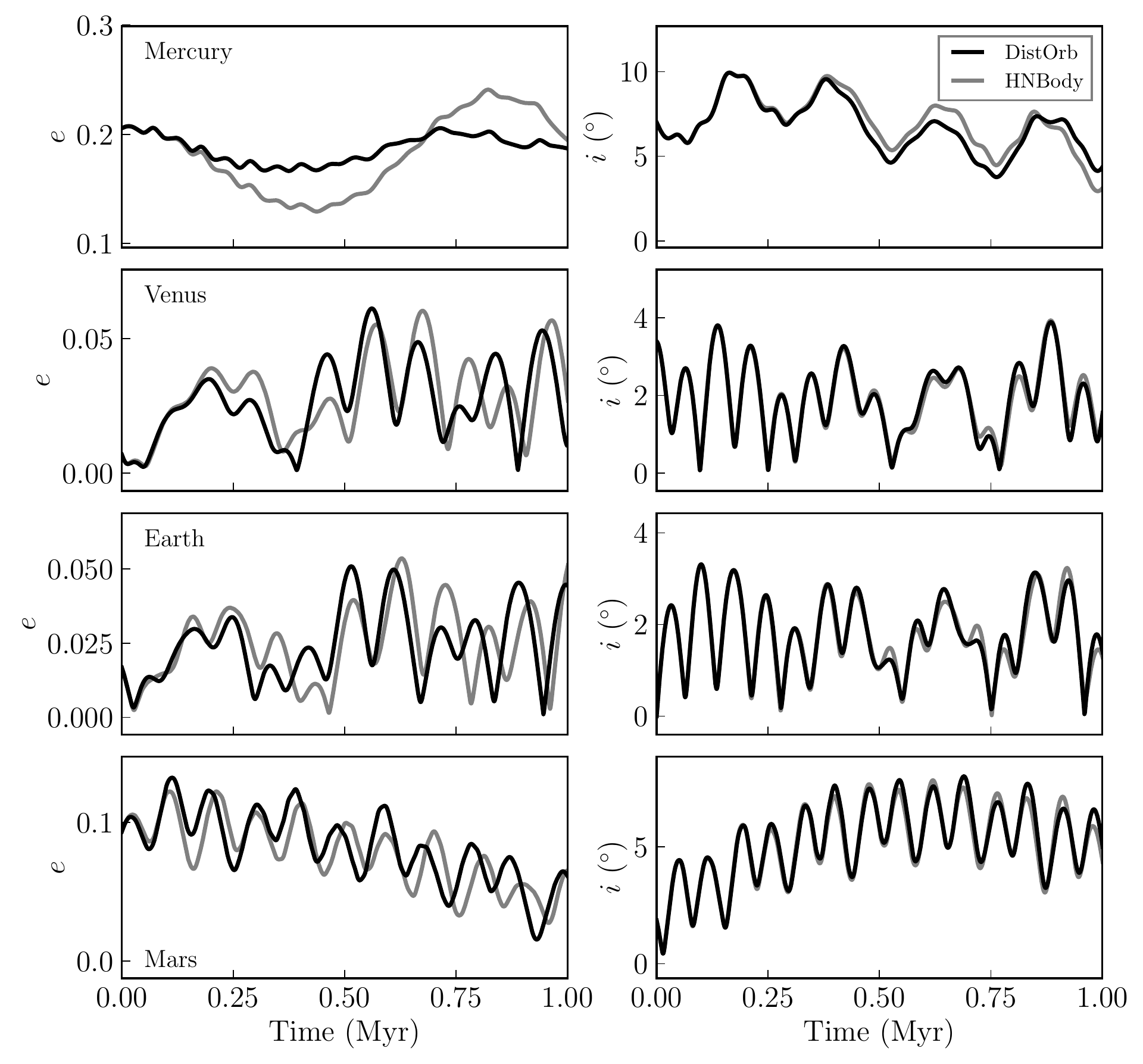}
    \caption{Eccentricity evolution (left) and inclination evolution (right) for the inner solar system planets over the next 1 Myr. Black is the result for \vplanet's \distorb module and grey is for the N-body code \hnbody. Initial conditions are taken from Appendix A of \cite{MurrayDermott99}. Approximate runtime: 5.5 minutes. \href{https://github.com/VirtualPlanetaryLaboratory/vplanet/tree/master/examples/SSDistOrbDistRot}{\link{examples/SSDistOrbDistRot}}}
    \label{fig:innersol}
\end{figure*}

\begin{figure*}
    \centering
    \includegraphics[width=\textwidth]{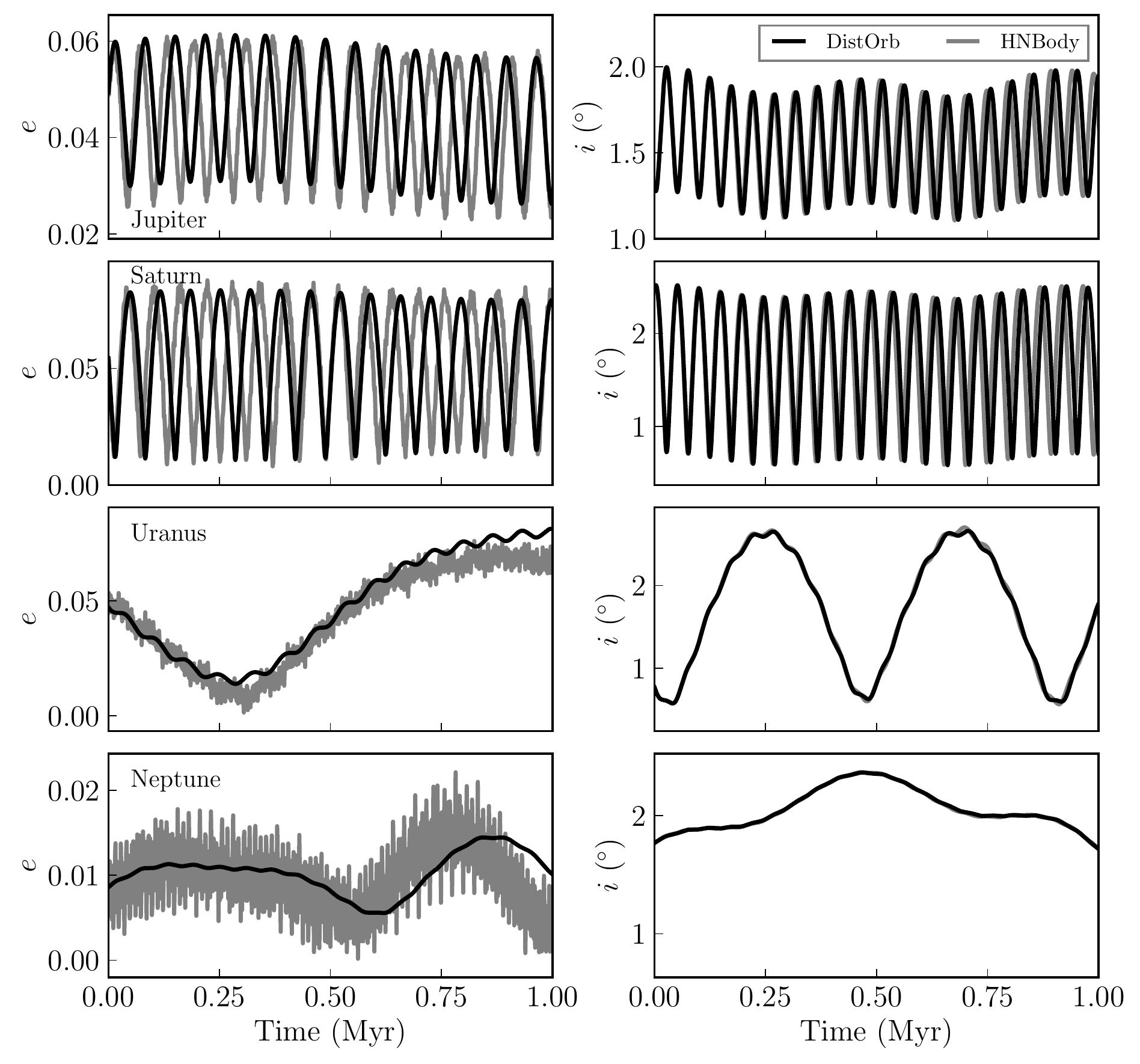}
    \caption{Eccentricity evolution (left) and inclination evolution (right) for the outer solar system planets over the next 1 Myr. Black is the result for \vplanet's \distorb module and grey is for the N-body code \hnbody. Initial conditions are taken from Appendix A of \cite{MurrayDermott99}. \href{https://github.com/VirtualPlanetaryLaboratory/vplanet/tree/master/examples/SSDistOrbDistRot}{\link{examples/SSDistOrbDistRot}}}
    \label{fig:outersol}
\end{figure*}

\begin{figure*}
\centering
\includegraphics[width=\textwidth]{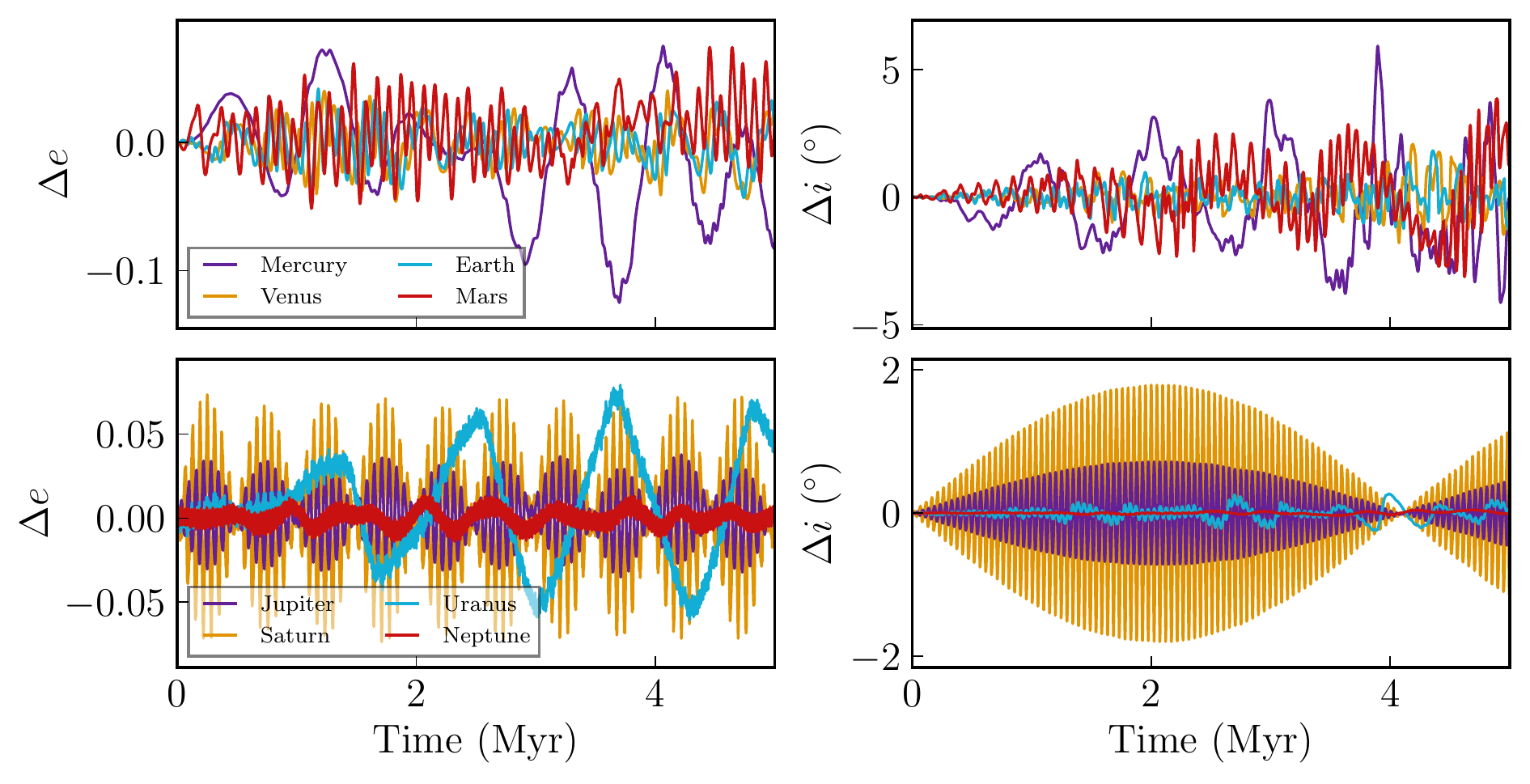}
\caption{Absolute errors in the eccentricity (left) and inclination (right) between \distorb and \hnbody over 5 Myr. Upper panels show the inner solar system planets and lower panels show the outer planets. \href{https://github.com/VirtualPlanetaryLaboratory/vplanet/tree/master/examples/SSDistOrbDistRot}{\link{examples/SSDistOrbDistRot}}}
    \label{fig:disterror}
\end{figure*}

An approximate solution to the orbital evolution of a planetary
system can be derived from a quantity known as the disturbing function, the non-Keplerian component of the gravitational potential in a multi-body system. The disturbing function is most useful when written as a Fourier expansion in the orbital elements. First derived by Lagrange and Laplace \citep[see, for example, chapter 7 of][]{MurrayDermott99}, this approach produces ODEs for the evolution of orbital parameters. Outside of mean motion resonances, an orbit-averaged, or secular, disturbing function can be used, which has the advantage that large
time steps (hundreds of years) can be taken. The \distorb module in \vplanet is based on the
fourth order (in eccentricity and inclination)
secular solution derived in \cite{MurrayDermott99} and \cite{Ellis2000}. The theory is described in detail
in Appendix~\ref{app:distorb}; here, we present results from the model.

Figures \ref{fig:innersol} and \ref{fig:outersol} show the orbital
evolution of the inner and outer solar system planets, respectively, as calculated by \distorb and
\hnbody. The latter software package is an N-body code that calculates the gravitational evolution from first principles \citep{RauchHamilton02}. For the \distorb runs, we used the fourth-order integration (a second order
Lagrange-Laplace solution can also be utilized
in the code; see Appendix~\ref{app:distorb}). Here we
compare to \hnbody because that model also
contains the general relativistic corrections,
however, for the solar system these effects are
small and \hnbody results appear almost identical
to other N-body integrators.

The inclination evolution in \distorb compares
extremely well with \hnbody. The eccentricity evolution compares reasonably well for most of
the planets; the largest error is in the amplitude
of Mercury's eccentricity variation. Figure \ref{fig:disterror} shows the absolute errors in the eccentricity and inclination evolution
between the two models. The errors are largest for Mercury and Mars after $\sim 3$ Myr. The errors in $e$ and $i$ for Mercury, Mars, and Uranus grow in time. For Earth and Venus, the error also grow in time but remain smaller. The errors for Jupiter, Saturn, and Neptune are periodic and are explained easily by a slight mis-match in frequencies---this leads to a drift in the relative phase between the solutions, producing errors that are periodic and stable.

We do not
expect to perfectly reproduce the N-body solution with this secular model, as the Solar System is affected by the proximity of
Jupiter and Saturn to a 5:2 mean-motion resonance \citep{Lovett1895}. The other source of error is
Mercury's relatively large eccentricity ($e\sim0.2$), which the fourth-order model
does not handle as well as the direct N-body solution. For the Solar System, \distorb performs as well as previous studies \citep{MurrayDermott99}. Since the orbital elements of the planets are known with a
high degree of precision, the N-body solution is clearly desirable. However, for exoplanetary systems, for which the errors in $e$ and $i$ are quite large, \distorb offers a computational advantage: the above simulation of eight planets runs in $\sim 200$ s on a modern CPU, about 10 times faster than \hnbody. This speed allows for broader explorations of parameter space. For systems which are near to mean-motion resonances or which have tight constraints on the orbital parameters (as in our Solar System), \spinbody may be preferred over \distorb.

\section{Rotational Evolution from Orbits and the Stellar Torque: \texttt{DistRot}\label{sec:distrot}}

\begin{figure*}
    \centering
    \includegraphics[width=\textwidth]{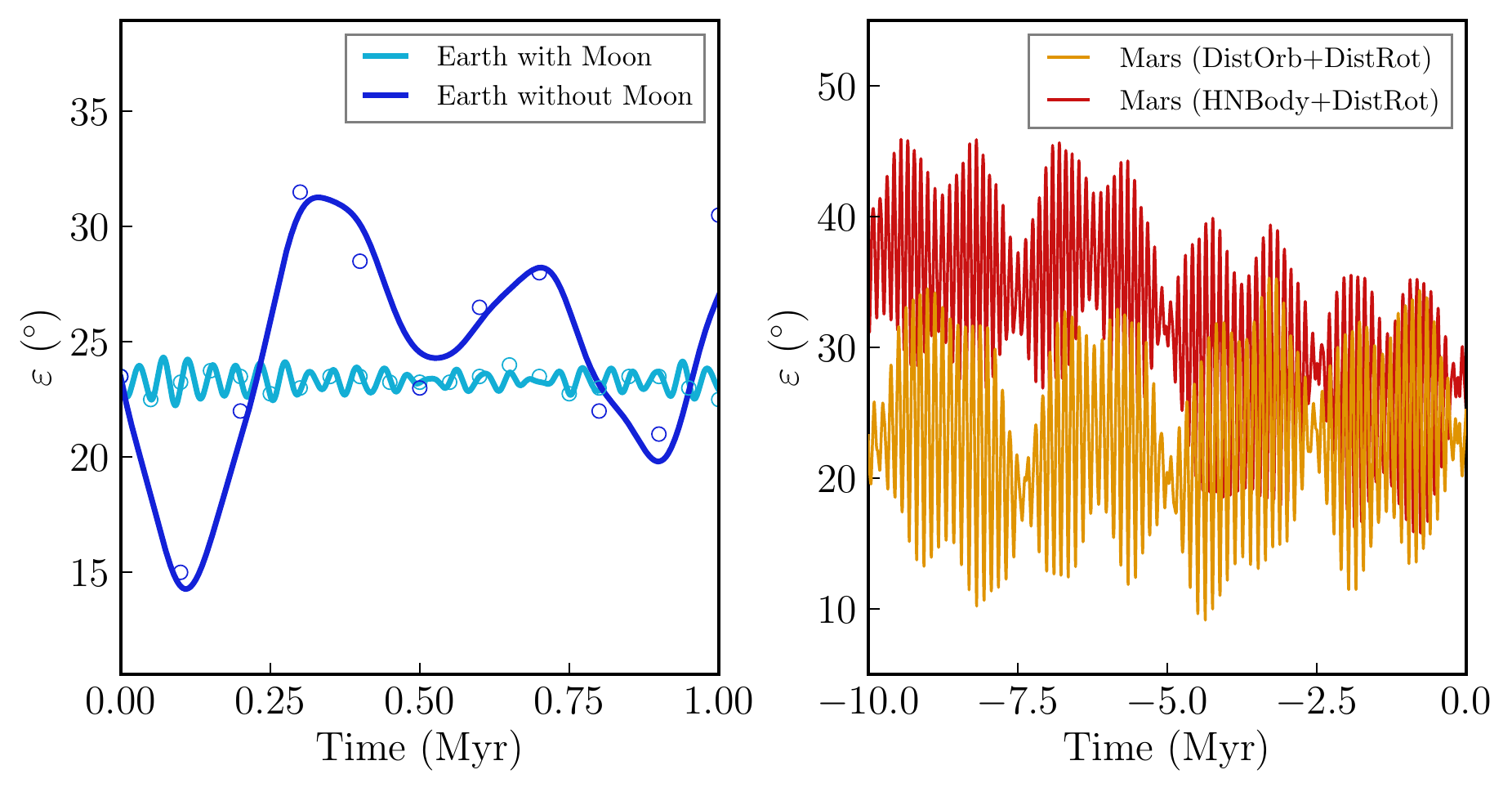}
    \caption{Obliquity evolution of Earth with and without the Moon over the next Myr (left) and Mars using secular and N-body models for the orbit over the last 10 Myr (right). The circular points are reconstructed from Figures 10 and 11 of \cite{Laskar1993}. \href{https://github.com/VirtualPlanetaryLaboratory/vplanet/tree/master/examples/SSDistOrbDistRot}{\link{examples/SSDistOrbDistRot}}}
    \label{fig:obliquity}
\end{figure*}

There are a number of physical processes that affect the position
of a planet's spin axis. The module \distrot captures the physics of two
processes: the torque acting on the equatorial bulge by the host star,
and the motion of the planet's orbital plane (\eg the change in inclination).
The model was derived in \cite{Kinoshita1975} and \cite{Kinoshita1977}; see
Appendix~\ref{app:distrot} for details.

Figure 6 shows the obliquity evolution for Earth and Mars using \distrot. We show the obliquity
for Earth over the next Myr with and without the effect of the
Moon, which compares extremely well with \cite{Laskar1993}, Fig.~11. For the case including the Moon, the relative error between our solution and that of \cite{Laskar1993} is $\sim 0.6\%$ on average, with a maximum of $\sim 2\%$. For the case without, the relative error is $\sim 5\%$ on average, with a maximum $\sim 13\%$ (on the final point at 1 Myr). The larger error for the latter comes from a slight mis-match in the frequencies and the larger amplitude of the oscillation. Note that we do not directly include the
effect of the Moon---here, the effect is mimicked by forcing Earth's precession rate to the known value,
$50.290966 ''/$yr \citep{Laskar1993}. This forced precession rate can be modified to recreate the effects of an arbitrary moon.

Mars is more challenging because
its obliquity is known to be very sensitive to
orbital frequencies \citep{Ward91,Ward1992,Touma1993,Laskar04}. We show the obliquity evolution utilizing two
methods for the orbital evolution over the last 10 Myr. The first couples \distrot directly to
\distorb. In this case, the obliquity evolution over the last $\sim 5$ Myr compares well with
previous studies \citep[\eg][Fig.~1]{Touma1993}, but it does not contain the well known shift to a higher obliquity state at
$\sim 5$ Ma, owing to an imperfect representation of the orbital frequencies (see below). Still, we match the $\sim 20^{\circ}$ oscillation with periods of $\sim 120$ kyr, $\sim 1.25$ Myr, and $\sim 2.5$ Myr \citep{Ward1992, Touma1993}. In the second case, we use the
orbital evolution from \hnbody as input into \distrot. This second case does produce the $5$ Ma
obliquity shift, indicating that the problem lies with the accuracy of the orbital model, not with
\distrot itself.

\section{Tidal Effects: \texttt{EqTide}\label{sec:eqtide}}

Tidal effects modify a planet's orbit, rotational properties, and internal power. The \eqtide module employs the equilibrium tide models originally developed by \cite{Darwin1880}. We specifically use formulations called the constant-phase-lag (CPL) and constant-time-lag (CTL) models developed by \cite{FerrazMello08} and \cite{Leconte10}, respectively. See Appendix~\ref{app:eqtide} for more details on these models.

\begin{figure*}[ht]
    \centering
    \includegraphics[width=\textwidth]{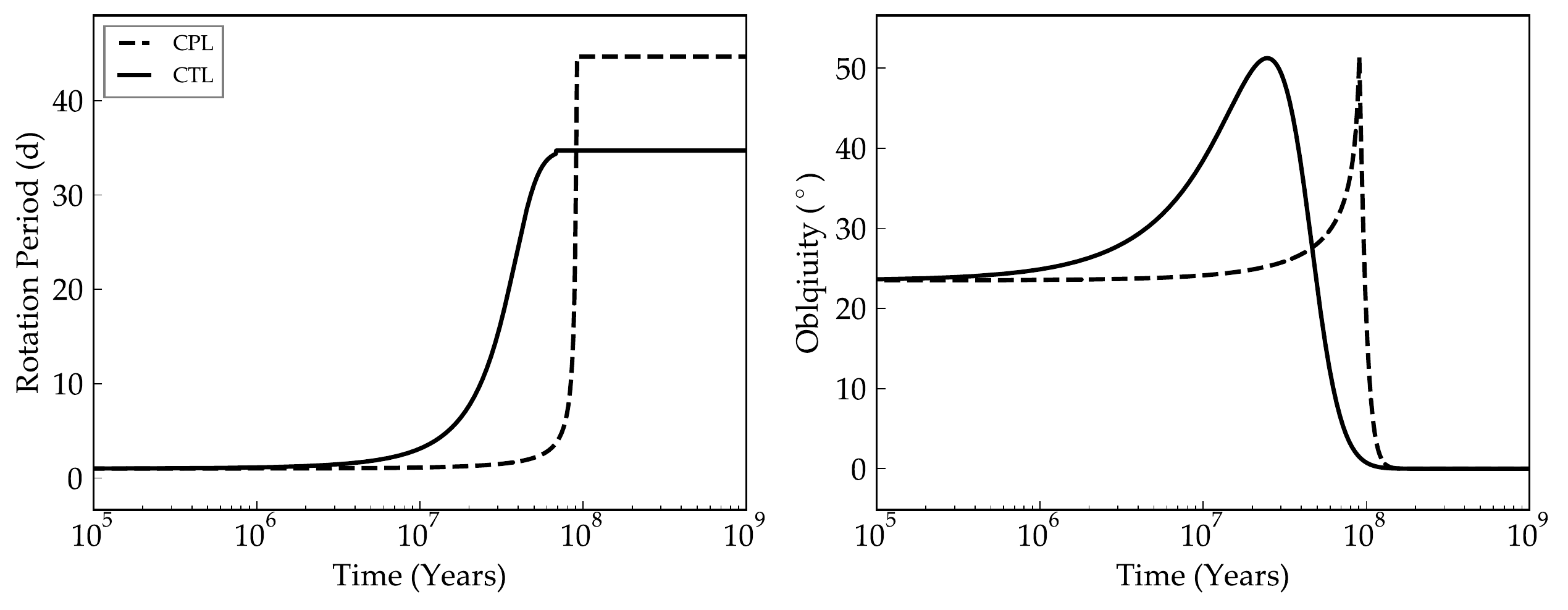}
    \caption{{\it Left:} Rotational period evolution of Gl 581 d for the CPL (dashed curve) and CTL (solid curve) equilibrium tide models and assuming an initial rotation period of 1 day, a tidal Q of 100, and an obliquity of 23.5 degrees. {\it Right:} Same, but for obliquity. Approximate runtime: 10 seconds. \href{https://github.com/VirtualPlanetaryLaboratory/vplanet/tree/master/examples/TideLock}{\link{examples/TideLock}}}
    \label{fig:TideLock}
\end{figure*}

In Fig.~\ref{fig:TideLock}, we show the rotational evolution of the putative exoplanet Gl 581 d \citep{Udry07}, which may in fact be an artifact and not actually a planet \citep{Robertson14}. We nonetheless consider this example as it was examined by \cite{Heller11} and thus provides a straightforward validation of \eqtide. Physical and orbital parameters are listed in Table~\ref{tab:tides} \citep{Udry07}.  Fig.~\ref{fig:TideLock} is very similar to Fig.~6 in \cite{Heller11} in which CPL is labeled ``FM08'' and CTL is ``Lec10,'' with differences less than 1\%, most likely due to updates of fundamental constants \citep{Prsa2016}.

\begin{deluxetable}{lcccc}
\tabletypesize{\small}
\tablecaption{Properties of Selected Tidally Perturbed Worlds \label{tab:eqtide}}
\tablewidth{0pt}
\tablehead{
\colhead{Parameter} & \colhead{Gl 581} & \colhead{Gl 581 d} & \colhead{Jupiter} & \colhead{Io$^a$}
}
\startdata
$m_p$ ($\mearth$) & 1.03 $\times 10^5$ & 5.6 & 317.828 & 0.015\\
$r_p$ ($\rearth$) & 31.6 & 1.6 & 11.209 & 0.286\\
$a$ ($\rearth$) & - & 5123 & - & 66.13\\
$e$ & - & 0.0549 & - & 0.38\\
$P$ (d) & 94.2 & 1 & 0.47 & \\
$\epsilon$ ($^\circ$) & 0 & 23.5 & 3.08$^a$ & 0.0023\\
$r_g$ & 0.5 & 0.628 & 0.5 & 0.27\\
$k_2$ & 0.5 & 0.024059 & 0.3 & 1.5\\
$Q$ & $10^6$ & 100 & $10^5$ & 100\\
\enddata \vspace*{0.1in}
\label{tab:tides}
\tablenotetext{a}{https://ssd.jpl.nasa.gov/?sat\_phys\_par}
\end{deluxetable}

\begin{figure}[hb]
    \centering
    \includegraphics[width=0.45\textwidth]{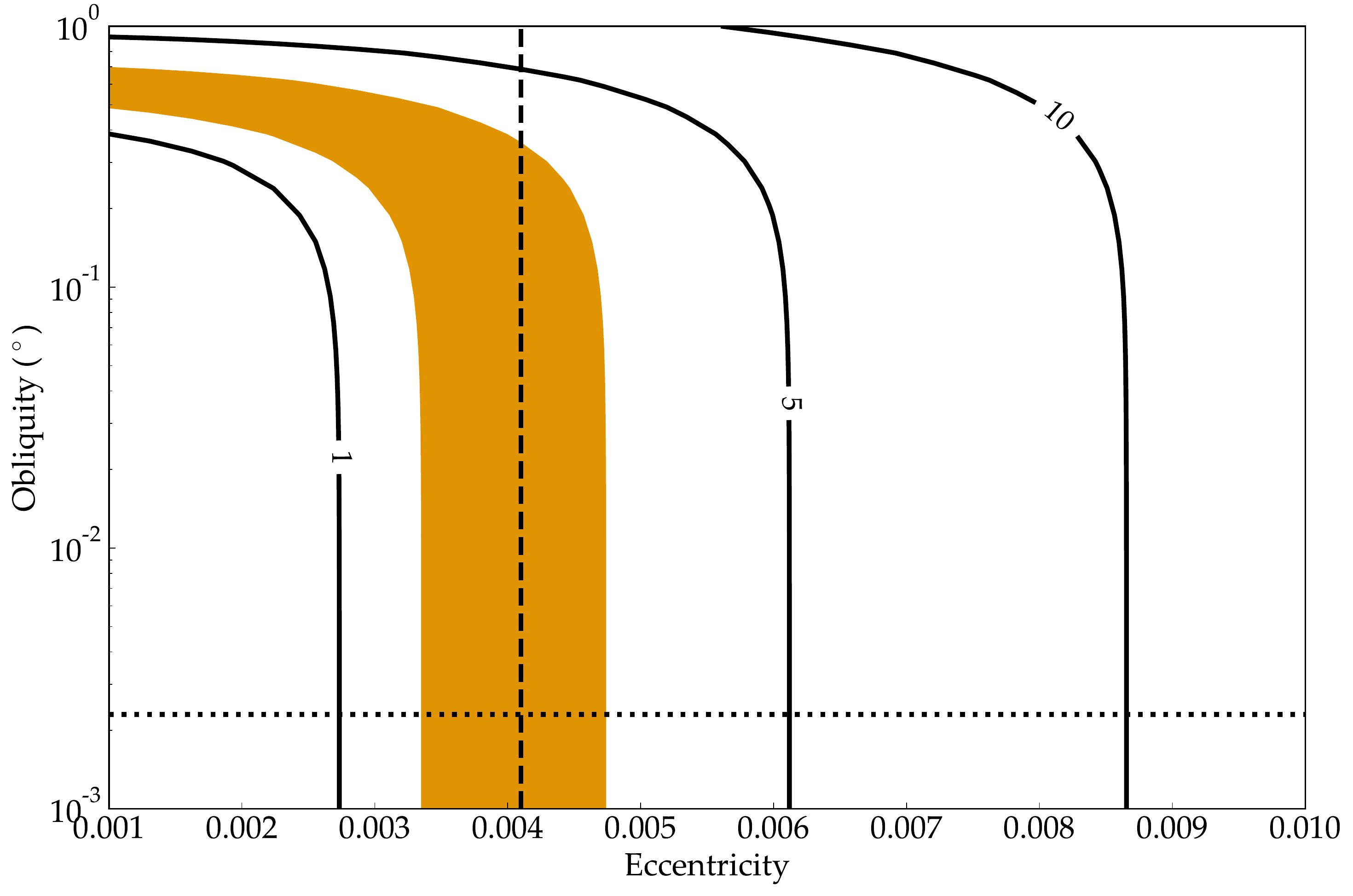}
    \caption{Surface tidal heat flux of Io as a function of $e$ and $\varepsilon$. Contour units are W/m$^2$, the vertical line corresponds to Io's observed eccentricity, and the horizontal line is the expected obliquity if Io is in a Cassini state \citep{BillsRay00}. The orange shaded region corresponds to the observed value of 1.5--3 W/m$^2$ \citep{Veeder12}. Approximate runtime: 1 minute. \href{https://github.com/VirtualPlanetaryLaboratory/vplanet/tree/master/examples/IoHeat}{\link{examples/IoHeat}}}
    \label{fig:TideHeatIo}
\end{figure}

In Fig.~\ref{fig:TideHeatIo} we show the tidal heating surface flux of Io as a function of $e$ and $\varepsilon$.  The current heat flux is 1.5 -- 3~\wpmsq~\citep{Veeder94,Veeder12}, and the physical and orbital parameters of Jupiter and Io are listed in Table \ref{tab:tides}. Io's orbital eccentricity has been damped to 0.004, and is likely in a Cassini state, \citep[see $\S$~\ref{sec:multidamp} and][]{BillsRay00}, with an obliquity of 0.0023$^\circ$, a displacement that remains below the detection threshold. The predicted heat flow of Io is about 2--3 times higher than observed, which has led some researchers to speculate that Io's heat flow is not in equilibrium \citep{Moore03}, although it is also likely that the equilibrium tide formalism does not accurately reflect Io's tidal response. Nonetheless, \vplanet successfully reproduces Io's surface energy flux.

\section{Galactic Evolution: \texttt{GalHabit}\label{sec:galhabit}}

\begin{figure}[hb]
    \centering
    \includegraphics[width=0.45\textwidth]{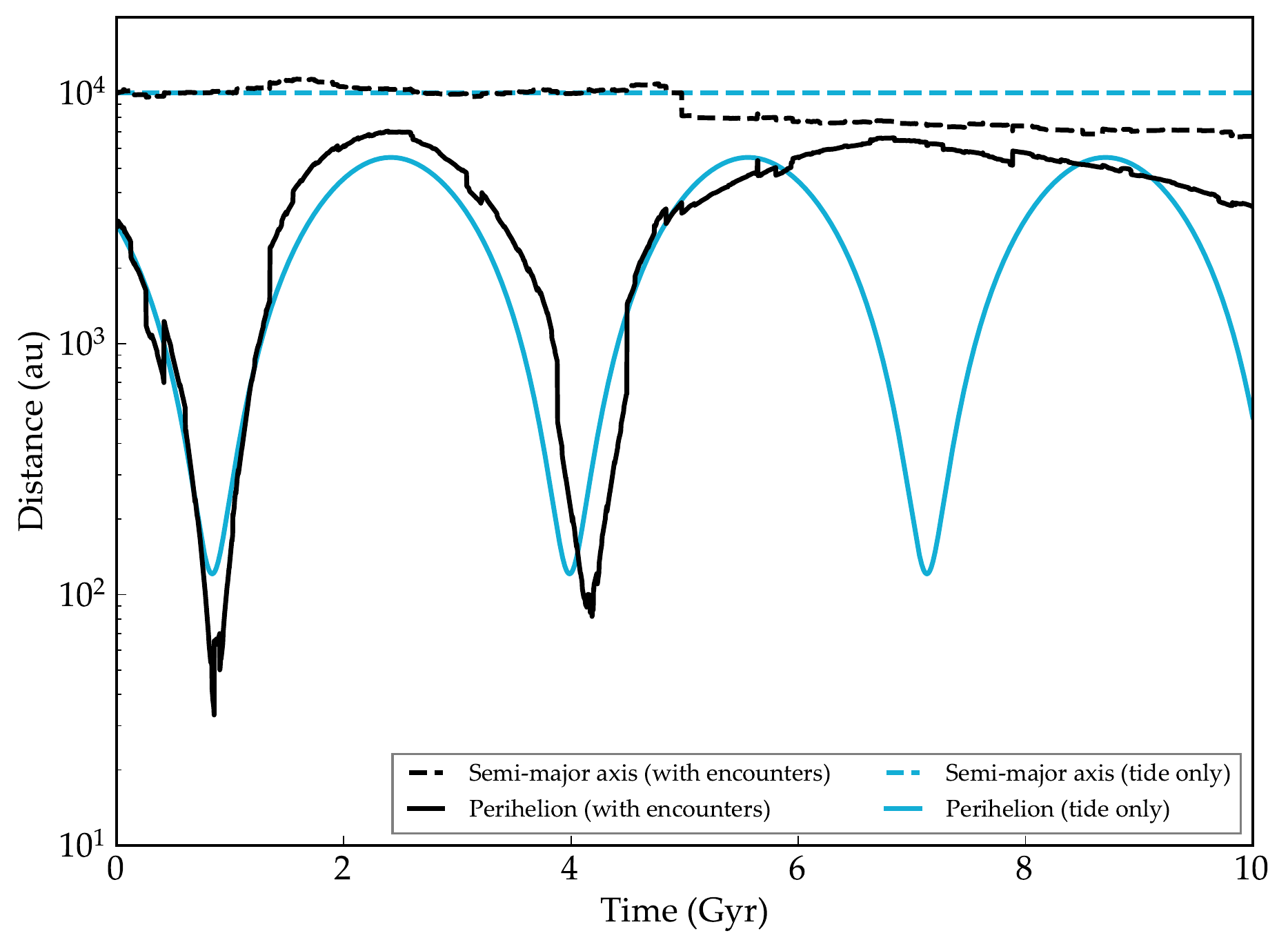}
    \caption{Evolution of an M dwarf orbiting the \text{Sun} under the influence of the galactic environment. The blue curves represent the evolution due to the galactic tide alone, while the black includes the effects of random stellar encounters. The dashed curves are the semi-major axis (unaffected by the tide); the solid curves are the perihelion distance. Note that stellar encounters are random, so the black curves are only qualitatively reproducible. Approximate runtime: 6 minutes. \href{https://github.com/VirtualPlanetaryLaboratory/vplanet/tree/master/examples/GalaxyEffects}{\link{examples/GalaxyEffects}}}
    \label{fig:galhabit}
\end{figure}

The \galhabit module accounts for two effects of the galactic environment on the orbits of binary star systems:
the galactic tide and perturbations from passing stars. Such effects have been shown to impact the
stability of planetary systems \citep{Kaib13} and so should be considered when studying planets in wide binary systems.
We utilize the secular approach of \cite{HeislerTremaine86} for the galactic tide and model stellar encounters following the Monte-Carlo formulations in \cite{Heisler87} and \cite{Rickman08}.

We model the evolution of an M dwarf orbiting the sun, with a mass of $0.12$ M$_{\odot}$, $a = 10000$ au, $e = 0.7$, and $i=80^{\circ}$. This approach is similar to the simulation used in Figure 1 of \cite{Kaib13}. Two examples are shown: the evolution of the orbit under the galactic tide alone and the evolution with both the tide and random stellar encounters. We cannot reproduce the evolution shown in \cite{Kaib13} exactly because our stellar encounters are drawn randomly. Nonetheless, we qualitatively recover the behavior of such a system. The timescale of tidal evolution is approximately 3 Gyr, similar to the simulation shown in Figure 1 of \cite{Kaib13}. The order of magnitude of the change in semi-major axis due to impulses is on the order of $100-1000$ au, also similar to \cite{Kaib13}.

We have tested the impulse approximation against an \hnbody~simulation with the Bulirsch-Stoer integration method in the case above (errors are expected to increase with decreasing semi-major axis), for a total of 337,235 comparison simulations. The scale of the change
in the periastron distance is on the order of 100--1000 AU, which agrees well with the errors found by using the impulse approximation by \cite{Rickman2005}.

\section{Climates of Habitable Planets: \texttt{POISE}\label{sec:poise}}

\begin{figure*}
    \centering
    \includegraphics[width=\textwidth]{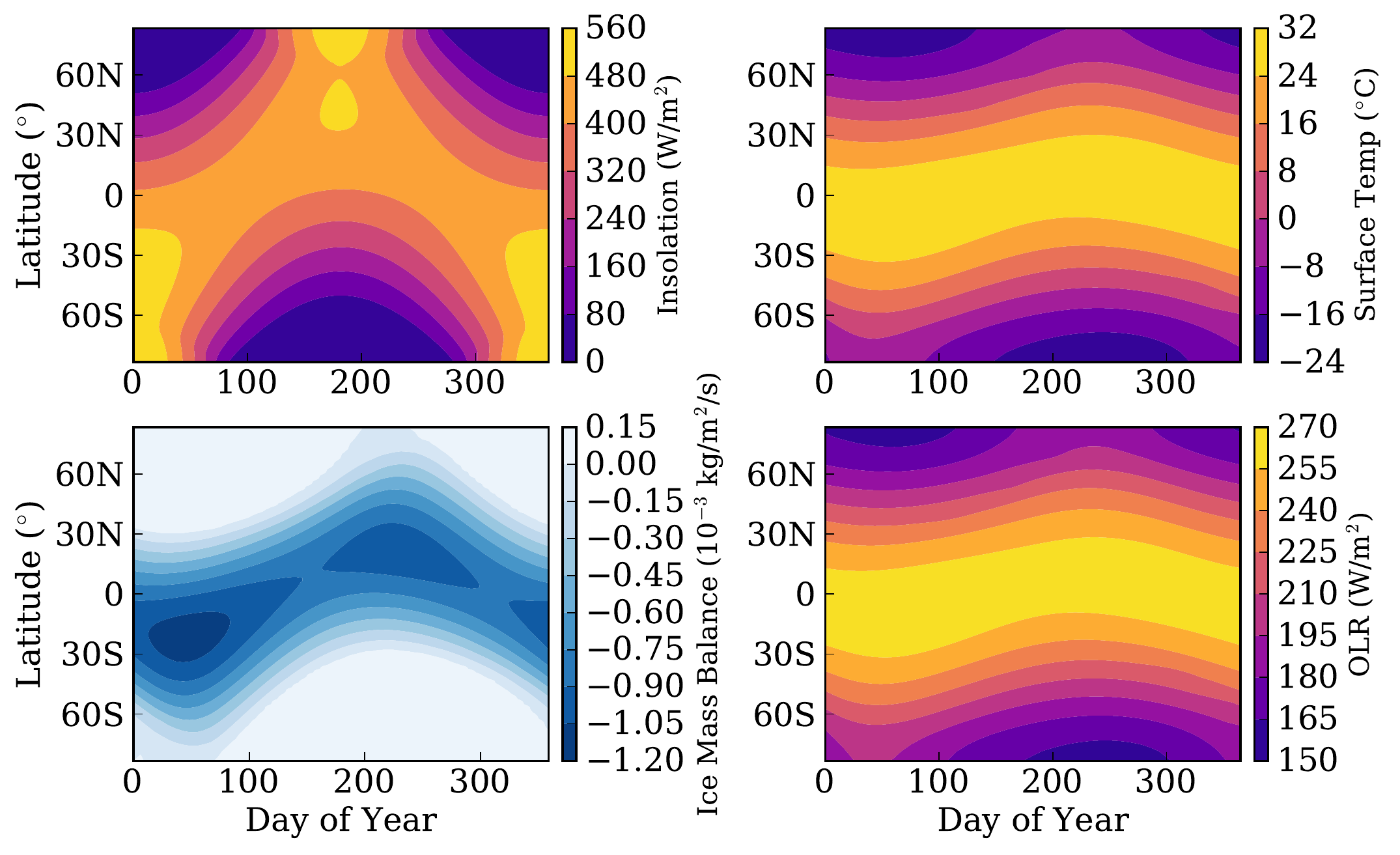}
    \caption{Insolation (upper left), surface temperature (upper right), ice mass balance (lower left), and out-going longwave radiation (OLR; lower right), for Earth over a single year, as modeled by \poise. Note that negative values in ice mass balance represent \emph{potential} melting, \ie this value is calculated even in the absence of ice on the surface. Approximate runtime: 5 seconds. \href{https://github.com/VirtualPlanetaryLaboratory/vplanet/tree/master/examples/EarthClimate}{\link{examples/EarthClimate}}}
    \label{fig:seasons}
\end{figure*}

To model climate, \vplanet utilizes an energy balance model (EBM) appropriate for Earth-like atmospheres. \poise is based on \cite{NorthCoakley1979} who solved for the temperature and albedo as a function of latitude and day of year and forced by the incoming instellation. An additional component, a model for the dynamics of ice sheets on land, is based on \cite{Huybers2008}. The instellation is calculated directly from the planet's orbital and rotational parameters, and the luminosity of the host star.

In Figure \ref{fig:seasons} we show the seasonal cycle for Earth as modeled by
\poise. Here, we have included ice sheet growth on land, but sea ice is modeled simply
as an increase in surface albedo when temperatures are below freezing. We have compared the outgoing longwave radiation and absorbed stellar fluxes to satellite data of Earth \citep{Barkstrom1990}; the full comparison is shown in \cite{Deitrick18b}. We find good agreement between the model and data for the OLR, with the largest errors ($\sim 20$\%) occurring near the poles and the tropics. The error at these locations is due to the relatively coarse nature of the model, which is unable to  capture the  effects of atmospheric circulation and water vapor there. The error for the absorbed stellar radiation is similar over the equator and mid-latitudes ($\leq 20$\%), though becomes much larger toward the poles. This is primarily a result of the simple parameterization of the albedo, which does not take into account the additional contribution from the atmosphere, and does not capture albedo variations with longitude. Overall, \poise~performs well despite the model's simplified nature. The ice mass
balance shows the sum of annual accumulation and melting, which equals ice flow convergence for a stable ice sheet. Here, the negative values
represent \emph{potential} melting---this value is calculated even in the absence of
ice. \poise can simulate an ice-covered, ice-free, or partially ice-covered planet, and variability among all three.

The hemispheric asymmetry (most easily seen in the left-hand panels of Fig. \ref{fig:seasons}) is
a result of Earth's eccentricity and the relative angle between perihelion and
equinox. Since perihelion occurs near the beginning of the calendar year, southern
summer is more intense than northern, and a greater amount of ice melt can occur.

\section{Radiogenic Heating: \texttt{RadHeat}\label{sec:radheat}}

\begin{figure*}[!ht]
    \centering
    \includegraphics[width=\textwidth]{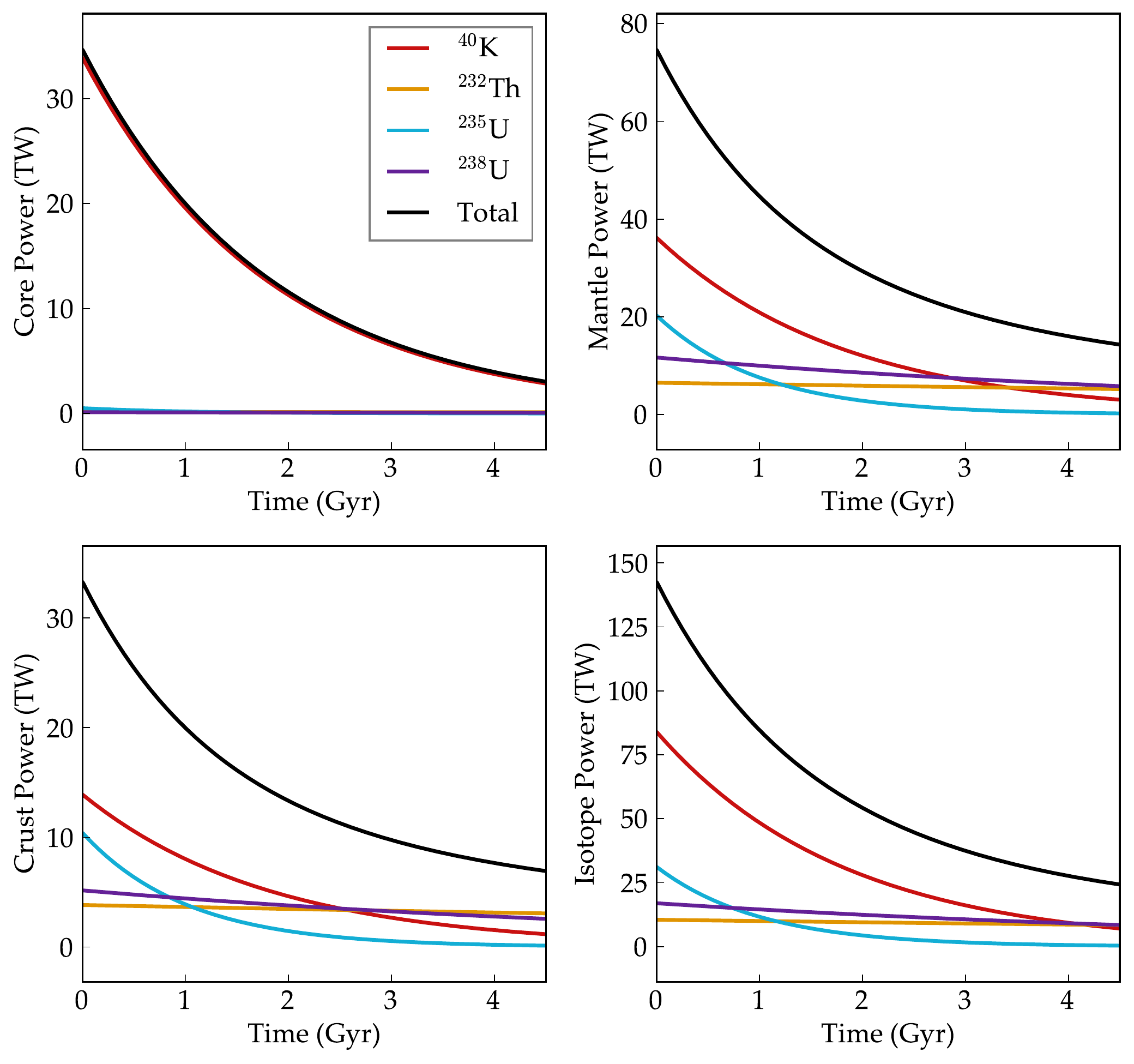}
    \caption{Radiogenic heating of the Earth since its formation. {\it Top Left:} Core power. Note that Th and both U powers are negligible in the core. {\it Top Right:} Mantle power. {\it Bottom Left:} Crust power. {\it Bottom Right:} Power by isotope. Approximate runtime: 1 second. \href{https://github.com/VirtualPlanetaryLaboratory/vplanet/tree/master/examples/RadHeat}{\link{examples/RadHeat}}}
    \label{fig:radheat}
\end{figure*}

The major sources of radiogenic heating are included in the module \radheat, specifically \U, \UU, \th232, \k40, and \al26. On Earth, and presumably other planets, these isotopes are distributed through planetary cores, mantles, and crusts. $^{26}$Al is relatively short-lived, but can provide enormous power on planets that form within a few Myr. The current amount of radiogenic power in Earth is poorly constrained \citep[see, \eg][]{Araki05}, and \vplanet assumes the current total power is 24.3 TW \citep{TurcotteSchubert02}. The evolution of these reservoirs on Earth, described in more details in $\S$~\ref{sec:thermint} and App.~\ref{app:thermint} is shown in Fig.~\ref{fig:radheat}. For more information on this module, consult Appendix~\ref{app:radheat}.

\section{Accurate Orbital Evolution: \texttt{SpiNBody}\label{sec:spinbody}}

\begin{figure*}[ht!]
    \centering
    \includegraphics[width=\textwidth]{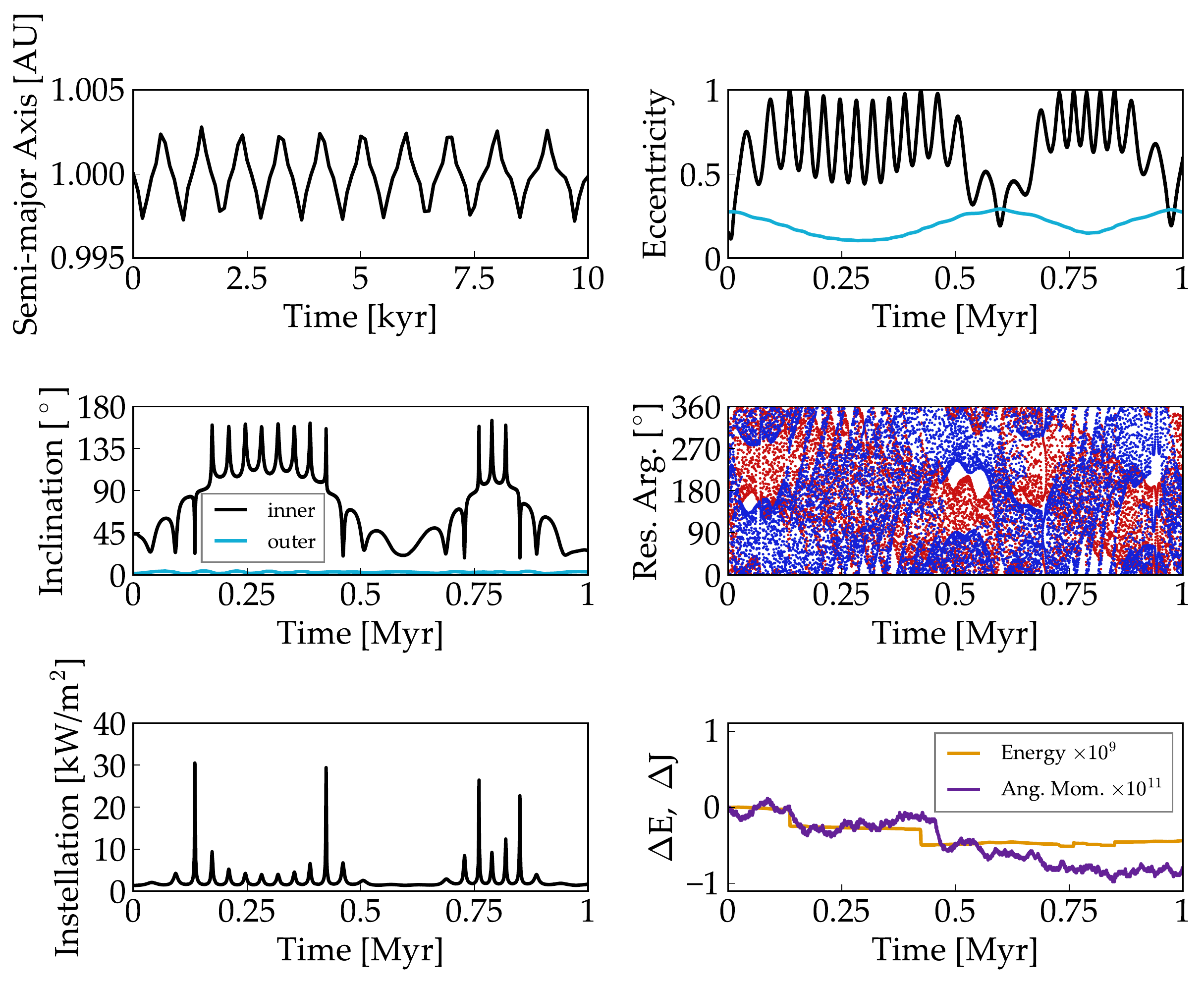}
    \caption{Evolution of a planetary system in a chaotic 3:1 eccentricity-inclination resonance. {\it Top left:} Semi-major axis of the inner planet. Note the $x$-axis timescale. {\it Top right:} Eccentricity. The black curve is for the inner planet, blue for outer. {\t Middle left:} Inclination. {\it Middle right:} The two resonant arguments: ($3\lambda' - \lambda  - 2\varpi$) in red and ($3\lambda' - \lambda  - 2\varpi'$) in blue, where primes indicate the outer planet. {\it Bottom left:} Orbit-averaged instellation of the inner planet. {\it Bottom right:} Evolution of total energy (orange) and angular momentum (purple). Approximate runtime: 14 hours. \href{https://github.com/VirtualPlanetaryLaboratory/vplanet/tree/master/examples/ChaoticResonances}{\link{examples/ChaoticResonances}}}
    \label{fig:ChaoticResonance}
\end{figure*}

The \vplanet N-body model, \spinbody, directly calculates the gravitational acceleration between massive objects. As this calculation is from first principles, \spinbody is valid for any configuration. Figure \ref{fig:ChaoticResonance} reproduces Fig.~13 from \cite{Barnes15_res}. This system consists of a solar-type star, an Earth-mass planet in the HZ, and outer Neptune-mass planet in a 3:1 resonance. The eccentricity occasionally surpasses 0.9999, but the system is stable for 10 Gyr. (Note that the N-body model assumes point masses, so the planet and star don't merge in this example, which is chosen to validate an extreme case.) The bottom right panel of Fig.~\ref{fig:ChaoticResonance} shows that energy and angular momentum are conserved to within 1 part in $10^9$, similar to \mercury~\citep{Chambers99}, and validates this \vplanet module. A simulation of the Solar System is also presented in Appendix~\ref{app:spinbody}.

\section{Stellar Evolution: \texttt{STELLAR}\label{sec:stellar}}
\begin{figure*}[ht!]
\begin{center}
\includegraphics[width=\textwidth]{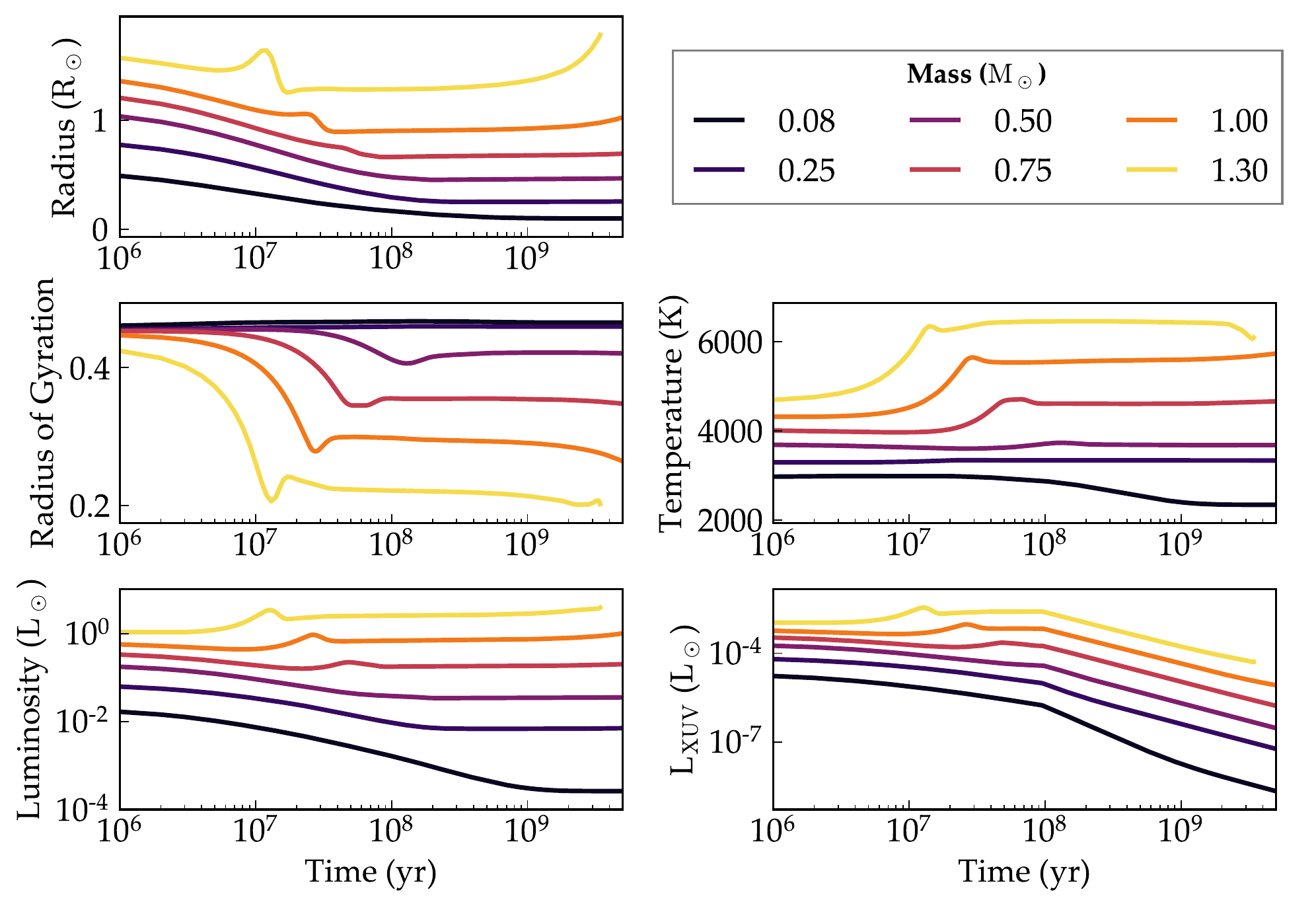}
\end{center}
\caption{ \label{fig:stellar} Evolution of the radius (top left), radius of gyration (middle left), luminosity (bottom left), effective temperature
(middle right), and XUV luminosity (bottom right) of stars of different masses predicted by the \stellar module. Approximate runtime: 2 minutes. \href{https://github.com/VirtualPlanetaryLaboratory/vplanet/tree/master/examples/StellarEvol}{\link{examples/StellarEvol}}}
\end{figure*}

Evolving stellar parameters shape the dynamics and habitability of stellar and exoplanetary systems. For example, exoplanets orbiting in the habitable zone of late M-dwarfs likely experienced an extended runaway greenhouse during the host star's superluminous pre-main sequence phase, potentially driving extreme water loss  \citep[\eg][]{LugerBarnes15}. The combination of evolving stellar radii and magnetic braking, the long-term removal of stellar angular momentum arising from the coupling of the stellar wind with the surface magnetic field, dictate the stellar angular momentum budget, molding observed stellar rotation period distributions as a function of stellar age and mass \citep[\eg][]{Skumanich1972,McQuillan2014,Matt2015}. Moreover, in stellar binaries, coupled stellar-tidal evolution depends sensitively on the evolving stellar radii and rotation state, driving orbital circularization on the pre-main sequence \citep[\eg][]{Zahn89}, potentially destablizing any circumbinary planets they may harbor \citep[\eg][]{Fleming18}, and can strongly impact the stellar rotation period evolution \citep[\eg][]{Fleming19}.

\vplanet's stellar evolution module, \stellar, tracks the evolution of the fundamental stellar parameters of low-mass ($M_\star \lsim 1.4 \mathrm{M_\odot}$) stars, including a star's radius, radius of gyration, $r_g$, effective temperature, luminosity, XUV luminosity, and rotation rate. \stellar models stellar evolution via a bicubic interpolation of the \citet{Baraffe15} models of solar metallicity stars over mass and time. \stellar computes the XUV luminosity according to the product of the luminosity and Eq.~(\ref{eq:lxuv}) for stars that drive atmospheric escape and water loss \citep[see $\S$~\ref{sec:atmesc}; ][]{LugerBarnes15}. Furthermore, \stellar allows the user to model the long-term angular momentum evolution of low-mass stars using one of three magnetic braking models \citep{Reiners2012,Repetto2014,Matt2015,Matt2019}. Here, we demonstrate the modeling capabilities of \stellar, while in  Appendix~\ref{app:stellar}, we describe the numerical and theoretical details of the module.

We demonstrate the evolution predicted by \stellar in Figure~\ref{fig:stellar} that depicts the evolution of the stellar radius, $r_g$, temperature, luminosity, and XUV luminosity, the latter computed using Eq.~(\ref{eq:lxuv}) and assuming $f_{sat} = 10^{-3}$, $t_{sat} = 1$ Gyr, and $\beta_{XUV} = -1.23$ \citep{Ribas05}, all as functions of time for several stars ranging in mass from late M dwarfs ($M_\star = 0.08 \mathrm{M_\odot}$) to late F dwarfs ($M_\star = 1.3 \mathrm{M_\odot}$). Our tracks agree with present-day solar values and display the well-known extended pre-main sequence phase of M dwarfs \citep[\eg][]{LugerBarnes15}.

\section{Geophysical Evolution: \texttt{ThermInt}\label{sec:thermint}}
\begin{figure*}[ht!]
\begin{center}
\includegraphics[width=0.8\textwidth]{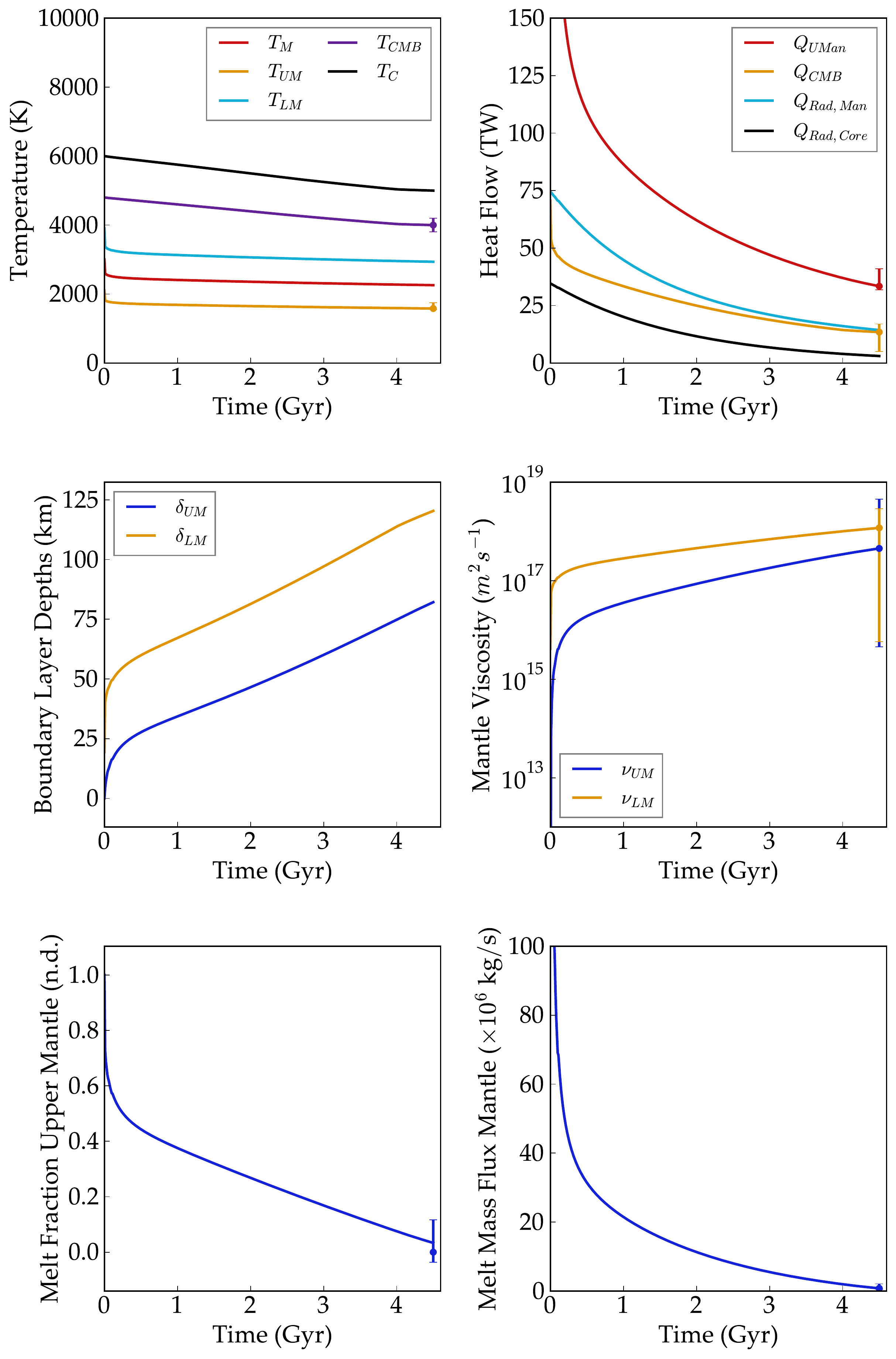}
\end{center}
\caption{\label{fig:earth1} Thermal evolution of Earth's mantle and some resultant material properties. Compare to Fig. 5 in \cite{DriscollBarnes15}. Approximate runtime: 1 second. \href{https://github.com/VirtualPlanetaryLaboratory/vplanet/tree/master/examples/EarthInterior}{\link{examples/EarthInterior}}}
\end{figure*}

\begin{figure*}[ht]
\begin{center}
\includegraphics[width=0.8\textwidth]{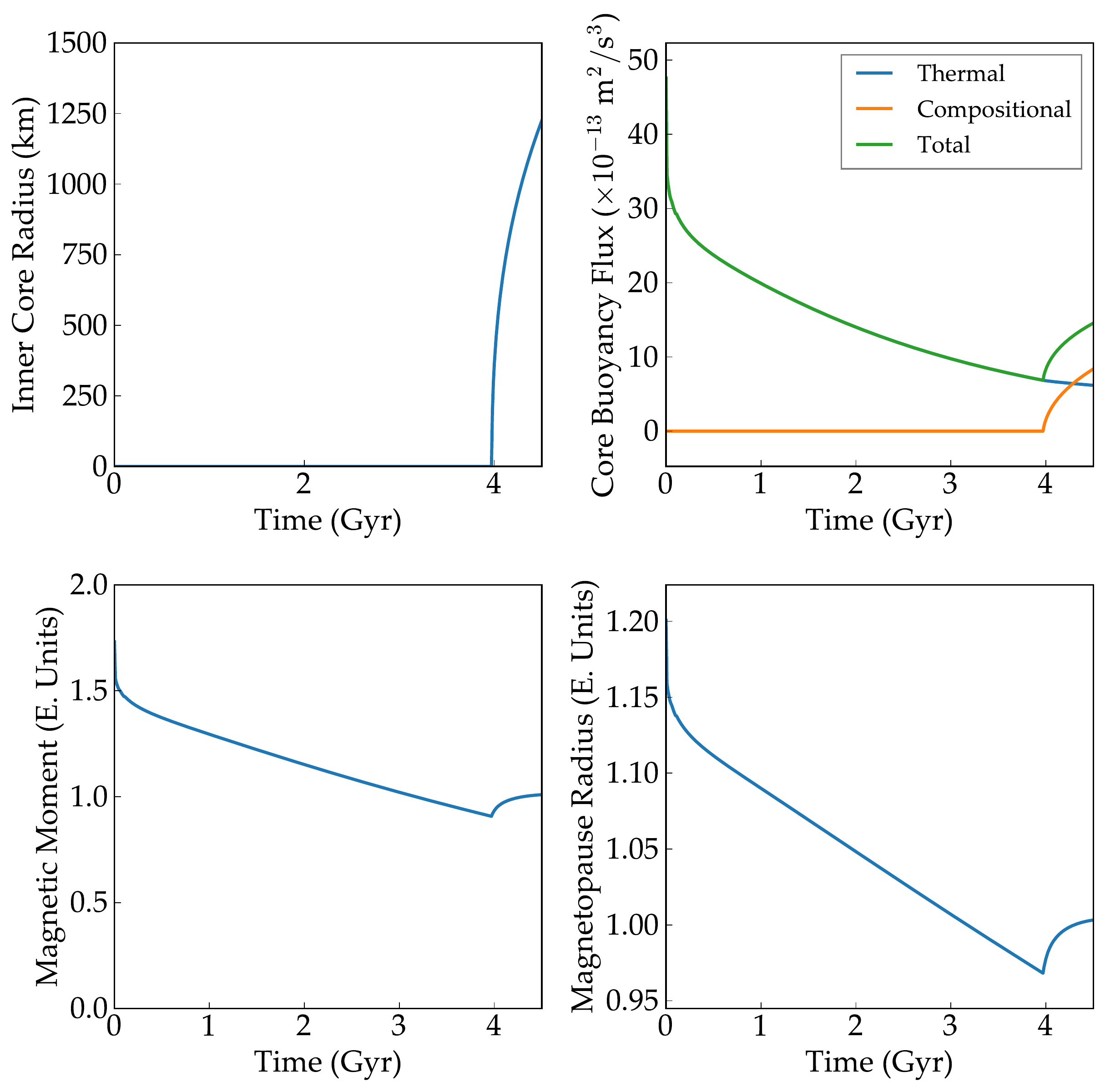}
\end{center}
\caption{\label{fig:earth2} Evolution of Earth's core and magnetic field. Compare to Fig. 5 in \cite{DriscollBarnes15}.  \href{https://github.com/VirtualPlanetaryLaboratory/vplanet/tree/master/examples/EarthInterior}{\link{examples/EarthInterior}}}
\end{figure*}

The thermal and magnetic evolution of the interior of rocky Earth-like planets  is modeled in \thermint by solving the coupled heat balance in the mantle and core.  Parameterized heat flow scalings and material properties appropriate for Earth are assumed.  A nominal Earth-model is calibrated to approximately reproduce the main constraints on the thermal and magnetic evolution: present-day mantle potential temperature of 1630 K, mantle surface heat flow of 38 TW, upwelling mantle melt fraction of 7\%, inner core radius of 1221 km, and a continuous core magnetic field.  The radiogenic abundances assumed in the mantle are based on bulk silicate Earth models \citep{arevalo2013,jaupart2015} and 3 TW of K$^{40}$ in the core today to maintain a continuous dynamo \citep{DriscollBercovici14}.
Using the default Earth values, \thermint+~\radheat produce an ``Earth interior model'' with the following values at 4.5 Gyr: $Q_{surf}=33.4(\pm5)$ TW, $Q_{cmb}=13.5(\pm4)$ TW, $T_{UM}=1587(-34,+161)$ K, $T_{CMB}=4000(\pm200)$ K, $R_{ic}=1224$ km, $M=80$ ZAm$^2$, and inner core nucleation at $3.97$ Gyr ($0.53$ Ga). The error bars in Figure \ref{fig:earth1} reflect the range of values in \cite{jaupart2015}. For the nominal Earth interior model, Figure \ref{fig:earth1} shows the thermal evolution of the mantle and core temperatures, heat flows, thermal boundary layer thicknesses, mantle viscosities, upwelling mantle melt fraction, and mantle melt mass flux over time. Figure \ref{fig:earth2} shows the thermal and magnetic evolution of the core in terms of inner core radius, core buoyancy fluxes, geodynamo magnetic moment, and magnetopause radius over time. There are no error bars in Fig.~\ref{fig:earth2} because $R_{ic}$ is from the ``preliminary reference Earth model \citep[PREM][]{DziewonskiAnderson81} and the other quantities (core buoyancy fluxes, magnetic moment, and magnetopause radius) are calibrated to give Earth-like values.  In both figures, the evolution concludes at the modern Earth and all parameters are consistent with observations.

\section{Multi-Module Applications\label{sec:multi}}

In this section we present results in which the previous modules are coupled together to reproduce previously published results.

\subsection{Milankovitch Cycles\label{sec:milankovitch}}

\begin{figure*}[ht!]
\begin{centering}
\includegraphics[width=\textwidth]{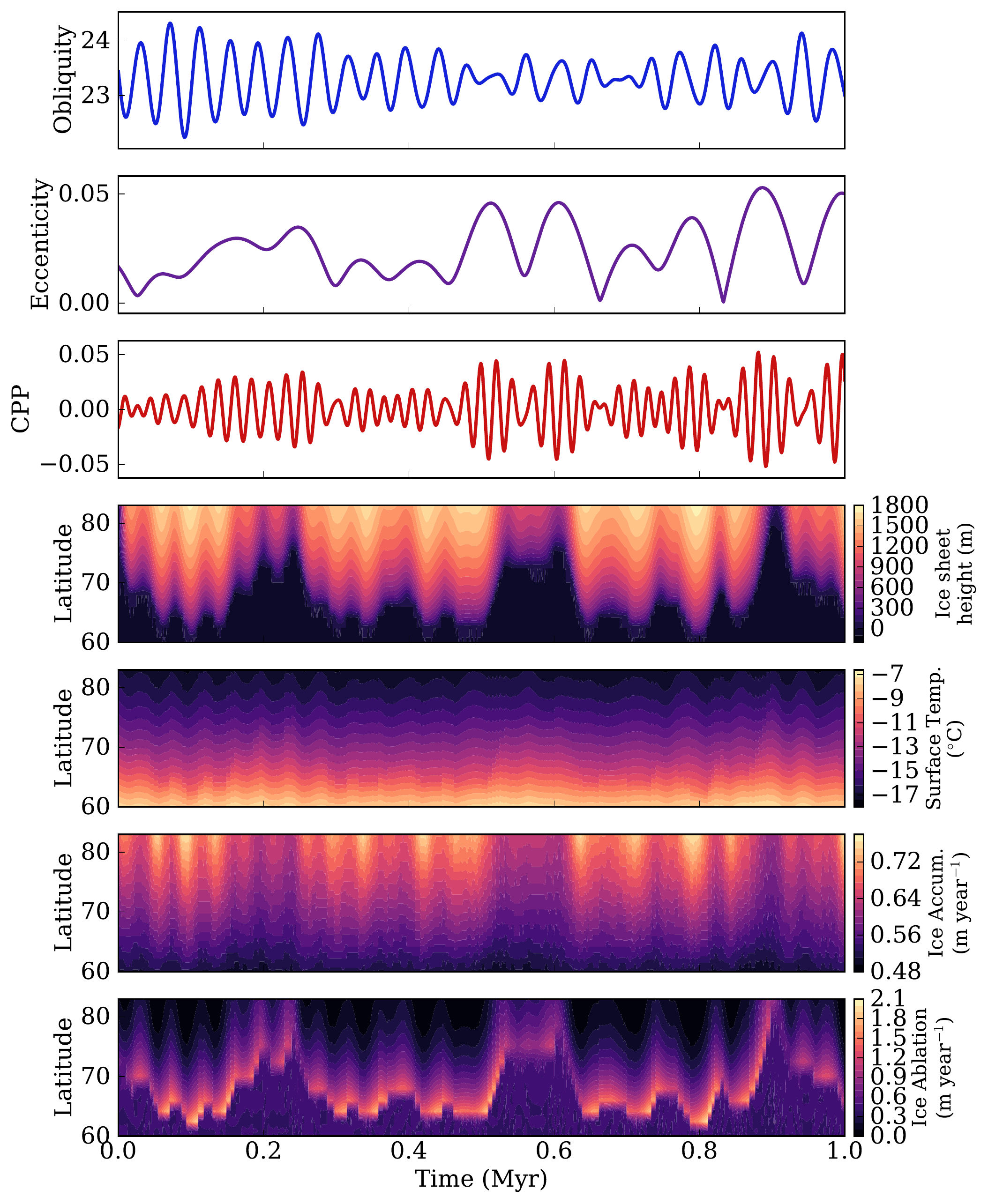}
\caption{\label{fig:huybers} Milankovitch cycles on Earth's
northern hemisphere. See
Figure 4 of \cite{Huybers2008} for comparison. From top to bottom: obliquity, eccentricity, CPP
$= e \sin{(\varpi+\psi)}$, ice sheet height (m),
annually averaged surface temperature ($^{\circ}$C), annual ice
accumulation rate (m yr$^{-1}$), and annual ice ablation rate (m
yr$^{-1}$). Approximate runtime: 5 minutes. \href{https://github.com/VirtualPlanetaryLaboratory/vplanet/tree/master/examples/EarthClimate}{\link{examples/EarthClimate}}}
\end{centering}
\end{figure*}

In this subsection we demonstrate \vplanet's ability to reproduce Milankovitch cycles on Earth.
This section utilizes \distorb, \distrot, and \poise.
Note that in order to reproduce the effect of Earth's moon on Earth's obliquity, we
force the precession rate to be $50.290966''$ year$^{-1}$
\citep{Laskar1993}. This choice does not perfectly match the dynamics
of the Earth-Moon-Sun system, but it is close enough to
replicate the physics of the ice age cycles.
The surface properties are shown in Figure \ref{fig:huybers}
\citep[see][Figure 4 for comparison]{Huybers2008}, for a
200,000 year window. The ice sheets in the northern hemisphere
high latitude region grow and retreat as the obliquity,
eccentricity, and climate-precession-parameter, or
CPP ($e\sin{(\varpi+\psi)}$), vary. The ice deposition rate is
less than that used by \cite{Huybers2008} and so the ice
accumulation per year is slightly smaller. The ice ablation
occurs primarily at the ice edge (around latitude $60^{\circ}$)
and is slightly smaller than \cite{Huybers2008}, peaking at $\sim 2.1$ m yr$^{-1}$, compared
to their $\sim 3$ m yr$^{-1}$.

We note that in this framework,
the net growth and retreat of ice sheets is highly sensitive to the
tuneable ice deposition rate, $r_{\text{snow}}$. With $r_{\text{snow}}$ too low, we do not
build up ice caps on Earth at all. With $r_{\text{snow}}$ too high, the ice sheets grow so large
that they become insensitive to orbital forcing. With $r_{\text{snow}}\sim2.25 \times 10^{-5}$ kg
m$^{-2}$ s$^{-1}$, we roughly reproduce the Earth's ice age cycles at
$\sim40,000$ years and $\sim100,000$ years over a 10 million
year simulation.

Note that \cite{Huybers2008} used a
deposition (precipitation) rate of 1 m per year, which is indeed roughly the global
average for Earth. However, precipitations rates are diminished in cold regions, which
suggests that a diminished value for ice deposition is justified. Additionally, we find in \vplanet that
when a value of 1 m per year is used, the ice sheets become too thick and are insensitive to
orbital forcing. Note as well that \cite{Huybers2008}'s
climate (EBM) components are coded on different grids and utilize different
parameterizations for radiative transfer.

There are a number of differences between our reproduction of
Milankovitch cycles and those of \cite{Huybers2008}. Most notably, our
ice sheets tend to persist for longer periods of time, taking up
to three obliquity cycles to fully retreat. As previously stated, we also require a
lower ice deposition (snowing) rate than \cite{Huybers2008} in
order to ensure a response from the ice sheets to the orbital
forcing. We attribute these differences primarily to the
difference in energy balance models used for the atmosphere.
For example, our model has a
single-layer atmosphere with a parameterization of the OLR tuned
to Earth, while \cite{Huybers2008} used a multi-layer atmosphere
with a simple radiative transfer scheme. Further, while the
\cite{Huybers2008} model contained only land, our model has both
land  and water which cover a fixed fraction of the surface. The
primary effect of having an ocean in this model is to change the
effective heat capacity of the surface. This dampens
the seasonal cycle and affects the ice sheet growth and
retreat. Thus, our seasonal cycle is somewhat muted compared to
theirs, and our ice sheets do not grow and retreat as
dramatically on orbital time scales. Ultimately, our ice age
cycles are more similar to the longer late-Pleistocene cycles
than to $\sim 40,000$ year cycles of the early-Pleistocene.

Even though we cannot perfectly match the results of
\cite{Huybers2008}, we find
the comparison acceptable. Both models make approximations to a
number of physical processes and thus have numerous parameters
that have to be tuned to reproduce the desired behavior. Thus, it is no surprise
that we do not reproduce their results precisely. Furthermore, despite all
of the crude assumptions made in the energy balance model (the parameterization of radiative transfer,
the reduction of the sphere to a single dimension, etc), we are nevertheless able
to produce the ice age frequencies at $\sim 40$ kyr and $\sim 100$ kyr, suggesting that
our simple model captures the basic physical processes involved. Finally, the details of
Earth's ice ages (such as the dominance of the $100$ kyr cycle in the late-Pleistocene), remain
difficult to capture and explain, even with sophisticated models \citep{AbeOuchi2013,SanchezGoni2019}.

\subsection{Evolution of Tight Binary Stars\label{sec:binaries}}

Stars form large and contract until the central pressure become large enough for fusion. For stars in tight binary systems, large tidal torques rapidly circularize the orbit early in the system's history. In a classic study of the evolution of short-period binary stars, \citet{ZahnBouchet89} found that orbits of binaries with an orbital period of less than about 8 days tidally circularized before they reach the Zero Age Main Sequence (ZAMS), consistent with contemporaneous observations. More recent surveys have found that circularization extends to binary orbital periods of about 10 days \citep[\eg][]{Meibom2005,Lurie2017}.  Here we reproduce the Zahn \& Bouchet result by coupling the stellar evolution models of \citet{Baraffe15}, incorporated in the \stellar module, and the equilibrium tide CPL model via \eqtide. See Sections \ref{sec:eqtide} and \ref{sec:stellar} and their corresponding Appendices, \ref{app:eqtide} and \ref{app:stellar}, respectively, for a more in-depth discussion of those models.

Adopting the initial conditions used to produce Figure 1 from \citet{ZahnBouchet89}, we model an equal mass $1\msun-1\msun$ binary with initial orbital eccentricity $e = 0.3$ and an orbital period of 5 days.  The initial stellar rotation rate to mean motion ratio is set at $\Omega/n = 3$ in line with the estimates from \cite{ZahnBouchet89}, who assumed conservation of angular momentum during the stellar accretion phase.  For our tidal model, we set $Q = 1.25 \times 10^5$ and $k_2 = 0.5$, both reasonable values for stars given the wide range of assumed values in the literature \citep[\eg][]{Barnes13,Fleming18}. The parameters $Q$ and $\tau$ can, and probably do, vary as a function of the forcing frequency \citep[see \eg][]{Penev18}, but \vplanet does not (yet) include this complication.

\begin{figure}[ht]
\includegraphics[width=0.45\textwidth]{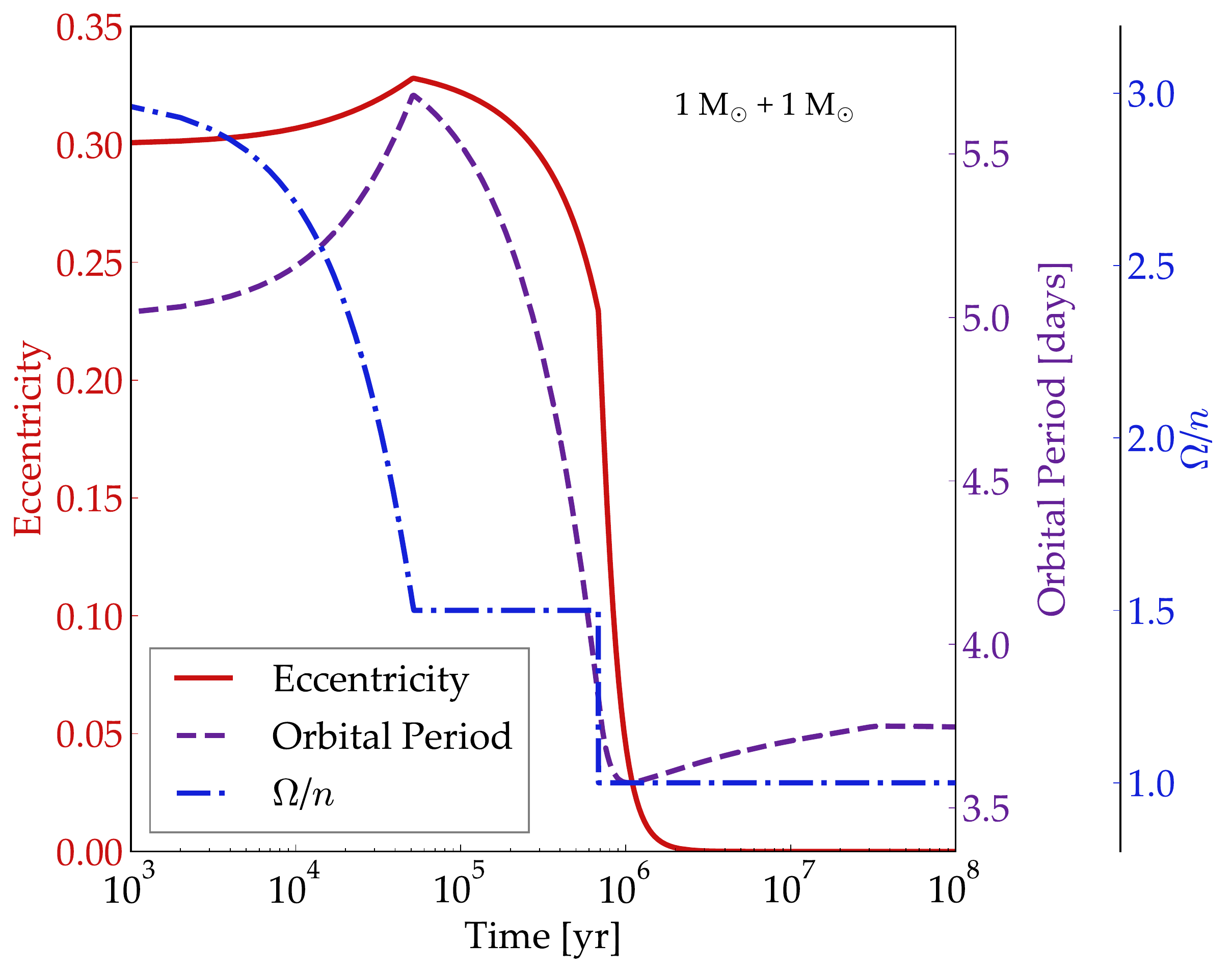}
\caption{Coupled stellar and tidal evolution of a solar twin binary from the pre-main sequence onward calculated in \vplanet using the \eqtide and \stellar modules. The stars have identical properties initially and throughout the simulation. Orbital eccentricity evolution is given by the red solid line, the orbital period by the purple dashed line, and the ratio of stellar rotation rate to binary mean motion ($\Omega/n$) evolution is given by the blue dot dashed curve. The binary's evolution matches with that of an identical system presented in Figure 1 of \citet{ZahnBouchet89}. Approximate runtime: 70 seconds. \href{https://github.com/VirtualPlanetaryLaboratory/vplanet/tree/master/examples/BinaryTides}{\link{examples/BinaryTides}}}
\label{fig:zahn89}
\end{figure}

The results of the simulation are depicted in Figure \ref{fig:zahn89}.  Our results are in good agreement with \citet{ZahnBouchet89}, Fig.~1. In this case, a quantitative comparison is unwarranted as the \cite{ZahnBouchet89} model used older stellar evolution models, so we do not expect the results to be a perfect match. After an initial increase in orbital eccentricity, the binary circularizes within the first ${\sim}10^6$ years before the ZAMS, in agreement with \citet{ZahnBouchet89}.  The transition between increasing and decreasing eccentricity occurs when $e = \sqrt{1/19}$ at the $\Omega/n = 1.5$ transition, as expected from the
CPL model (see Section~\ref{app:eqtide:cpl}).  The orbital period peaks at $t = 10^5$ years and then decreases as the orbit circularizes. One difference between the two model predictions is that \citet{ZahnBouchet89} find an increase in $\Omega/n$ from unity to over 2 near $10^6$ years, before the stars tidally lock again after about $10^9$ years. In our model, the stars remain tidally locked after $10^5$ years as we force stars with rotation periods close to the orbital period to remain tidally locked to prevent numerical instabilities (see the appendix of \citet{Barnes13} for a more in-depth discussion of this numerical necessity). This tidal locking formalism can still model complex physical interactions near the tidally locked state \citep[e.g. subsynchronous rotation,][]{Fleming19}, but it can easily be disabled.

\subsection{Interiors of Tidally Heated Planets\label{sec:tidalheat}}
In this multi-module application we model the gravitational tidal dissipation in the interior of an Earth-like planet and its orbit.  The modules used are \thermint, \radheat, and \eqtide.  The tidal dissipation equations used in this application is the ``orbit-only'' model from \cite{DriscollBarnes15}, see also Appendix~\ref{app:eqtide:orbit}, and
the dissipation efficiency depends on the temperature of the mantle.  To reproduce the results of \cite{DriscollBarnes15} the dissipation efficiency in the orbital equations ($\dot{e}$ and $\dot{a}$) is approximated by $Im(k_2)\approx k_2/\mathcal{Q}$, where $\mathcal{Q}=\eta \omega/\mu$ is the Maxwell tidal efficiency.

\begin{figure*}[h!]
\begin{centering}
\includegraphics[width=0.8\textwidth]{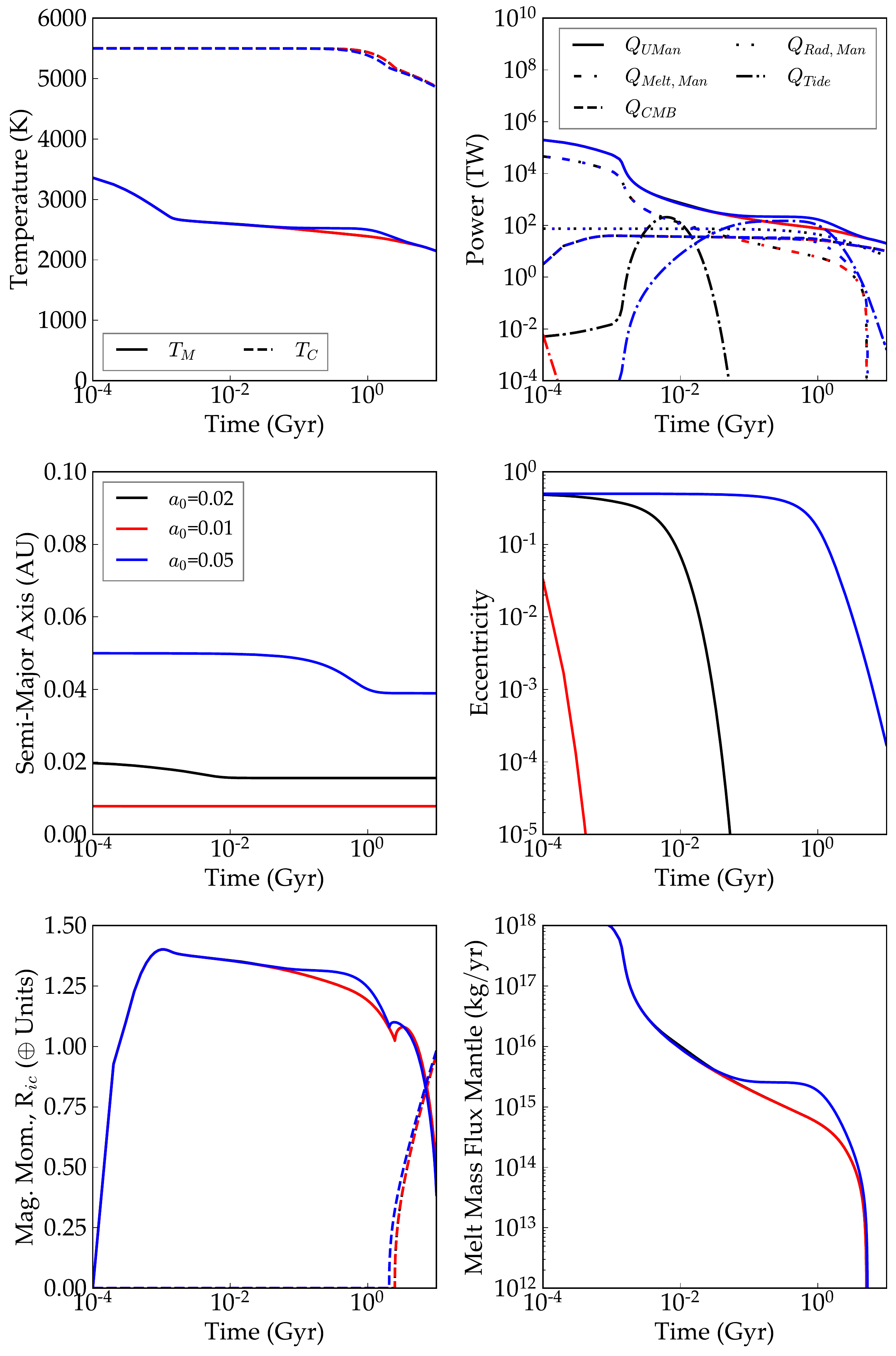}
\caption{\label{fig:tidalearth1} Internal evolution of three Earth-like planets orbiting a $0.1$ solar-mass star with initial eccentricities of $0.5$ and semi-major axes of $0.01$ (red), $0.02$ (black), and $0.05$ (blue).  Bottom left panel: Magnetic moment is scaled to Earth's current value of $80$ ZAm$^2$ and inner core radius scaled to core-mantle boundary radius $R_{ic}/R_{cmb}$ is shown as dashed curves, with the black curve hidden by the red curve. Approximate runtime: 10 minutes. \href{https://github.com/VirtualPlanetaryLaboratory/vplanet/tree/master/examples/TidalEarth}{\link{examples/TidalEarth}}}
\end{centering}
\end{figure*}

\begin{figure*}[ht]
\begin{centering}
\includegraphics[width=0.8\textwidth]{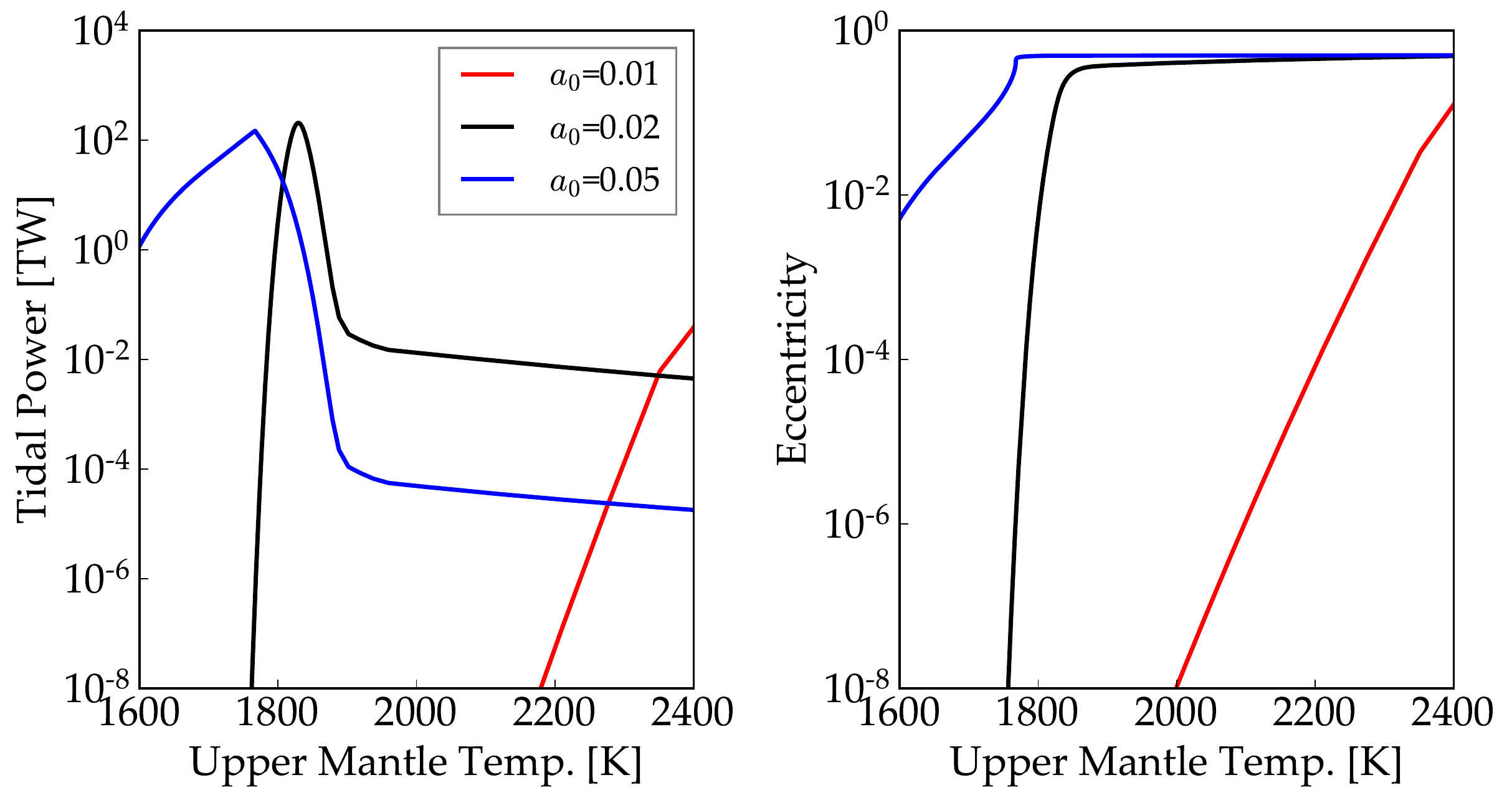}
\caption{\label{fig:tidalearth2} Internal properties for the same models as Figure \ref{fig:tidalearth1} shown here as functions of $T_{UM}$. \href{https://github.com/VirtualPlanetaryLaboratory/vplanet/tree/master/examples/TidalEarth}{\link{examples/TidalEarth}}}
\end{centering}
\end{figure*}

This example reproduces the results of \cite{DriscollBarnes15} for three tidally evolving planets orbiting a $0.1$ solar-mass star to within a few percent. (Note that some of the underlying physics in their interior model has been updated in \thermint, so we do not expect an exact match.)  Figures \ref{fig:tidalearth1} and \ref{fig:tidalearth2} compare the thermal, magnetic, and orbital evolution of three Earth-like planets each with the same initial eccentricity of 0.5 and initial Semi-major axes of $0.01$, $0.02$, $0.05$ all in AU.  Figure \ref{fig:tidalearth1} reproduces \cite{DriscollBarnes15}, Fig. 5 and Fig. \ref{fig:tidalearth2} reproduces \cite{DriscollBarnes15}, Fig. 4. The main features that are reproduced from \cite{DriscollBarnes15} are that the tidal power peaks at $T_{UM}=1800$ K, the orbits circularize around $10^{-4}$ Gyr for $e_0=0.01$, $10^{-2}$ Gyr for $e_0=0.02$, and $>10$ Gyr for $e_0=0.05$, the inner core nucleates around 2 Gyr, and mantle melt mass flux goes to zero around 5 Gyr.

\subsection{Tidal Damping in Multi-Planet Systems\label{sec:multidamp}}

In multi-planet systems in which one or more planets is close enough to the host star for tides to damp the orbit, orbital and rotational properties can reach an approximately fixed state as tides remove energy and angular momentum from planetary orbits and rotations. In this subsection, we reproduce the damping of orbital evolution into the ``fixed point solution'' \citep{WuGoldreich02,ZhangHamilton08} and the damping of rotational cycles into a Cassini state \citep{Colombo66,WardHamilton04,Brasser2014,Deitrick18a}.

One difficulty arises from the fact that the semi-major axis decays, and its evolution is not accounted for in \distorb, which ignores terms involving the mean longitude. The functions, $f_i$,
in the disturbing
function (Table \ref{distfxn}) are computationally expensive. In the secular approximation
used in \distorb, the semi-major axes of the planets do not change, thus the $f_i$ do not change.
So when \distorb is used without \eqtide, we calculate all of the $f_i$ values at the start of the
simulation and store them in an array. This accelerates the model by over a factor of 100
compared to recalculating $f_i$ every time step.

However, the tidal forces in \eqtide do change the semi-major axes, and so when the two models are
coupled, we must recalculate $f_i$. Rather than recalculate every time step, we additionally store
the derivatives, $df_i/d\alpha$, and the value of $\alpha$ (the semi-major axis ratio for each pair
of planets) at which $f_i$ was calculated. Every time step, then, we can calculate the change in $\alpha$, and recalculate $f_i$ only when
\begin{equation}
    \Delta \alpha > \eta f_i \left( \frac{df_i}{d\alpha} \right) ^{-1},
\end{equation}
where $\eta$ is a user-defined tolerance factor. The smaller $\eta$ is, the more accurate the
simulation will be, at the expense of computation time.

\subsubsection{Apsidal Locking\label{sec:apsidal}}

\begin{figure*}[ht!]
\begin{centering}
\includegraphics[width=\textwidth]{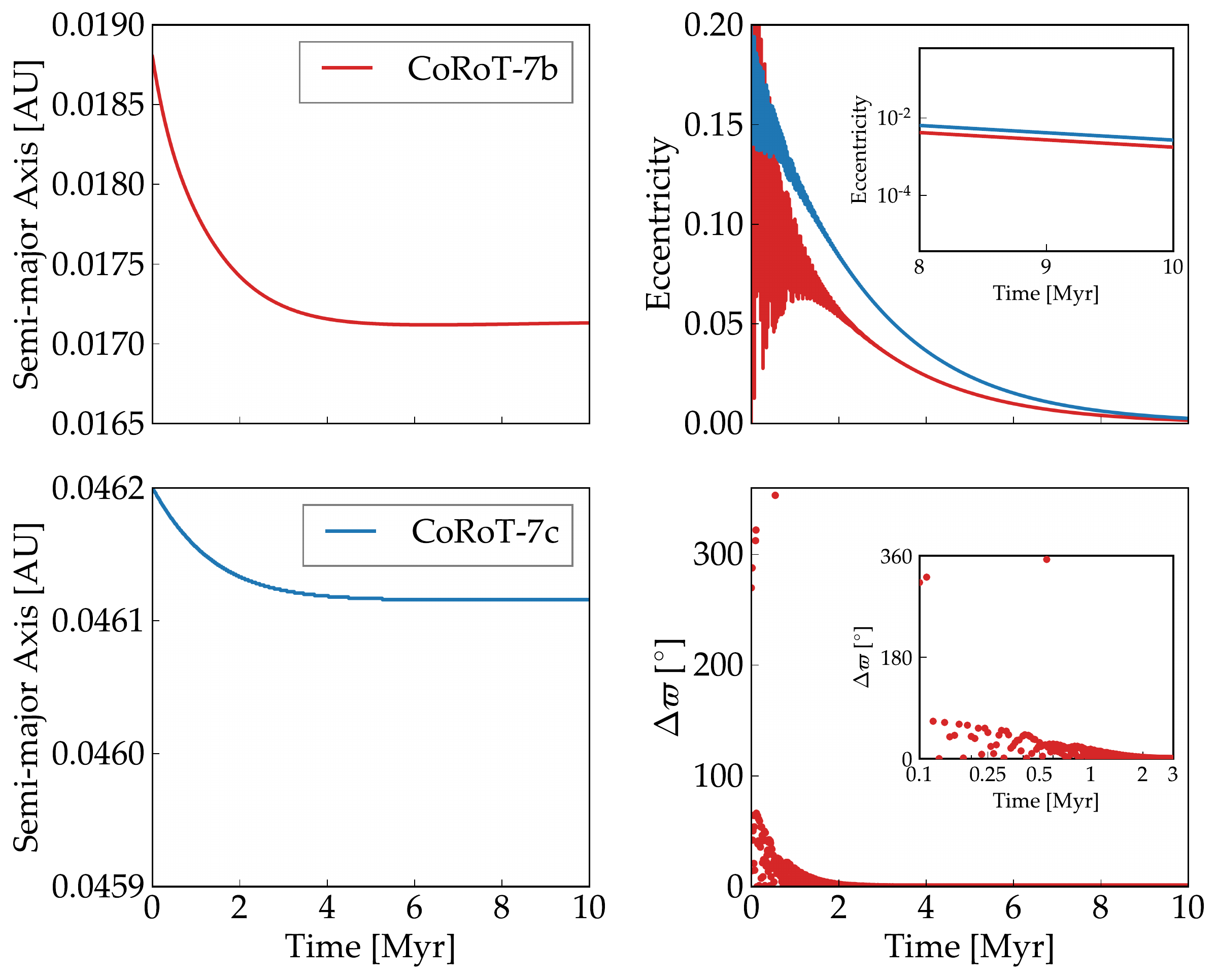}
\caption{\label{fig:corot7} Tidally damped orbital evolution of CoRoT-7 b and c as computed by \eqtide and \distorb. Compare to Figures 2 and 3 from \citet{Rodriguez11}. {\it Top left:} Semi-major axis evolution of CoRoT-7 b. {\it Top Right:} Eccentricity evolution for both CoRoT-7 b and c.  CoRoT-7 b's eccentricity is initially excited by gravitational perturbations from CoRoT-7 c, but eventually damps towards 0 due to tidal forces.  The inset shows the slightly non-zero eccentricities for planet b and c decaying towards 0 near the end of the simulation. {\it Lower left:} Semi-major axis evolution of CoRoT-7 c. {\it Lower right:} Difference between the longitudes of pericenters of CoRoT-7 b and c, $\Delta \varpi = \varpi_b - \varpi_c$.  In the inset, we display both planets becoming apsidally locked within 3 Myr due to tidal damping. Approximate runtime: 8 minutes. \href{https://github.com/VirtualPlanetaryLaboratory/vplanet/tree/master/examples/ApseLock}{\link{examples/ApseLock}}}
\end{centering}
\end{figure*}

Multi-planet systems in which one or more planets experience strong tidal damping of $e$ can reach a so-called fixed point state in which the angular momentum exchange between the planets ends and the longitudes of periastron circulate with identical frequency \citep{WuGoldreich02,Rodriguez11}. In Fig.~\ref{fig:corot7}, we use \vplanet's \distorb and \eqtide modules to reproduce the evolution of $e$ and $\varpi$ for the CoRoT-7 system examined by \citet{Rodriguez11}. This figure should be compared to Figs. 2--3 from \citet{Rodriguez11}. Note that our model is purely secular whereas \citet{Rodriguez11} directly integrated the equations of motion, including resonant effects, so our results may slightly differ. We adopted the initial conditions from \citet{Rodriguez11} and list them in Table \ref{tab:corot7}. Our \vplanet simulations qualitatively reproduce the evolution examined in the original work by \citet{Rodriguez11}. Within the first few Myrs, the inner planet, CoRoT-7 b, experiences large eccentricity oscillations that drive its inward tidal migration. As these eccentricity oscillations damp towards 0, both planets enter the fixed point state where the differences between their longitudes of periastron damp to 0, after which their $\varpi$'s circulate together with the same frequency. The tidal eccentricity damping for both planets occurs rapidly within the first 10 Myr, similar to the damping time of about 7 Myr found in \cite{Rodriguez11}.

\begin{deluxetable}{lc}
\tabletypesize{\small}
\tablecaption{CoRoT-7 System Parameters \label{tab:corot7}}
\tablewidth{0pt}
\tablehead{
\colhead{Parameter} & \colhead{Value}
}
\startdata
$M_\star$ [$M_{\odot}$] & 0.93\\
$m_b$ [$M_{\oplus}$] & $8$ \\
$m_c$ [$M_{\oplus}$] & $13.6$ \\
$a_{b,\textrm{initial}}$ [AU] & $0.0188$ \\
$a_{c,\textrm{initial}}$ [AU] & $0.0462$ \\
$e_{b,\textrm{initial}}$ & $0$ \\
$e_{c,\textrm{initial}}$ & $0.2$ \\
$Q_b$ & $100$ \\
$Q_c$ & $100$ \\
\enddata \vspace*{0.1in}
\end{deluxetable}

\subsubsection{Cassini States\label{sec:cassini}}

In this section we consider the damping of a planet's obliquity into a Cassini state \citep{Colombo66,WardHamilton04,WinnHolman05,Deitrick18a}. In a two-planet system, such a configuration occurs when a planet's orbital and rotational angular momentum vectors remain coplanar with the system's total angular momentum vector. Such an alignment could occur by chance, but it is most likely for a world to reach a Cassini state if its obliquity experiences both damping and excitation, \ie a damped driven configuration. In that case, the obliquity reaches a non-zero equilibrium value, such as the $\sim 6^\circ$ obliquity of the moon, as first noted by Giovanni Cassini himself.

The physics and mathematics of Cassini states have been discussed at length in the literature, but, briefly, any given three (or more) body system will have Cassini states available to its members. Each member has up to 4 possible Cassini states available, but only up to 2 can be stable. One or more separatrices exist in the phase space, and if damping is present, the stable Cassini states represent attractors.

A Cassini state can be quantitatively identified using the following relations from \cite{WardHamilton04}, see also \citep{Deitrick18a}:
\begin{equation}
\sin{\Psi} = \left \| \frac{(\mathbf{k} \times \mathbf{n})\times(\mathbf{s}\times\mathbf{n})}{|\mathbf{k} \times \mathbf{n}||\mathbf{s}\times\mathbf{n}|} \right \|, \label{eq:cass1}
\end{equation}
where $\mathbf{k}$, $\mathbf{n}$, and $\mathbf{s}$ are the vectors associated with the perpendicular to the appropriate reference plane (the invariable plane or Laplace plane, for example), the angular momentum of the body's orbit, and the angular momentum of the body's rotation. Alternatively, the complimentary relation can be used:
\begin{equation}
\cos{\Psi} = \frac{(\mathbf{k} \times \mathbf{n})\cdot(\mathbf{s}\times\mathbf{n})}{|\mathbf{k} \times \mathbf{n}||\mathbf{s}\times\mathbf{n}|}. \label{eq:cass2}
\end{equation}
A Cassini state occurs when $\cos\Psi$ and/or $\sin\Psi$ oscillate about 1, 0, or -1, with the equilibrium value depending on the particular Cassini state.

\begin{figure*}[ht!]
\begin{centering}
\includegraphics[width=\textwidth]{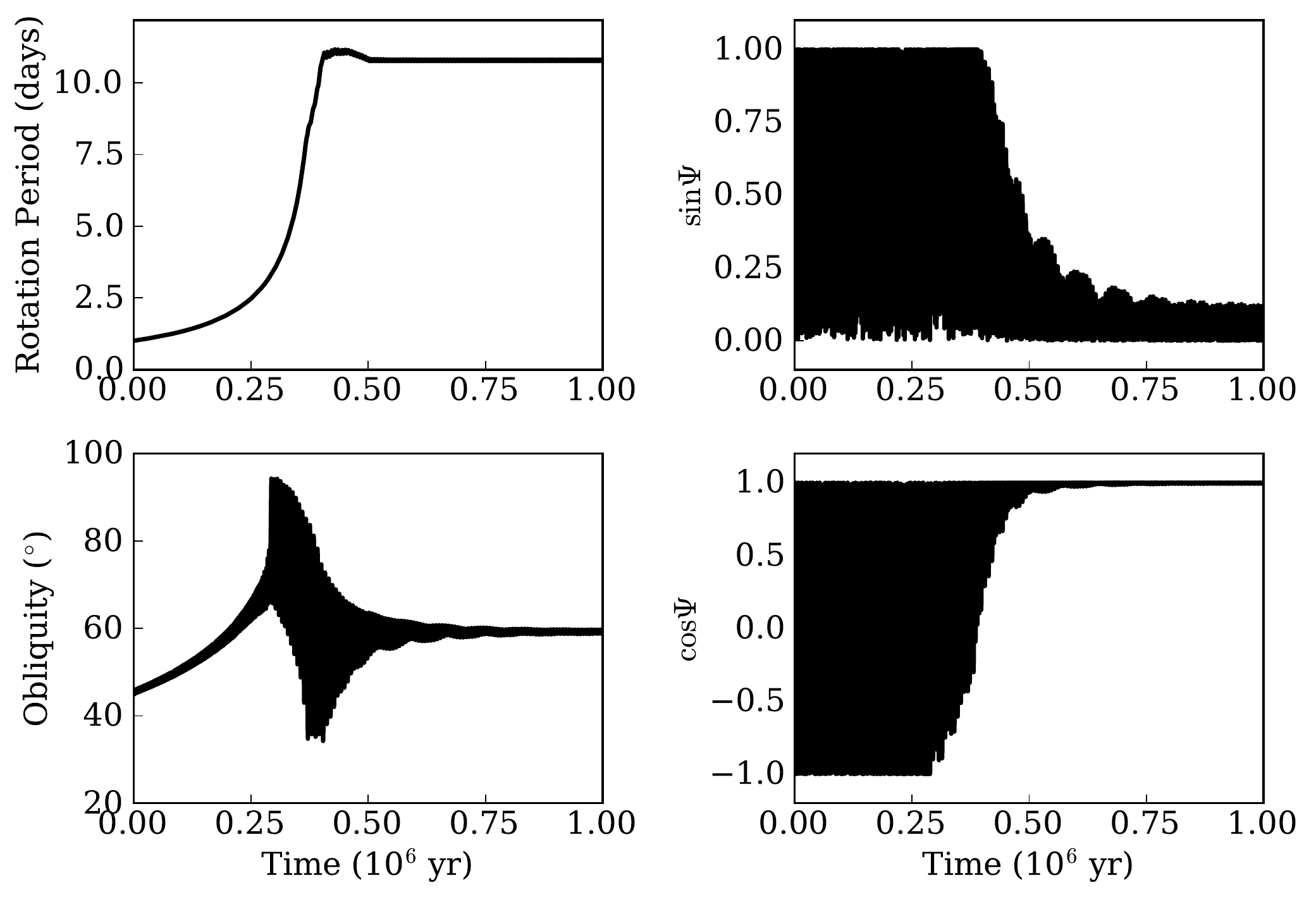}
\caption{\label{fig:cassini}An Earth-mass planet damping into Cassini state 2 under the influence of tides and perturbations from an 18 M$_{\oplus}$ companion planet. This example is constructed to be similar to Fig. 2 of \cite{WinnHolman05}. Approximate runtime: 1 minute. \href{https://github.com/VirtualPlanetaryLaboratory/vplanet/tree/master/examples/CassiniStates}{\link{examples/CassiniStates}}}
\end{centering}
\end{figure*}

To demonstrate evolution into a Cassini state, we construct a simulation based on Figure 2 of \cite{WinnHolman05}. The planetary system parameters are listed in Table~\ref{tab:cassini}, and with a stellar mass of $1\msun$. We use the \eqtide, \distorb, and \distrot modules to perform this experiment. The evolution of the system is shown in Fig.~\ref{fig:cassini} and the obliquity settles into an equilibrium value of $\sim 59^\circ$. In this case, we include first order GR corrections, see $\S$~\ref{app:distorb}. 

\begin{deluxetable}{lcc}
\tabletypesize{\small}
\tablecaption{Physical and orbital parameters for the system shown in Figs.~\ref{fig:cassini}--\ref{fig:cassini_ham}.}
\tablewidth{0pt}
\tablehead{
\colhead{Parameter} & \colhead{Planet b} & \colhead{Planet c}}
\startdata
    $m_p$ ($\mearth$) & 1 & 18\\
    $r_p$ ($\rearth$) & 1 & 1.5\\
    $a$ (AU) & 0.125 & 0.2246\\
    $e$ & 0.1 & 0.1\\
    $i$ ($^\circ$) & 0.5 & 0.001\\
    $\varpi$ ($^\circ$) & 248.87 & 356.71\\
    $\Omega$ ($^\circ$) & 20.68 & 20\\
    $\epsilon$ ($^\circ$) & 45 & -\\
    $p_A$ ($^\circ$) & 0 & -\\
    $Q$ & 100 & -\\
    $k_2$ & 0.3 & -\\
    \label{tab:cassini}
\enddata \vspace*{0.1in}
\end{deluxetable}

\begin{figure}[h!]
\begin{centering}
\includegraphics[width=0.45\textwidth]{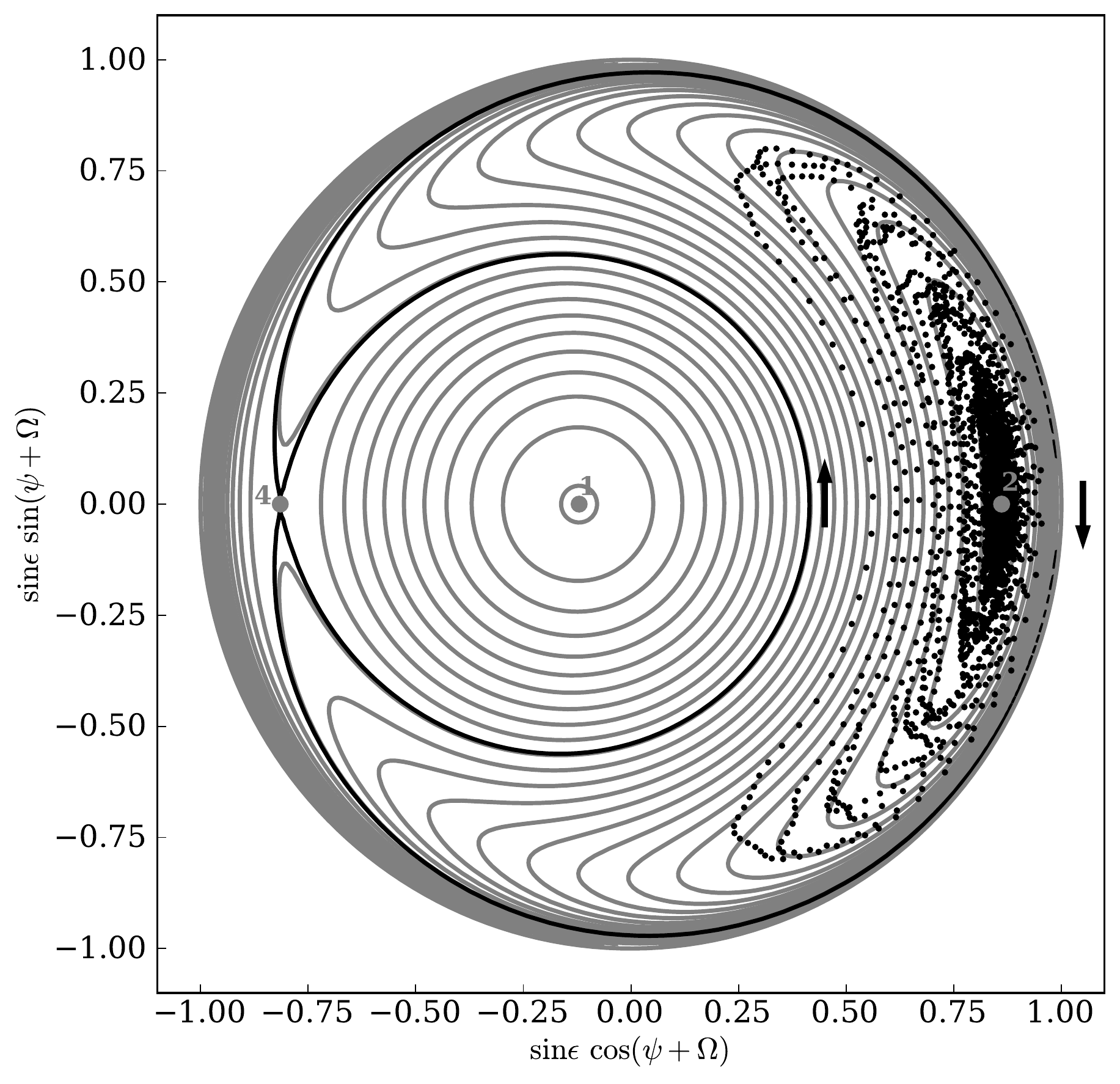}
\caption{\label{fig:cassini_ham}Phase space of the three prograde Cassini states (numbered). The gray curves show lines of constant Hamiltonian (Equation 5 in \citealt{WinnHolman05}) and the black curve shows the separatrix between states 1 and 2. The three possible Cassini states for this system are denoted by the light gray numbers 1, 2, and 4. This case is constructed to be similar to the illustrative case shown in Figure 2 of \cite{WinnHolman05}, with $-g/\alpha \sim 0.75$ and $i = 0.5^{\circ}$ ($g$ is the dominant inclination frequency, $\alpha$ the precession frequency, and $i$ the mean inclination). The location of the planet's pole after $\sim$ 400 kyr is shown as the black points, with dots separated by 100 years. As the planet's spin is damped by tidal torques, its obliquity sinks into Cassini state 2. The direction of motion in time is indicated by the black arrows. \href{https://github.com/VirtualPlanetaryLaboratory/vplanet/tree/master/examples/CassiniStates}{\link{examples/CassiniStates}}}
\end{centering}
\end{figure}

Fig.~\ref{fig:cassini_ham} shows the phase space of this configuration, as well as the evolution of this test system. Cassini states 1 and 2 are stable, but state 4 is a saddle point and hence is unstable. The gray curves show lines of constant Hamiltonian (Equation 5 in \citealt{WinnHolman05}) and the black curve shows the separatrix between states 1 and 2. In this case, the system is attracted to Cassini state 2 after $\sim 500$ kyr.

The results here for Cassini states should be regarded as approximate. The addition of triaxiality, due to rigidly-supported features or tidal forces, modifies the precession constant and becomes particularly important in spin-orbit resonance \citep{Goldreich1966,Peale1969,Hubbard1978,Jankowski1989,Bills2005a,Baland2016}. This changes the precise locations and evolution of the states. \vplanet does not yet include the scaling for triaxiality due to tides or the resonant terms in the obliquity evolution. Importantly, several studies have shown that state 2 may be destabilized by these effects or other physical processes \citep{Gladman1996,Levrard2007,Fabrycky2007}.

\subsection{Evolution of Venus\label{sec:venus}}

\subsubsection{Interior\label{sec:venus:interior}}
The interior of Venus is modeled using \thermint + \radheat, where Venus is assumed to have the same core mass fraction ($0.32$), radiogenic budget in the mantle ($20$ TW today) and core ($3$ TW today), and mantle and core melting curves.  The difference between the Venus and  Earth models is Venus' mantle is assumed to be in a stagnant lid which reduces the mantle heat flow.  The main constraint on the interior is to ensure no dynamo today.  To achieve this the bulk mantle viscosity and activation energy are assumed to be $2\times10^9$ m$^2$ s$^{-1}$ and $3.5\times10^5$ J mol$^{-1}$.  We note that this model is similar to that in \cite{DriscollBercovici14}, with the main difference here being that the viscosity depends on melt fraction.

\begin{figure*}[h!]
\begin{center}
\includegraphics[width=0.75\textwidth]{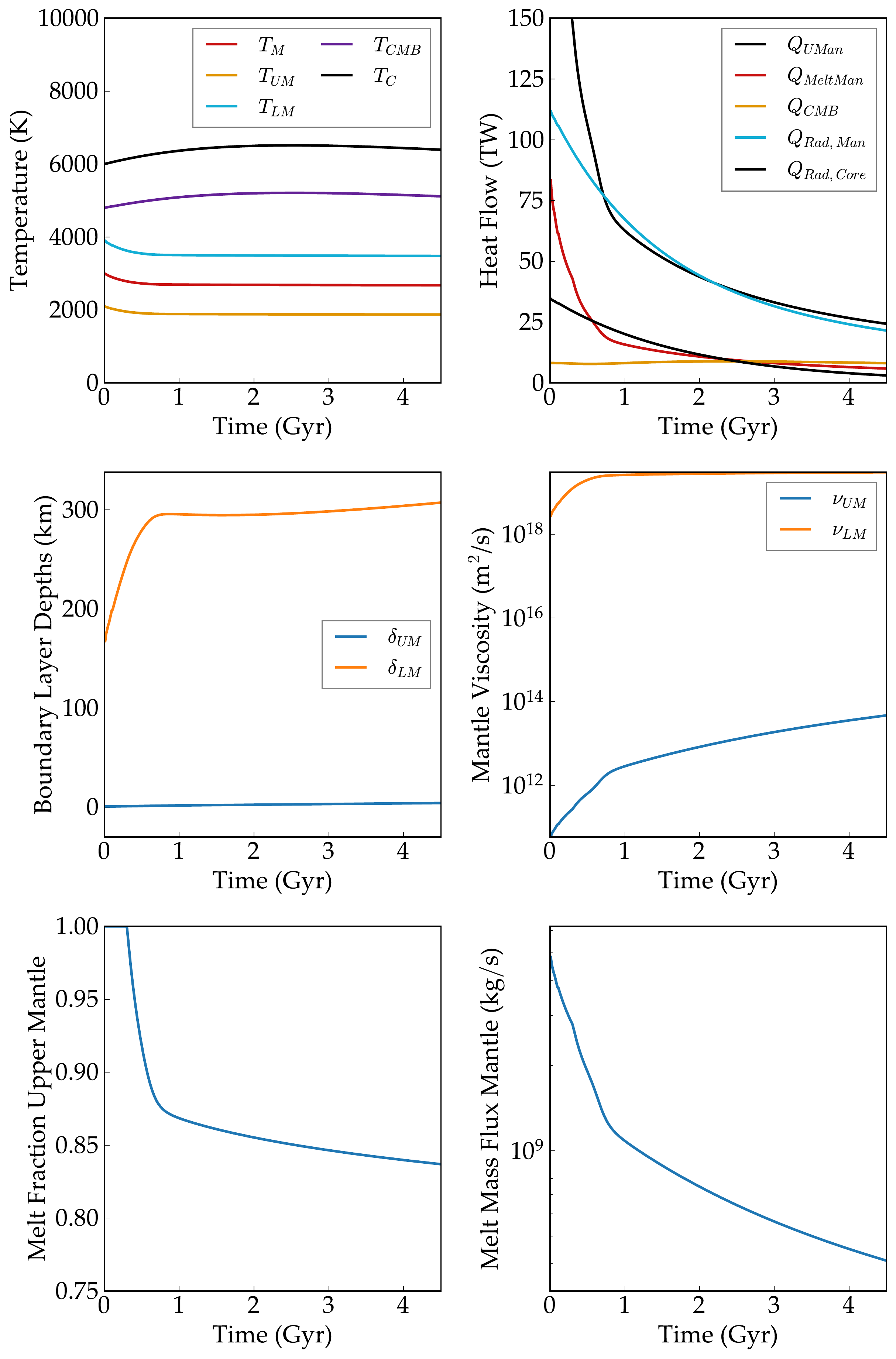}
\end{center}
\caption{\label{fig:venus1} Thermal evolution of the Venus interior model.  All parameters of the model are the same as the Earth model except the following: $\nu_r=2.2\times10^9$, $A_\nu=3.5\times10^5$, $\epsilon_{erupt}=0.01$, $Q_{rad,man}^*=20$ TW, and  to account for the stagnant lid $Q_{conv}$ is multiplied by a factor of $1/25$ \citep{solomatov1995,DriscollBercovici14}.  Compare to \cite{DriscollBercovici14} Fig 6. Approximate runtime: 1 second. \href{https://github.com/VirtualPlanetaryLaboratory/vplanet/tree/master/examples/VenusApproxInterior}{\link{examples/VenusApproxInterior}}}
\end{figure*}

\begin{figure*}[ht!]
\begin{center}
\includegraphics[width=0.75\textwidth]{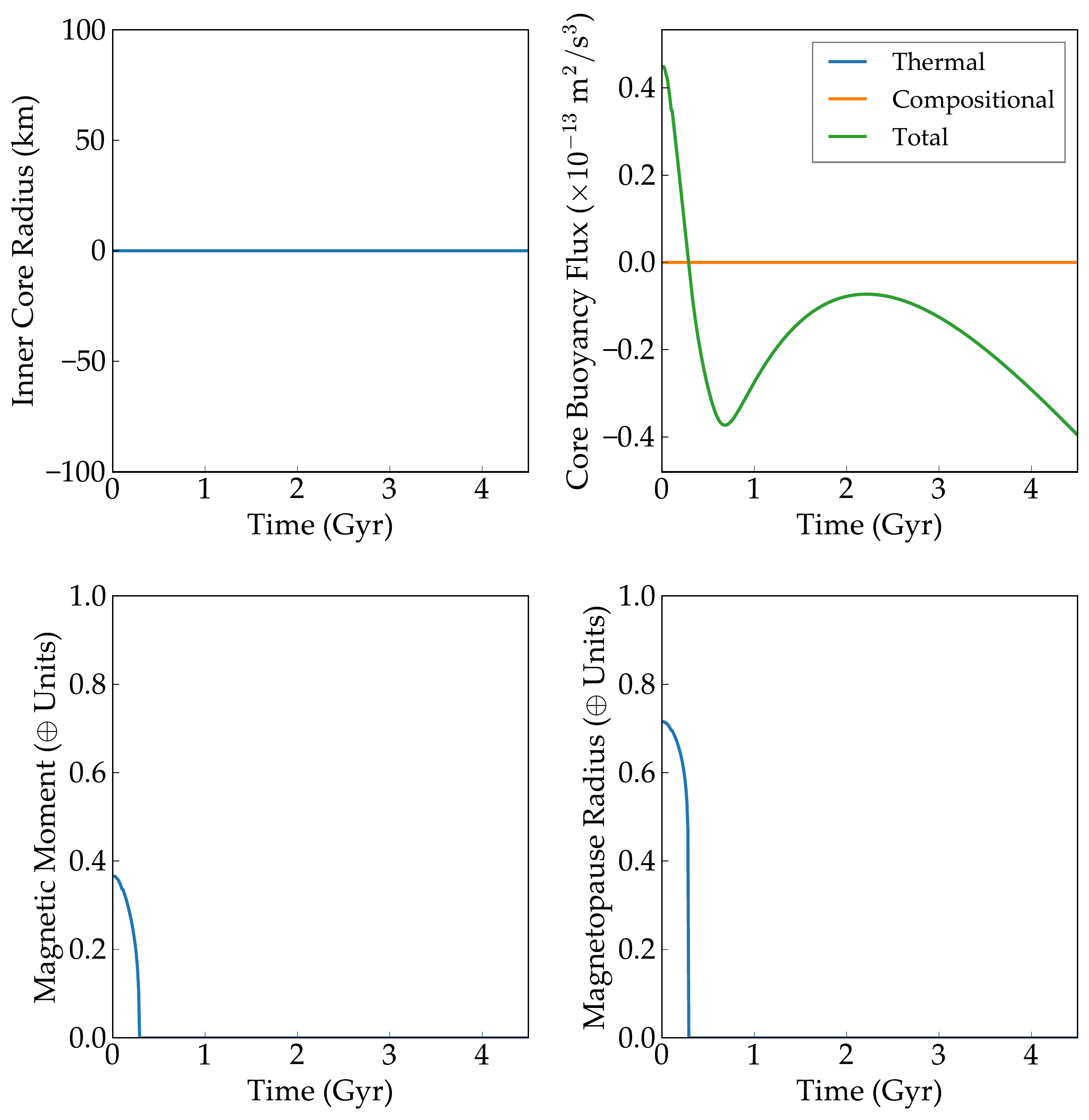}
\end{center}
\caption{\label{fig:venus2} Thermal and magnetic evolution of the Venus interior model in Figure \ref{fig:venus1}. Compare to \cite{DriscollBercovici14}, Fig. 6. \href{https://github.com/VirtualPlanetaryLaboratory/vplanet/tree/master/examples/VenusApproxInterior}{\link{examples/VenusApproxInterior}}}
\end{figure*}

Figures \ref{fig:venus1} and \ref{fig:venus2} show the thermal evolution of the mantle and core. Due to the isolation of the stagnant lid the core heats up over time and the mantle cools very little, which causes thermal convection and dynamo action to cease in the Venusian core around $0.5$ Gyr. Decreasing the mantle viscosity would cause the dynamo to die later or not at all.  These results are very similar to those of \cite{DriscollBercovici14}, Fig. 6, but are not an exact match as our model has been updated slightly.

\subsubsection{Atmospheric Loss\label{sec:venus:atm}}

\begin{figure*}[ht!]
\begin{center}
\includegraphics[width=0.95\textwidth]{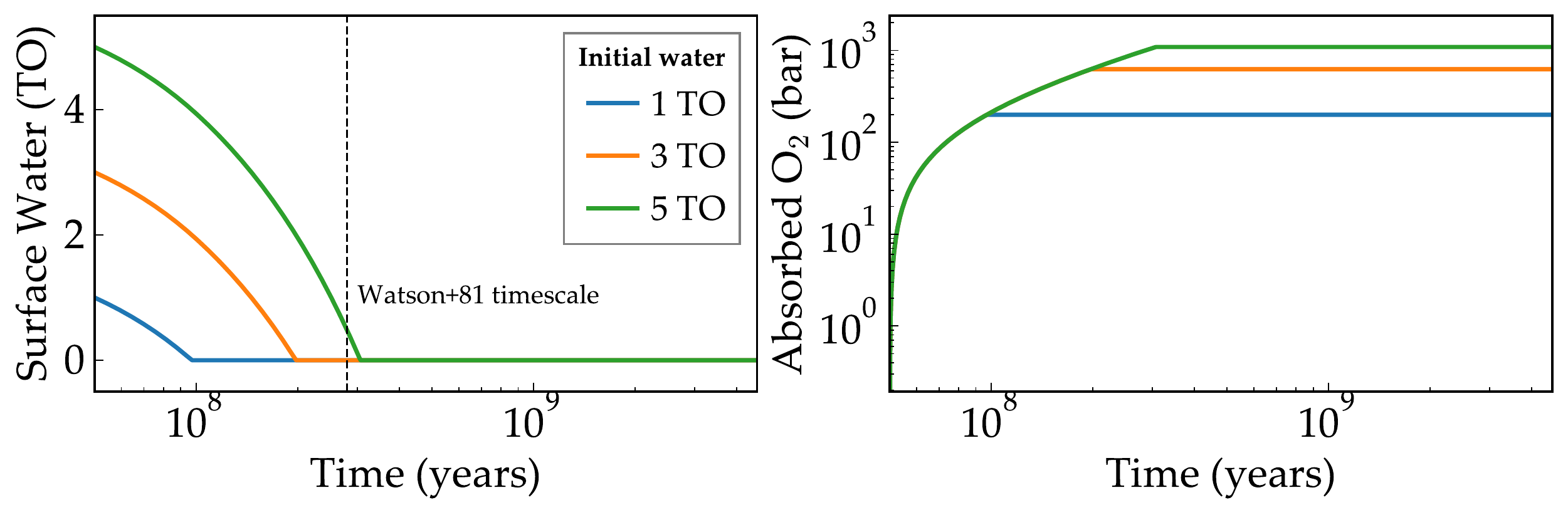}
\end{center}
\caption{\label{fig:venus} Atmospheric escape on Venus as predicted by \vplanet. The evolution of the surface water content (left) and the amount of photolytically-produced oxygen absorbed by the surface (right) are plotted for three
different initial surface water inventories. For an initial
water inventory equal to that of Earth, Venus could have
been completely desiccated within the first 100 Myr following
its formation. \href{https://github.com/VirtualPlanetaryLaboratory/vplanet/tree/master/examples/VenusWaterLoss}{\link{examples/VenusWaterLoss}}}
\end{figure*}

Venus is widely believed to have
once had a substantial surface water inventory (comparable to that of Earth) that was subsequently lost to both thermal and non-thermal escape processes \citep{Watson81,Kasting84,Kasting88,Chassefiere96,Chassefiere96Icarus,Gillmann09}. However, the total initial amount of water and the rate at which hydrogen escaped are extremely uncertain, as measurements of D/H fractionation \citep[\eg][]{Donahue82} only place a lower limit on the total amount lost.

Figure~\ref{fig:venus} shows the evolution of Venus' water content due to hydrodynamic escape in the first ${\sim}200$ Myr following its formation, assuming the XUV evolution law of \citet{Ribas05} and the escape efficiency model of \citet{Bolmont12}. The results in the left panel of Figure~\ref{fig:venus} are consistent with estimates that Venus may have lost on the order of one to a few terrestrial oceans of water in the first several hundred Myr. The dashed vertical line at $t=280$ Myr is the timescale for the loss of 1 TO predicted by \citet{Watson81}. As can be seen from the figure, our model predicts that nearly 5 TO can be lost in that amount of time. This discrepancy is due to two reasons. First, \citet{Watson81} assumed a constant value of $\epsilon_\mathrm{XUV} \mathcal{F}_\mathrm{XUV} \approx 2\, \mathrm{erg/cm^2/s}$ for Venus, which they computed from estimates of the current XUV flux at Earth and an efficiency $\epsilon_\mathrm{XUV} = 0.15$. Since then, studies have shown that the XUV flux from the Sun was about two orders of magnitude higher during the first 100 Myr \citep{Ribas05}, resulting in a much shorter timescale for ocean loss. Second, \citet{Watson81} did not account for the hydrodynamic drag of oxygen, which strongly damps the net rate of ocean loss during the first several tens of Myr \citep{LugerBarnes15}. Together, these effects lead to a timescale for the loss of 1 TO that is approximately 3 times shorter: about 100 Myr.

For reference, the right panel in the Figure shows the amount of photolytically produced oxygen that is retained by the planet, ranging from a few to several hundred bars. The vast majority of this oxygen would have gone into oxidation of the surface.

\subsection{Water Loss During the Pre-Main Sequence\label{sec:abiotico2}}
\begin{figure*}[ht!]
\begin{center}
\includegraphics[width=\textwidth]{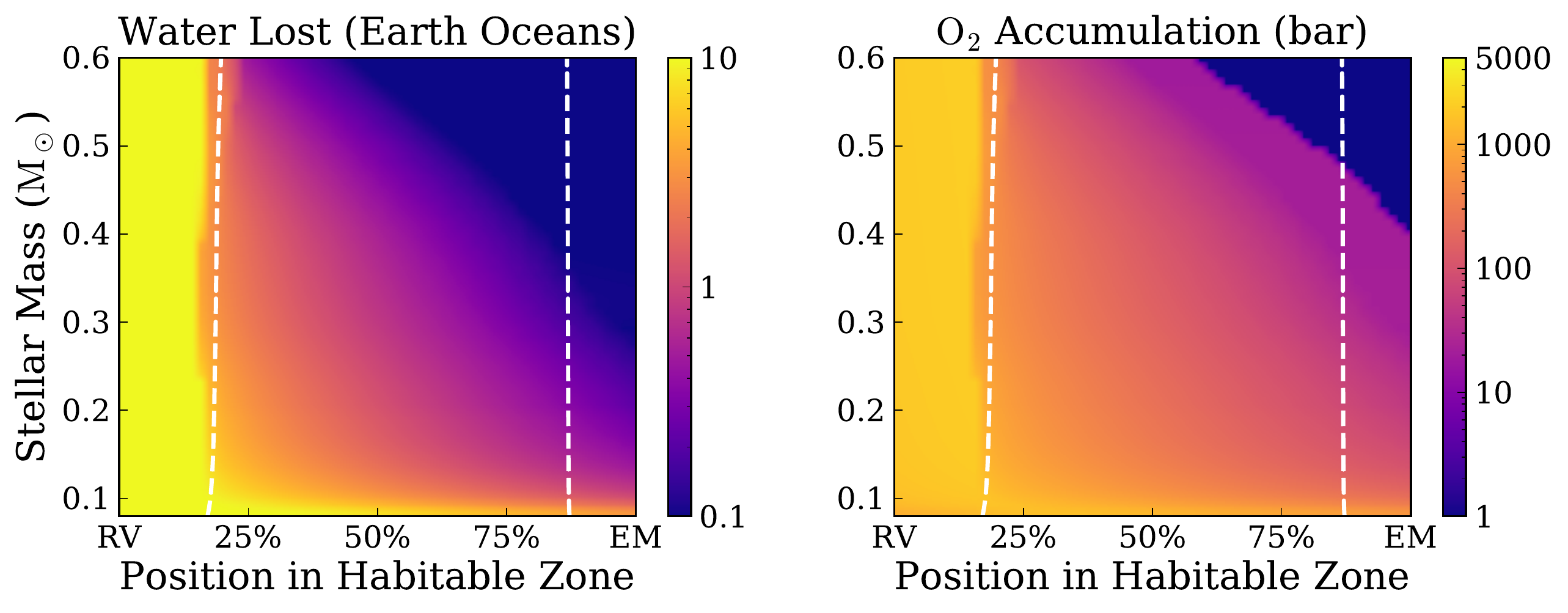}
\end{center}
\caption{\label{fig:o2buildup} Reproduction of Figure 7 in \citet{LugerBarnes15} using \vplanet. Shown here is the total amount of water lost (left) and the total amount of atmospheric oxygen that builds up (right) due to atmospheric escape on an Earth-mass planet orbiting different mass M dwarfs (vertical axes) and at different relative positions in the HZ (horizontal axes). The HZ is bounded by the recent Venus (RV) limit at the inner edge and the early Mars (EM) limit at the outer edge; the runaway greenhouse and maximum greenhouse limits are shown as dashed lines. The amount of water lost is somewhat lower than in \citet{LugerBarnes15} due primarily to the lower escape efficiency assumed here. Approximate runtime: 10 minutes. \href{https://github.com/VirtualPlanetaryLaboratory/vplanet/tree/master/examples/AbioticO2}{\link{examples/AbioticO2}}}
\end{figure*}

We recover the result that planets near the inner edge of the HZ of low mass M dwarfs can lose one to several oceans of water and produce hundreds to a few thousand bars of atmospheric O$_2$. In Figure~\ref{fig:o2buildup} we reproduce Figure~7 in \citet{LugerBarnes15}, showing the amount of water lost and the amount of atmospheric oxygen that builds up for a water-rich Earth-mass planet orbiting an M dwarf. Our water loss estimates are somewhat lower than those in \citet{LugerBarnes15} (by up to a factor of 2), primarily because of the lower escape efficiency predicted by the model of \citet{Bolmont17}, which
we use here, see Appendix~\ref{app:atmesc}. We also account for the increasing mixing ratio of oxygen as water is lost, which acts to slow the escape of hydrogen.

\section{Discussion and Conclusions\label{sec:conclusions}}

In the previous sections we described the \vplanet algorithm ($\S$~\ref{sec:vplanet}) and how individual modules ($\S\S$~\ref{sec:atmesc}--\ref{sec:thermint}) and module combinations ($\S$~\ref{sec:multi}) reproduce various previous results. Moreover, \vplanet has already been used for in novel investigations \citep{Deitrick18a,Deitrick18b,Fleming18,Lincowski18, Fleming19}, demonstrating that this approach can can provide new insight into planetary system evolution and planetary habitability. Furthermore, these insights can be tested; for example, \cite{Fleming18} coupled \stellar and \eqtide to derive a mechanism that removes circumbinary planets orbiting tight binaries, an hypothesis that can be falsified by upcoming \tess observations --- should it discover such planets, then the model is incorrect.

While the previous sections showed the coupling of many modules, not all module couplings have been tested yet. This situation is partly due to some modules being incompatible, \eg \distorb and \spinbody, but also due to the sheer number of combinations that are possible. With this first release, 20 module combinations have been validated against observations and previous results, see $\S\S$~\ref{sec:atmesc}--\ref{sec:multi}. Future research will explore more combinations, but this first version of the code includes a large number of processes that affect planetary system evolution. Future versions will include new physics and additional couplings between modules.

\vplanet can simulate a wide range of planetary systems, but it is still an incomplete model of planetary evolution. The relatively simple modules have important limitations and caveats, which are discussed at length in the previous sections and appendices. Users should consult these sections prior to performing simulations to ensure that they are not pushing the models into unrealistic regions of parameter space. Furthermore, we also urge caution when coupling module combinations not explicitly validated here as their stability and/or accuracy cannot be guaranteed.

The modularity of \vplanet and the spread of modules in this first release provide a framework to build ever more sophisticated models. Future versions could be tailored to particularly interesting systems such as TRAPPIST-1, or particular observations such as planetary spectra. For example, a magma ocean module could be created based on the \cite{Schaefer16} model for GJ 1132 b, and could be combined with tidal heating and a range of radiogenic heating by including the \radheat and \eqtide modules. Or stellar flaring could be added to provide more realistic simulations of atmospheric mass loss.

The discovery of life beyond the Solar System is challenging, in part because resources are scarce and planets are complicated systems. \vplanet's flexibility and speed permits parameter sweeps that can help allocate those resources efficiently, be they telescopes or computer time for more sophisticated, \ie computationally expensive, software packages. While numerous models and codes have been created to simulate planetary evolution, we are aware of none that is as broad and flexible as \vplanet. This paper has described not just its physics modules, but also a novel software design that facilitates interdisciplinary science: the function pointer matrix (see $\S$~\ref{sec:vplanet}). Furthermore the open source nature of the code, extensive documentation, and code integrity checks (see Appendix~\ref{app:support}) ensure transparency and reproducible results. These software engineering practices combined with the rigorous validations described in $\S\S$~\ref{sec:atmesc}--\ref{sec:multi} ensure that \vplanet is a reliable platform for the study of planetary system evolution and planetary habitability.

\vspace{1cm}
This work was supported by the NASA Virtual Planetary Laboratory Team which is funded under NASA Astrobiology Institute Cooperative Agreement Number NNA13AA93A, and Grant Number 80NSSC18K0829. Additional support was provided by NASA grants NNX15AN35G, and 13-13NAI7\_0024. DPF is supported by NASA Headquarters under the NASA Earth and Space Science Fellowship Program - Grant 80NSSC17K0482. This work also benefited from participation in the NASA Nexus for Exoplanet Systems Science (NExSS) research coordination network. We thank an anonymous referee whose comments greatly improved the quality of this manuscript. We are also grateful for stimulating conversations with Brian Jackson, H{\'e}ctor Martinez-Rodriguez, Terry Hurford, Ludmila Carone, Juliette Becker, John Ahlers, Quadry Chance, and Nathan Kaib.

\appendix

\section{The \texttt{AtmEsc} Module\label{app:atmesc}}

The escape of a planet's atmosphere to space is an extremely complex process. The rate at which a gaseous species escapes from a planet strongly depends on factors including, but not limited to, the magnetic properties of the planet and the host star, the space weather the planet is exposed to, the wavelength-dependent irradiation of the planet's atmosphere, as well as the temperature-pressure profile of the atmosphere and its detailed composition, down to the abundance of trace gases that can act as coolants. Decades of work on solar system bodies have enabled the measurement and modeling of the escape fluxes from the Earth, Mars, and Venus using complex hydrodynamic and kinetic models \citep{Hunten73,Watson81,Donahue82,Kasting1983,Hunten87,Zahnle1988,Chassefiere2004}. For extrasolar planets, however, the situation is drastically different. Even for the most well-studied exoplanets, little is known at present about their bulk properties other than their radii, their instellations, and occasionally their masses. Some constraints have been placed on the bulk atmospheric composition of some hot exoplanets via transit transmission spectroscopy, but even in the most favorable cases, little is known other than the presence or absence of a large hydrogen/helium envelope \citep[\eg][]{Nortmann18,Allart19} or loose constraints on the presence of simple molecules such as CO$_2$ and H$_2$O \citep[\eg][]{Line14,MacDonald19}. Stellar activity measurements can yield information about the space weather that some of these exoplanets are exposed to, but the measurement of an exoplanet's magnetic properties is yet to be made \citep[\eg][]{DriscollOlson11,Lynch18}. On the observation front, hydrogen escape fluxes have been inferred for only a few large, hot exoplanets from Lyman-alpha absorption measurements \citep[\eg][]{Odert19}.

However, while precious little is known about the atmospheric escape process from most (individual) exoplanets, recent studies have leveraged the statistical information from the ensemble of all known exoplanets to infer trends in atmospheric escape as a function of planet size and irradiation \citep{LopezRice16,OwenWu17}. These studies show that the distribution of radii of hot exoplanets discovered by the \emph{Kepler} mission are well explained, on average, by a surprisingly simple atmospheric escape model, introduced by \citet{Watson81} and based on investigations of the solar wind by \citet{Parker1964}. In what is commonly referred to as an \emph{energy-limited} model, the escape from a planetary atmosphere is driven by the supply of energy to the upper atmosphere by stellar extreme ultraviolet (XUV; 1--1000\AA) photons, which are absorbed by hydrogen atoms and converted into kinetic energy. In the simplest form of the model, a fixed fraction $\epsilon_\mathrm{XUV}$ of the incoming XUV energy goes into driving the escape
\citep{Watson81,Erkaev07,Lammer2013,Volkov13,Johnson13}. For a hydrogen-dominated atmosphere, the energy-limited particle escape rate $F_\mathrm{EL}$ is obtained by equating the energy provided by XUV photons to the energy required to lift the atmosphere out of the gravitational potential well:
\begin{align}
F_\mathrm{EL} = \frac{\epsilon_\mathrm{XUV}\mathcal{F}_\mathrm{XUV}R_\mathrm{p}}{4GM_\mathrm{p}K_\mathrm{tide}m_\mathrm{H}},
\label{eq:dfhdt}
\end{align}
where $\mathcal{F}_\mathrm{XUV}$ is the XUV energy flux, $M_\mathrm{p}$ is the mass of the planet, $R_\mathrm{p}$ is the planet radius, $\epsilon_\mathrm{XUV} \approx 0.1$ is the XUV absorption efficiency, and $K_\mathrm{tide}$ is a tidal correction term of order unity \citep{Erkaev07}. The total escape rate is this quantity integrated over the surface area of the planet,
whose effective radius to incoming XUV energy is $R_\mathrm{XUV} \approx R_p$. For terrestrial planets, we compute this quantity as
\begin{align}
    \label{eq:lehmer_rad_xuv}
    R_{XUV} = \frac{R_{p}^2}{H \ln(p_{XUV}/p_s) + R_p}
\end{align}
\citep{Lehmer17}, where $H$ is the atmospheric scale height,
$p_{XUV}$ is the pressure at the effective XUV absorption level, and
$p_s$ is the pressure at the surface.

In the absence of detailed information about the properties of an exoplanet that can control or modulate the atmospheric escape rate, we implement this simple model for atmospheric escape in \vplanet, with a few modifications to explicitly model the escape rate from potentially habitable terrestrial planets. Our model closely follows that of \cite{Luger15} and \cite{LugerBarnes15}. Here we briefly discuss the principal equations and slight  modifications to the models presented in those papers.

In \vplanet, we model atmospheric escape from two basic types of atmospheres: hydrogen-dominated atmospheres, such as that of an Earth or super-Earth with a thin primordial hydrogen/helium envelope, and water vapor-dominated atmospheres, such as that of a terrestrial planet in a runaway greenhouse. In the former case, we compute the escape in the energy-limited regime, Equation~(\ref{eq:dfhdt}), and assume that the hydrogen envelope must fully escape before any other volatiles can be lost to space, given the expected large diffusive separation between light H atoms and other atmospheric constituents. If the envelope is not lost by the time the star reaches the main sequence, we shut off the escape process to account for the transition to ballistic escape predicted by \cite{OwenMohanty16}. We model the planet's radius with the evolutionary tracks for super-Earths of \cite{Lopez12} and \cite{LopezFortney14}. If an exoplanet loses its H/He envelope, we compute its solid radius using the \citet{Sotin07} mass-radius relation. The XUV flux is computed from stellar evolution tracks (Appendix~\ref{app:stellar}) and the XUV absorption efficiency parameter $\epsilon_\mathrm{XUV}$ is a tunable constant.

\begin{figure}[tbh]
    \centering
    \includegraphics[width=\textwidth]{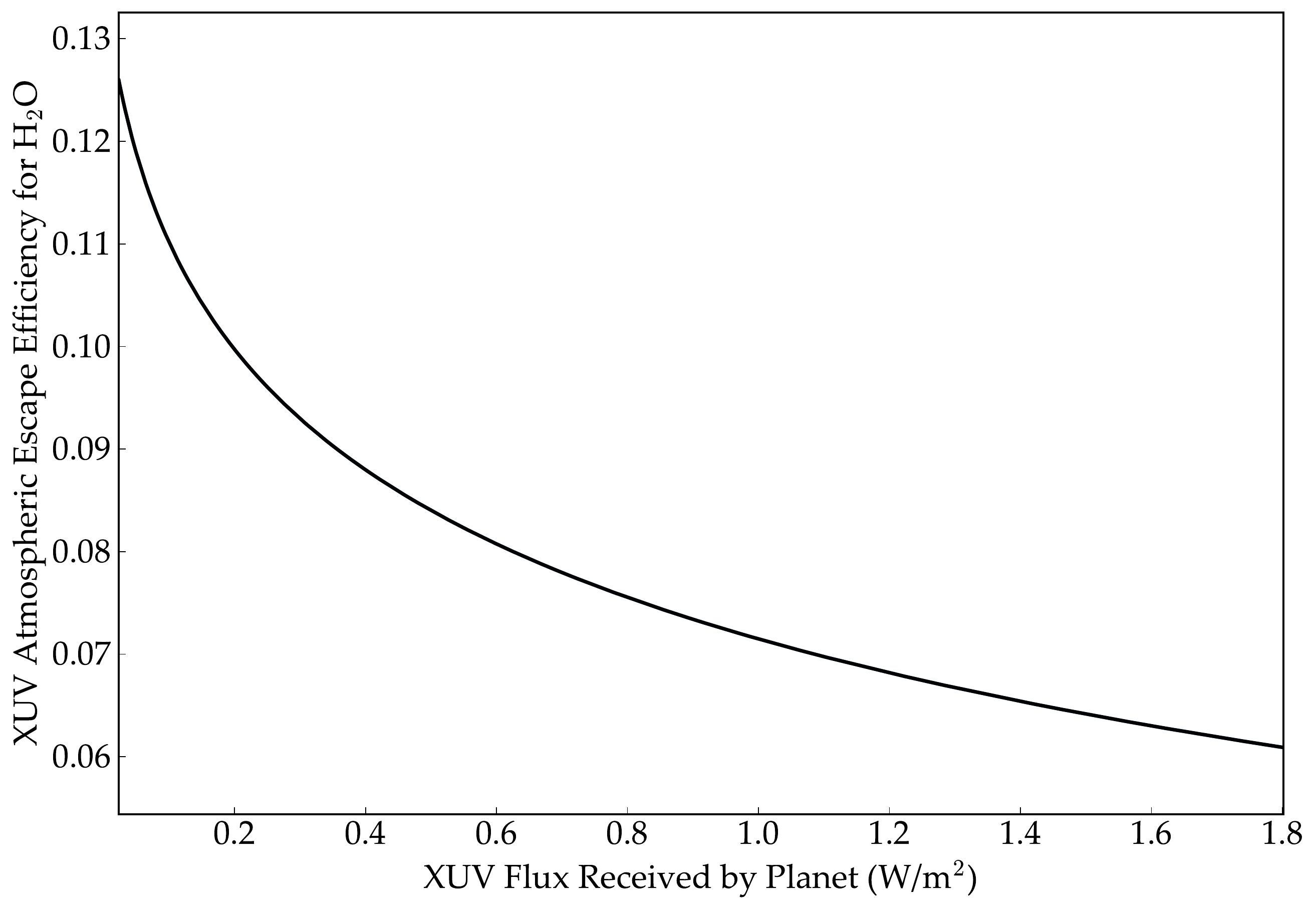}
    \caption{Relationship between the XUV escape efficiency parameter $\epsilon_\mathrm{xuv}$ as found by \cite{Bolmont17}. \href{https://github.com/VirtualPlanetaryLaboratory/vplanet/tree/master/examples/VenusWaterLoss}{\link{examples/VenusWaterLoss}}}
    \label{fig:spinbody_cons}
\end{figure}

In the case of a terrestrial planet with no hydrogen/helium envelope, we assume atmospheric escape only takes place if the total flux incident on the planet exceeds
the runaway greenhouse threshold, computed from
the equations in \citet{Kopparapu13}. Typically, the fluxes experienced by planets in or near the habitable zone are not high enough to drive the hydrodynamic escape of the high mean molecular weight bulk atmosphere. Although water vapor can be photolyzed by stellar ultraviolet photons, liberating hydrogen atoms that can go on to escape hydrodynamically, on Earth this process is strongly inhibited by the stratospheric cold trap, which prevents water molecules from reaching the upper atmosphere. However, during a runaway greenhouse, the surface temperature exceeds the temperature of the critical point of water (647 K) and the surface oceans fully evaporate, leading to an upper atmosphere that is dominated by water vapor \citep[\eg][]{Kasting88}. As in \citet{LugerBarnes15}, we use the energy-limited formalism to compute the loss rate of a planet's surface water via escape of hydrogen to space, with modifications to allow for the hydrodynamic drag of oxygen by the escaping hydrogen atoms. The total hydrogen particle escape rate is \citep{LugerBarnes15}:
\begin{align}
\label{eq:fhfhref}
F_\mathrm{H} =
  \begin{dcases}
   F_\mathrm{EL} & \text{if } F_\mathrm{EL} < F_\mathrm{diff}\\
   F_\mathrm{EL}\left( 1 + \frac{X_\mathrm{O}}{1 - X_\mathrm{O}}\frac{m_\mathrm{O}}{m_\mathrm{H}}\frac{m_\mathrm{c}-m_\mathrm{O}}{m_\mathrm{c}-m_\mathrm{H}} \right)^{-1} & \text{if } F_\mathrm{EL} \ge F_\mathrm{diff},
  \end{dcases}
\end{align}
where
\begin{align}
    F_\mathrm{diff} =  \frac{(m_\mathrm{O} - m_\mathrm{H})(1 - X_\mathrm{O})b_\mathrm{diff}gm_\mathrm{H}}{k_\mathrm{boltz} T_\mathrm{flow}}
\end{align}
is the diffusion-limited flux of oxygen atoms through a background atmosphere of hydrogen and
\begin{align}
\label{eq:mcfhref}
m_\mathrm{c} =
    \frac{1 + \frac{m_\mathrm{O}^2}{m_\mathrm{H}^2}\frac{X_\mathrm{O}}{1 - X_\mathrm{O}}}
        {1 + \frac{m_\mathrm{O}}{m_\mathrm{H}}\frac{X_\mathrm{O}}{1 - X_\mathrm{O}}} m_\mathrm{H} +
    \frac{k_\mathrm{boltz}T_\mathrm{flow}F_\mathrm{H}^\mathrm{ref}}{\left(1 + X_\mathrm{O}\left(\frac{m_\mathrm{O}}{m_\mathrm{H}} - 1\right)\right)b_\mathrm{diff}g}
\end{align}
is the \emph{crossover mass}, the largest particle mass that can be dragged upward in the flow. In the expressions above,
$m_\mathrm{H}$ and $m_\mathrm{O}$ are the masses of the hydrogen and oxygen atoms, respectively, $k_\mathrm{boltz}$ is the Boltzmann constant, $T_\mathrm{flow}$ is the temperature of the hydrodynamic flow, set here to 400 K \citep{Hunten87,Chassefiere96Icarus}, $b_\mathrm{diff} = 4.8\times 10^{17}(T_\mathrm{flow}/\mathrm{K})^{0.75}\ \mathrm{cm^{-1}s^{-1}}$ \citep{Zahnle86} is the binary diffusion coefficient for the two species, $g$ is the acceleration of gravity, and $X_\mathrm{O}$ is the oxygen molar mixing ratio at the base of the flow, equal to $\frac{1}{3}$ when the upper atmosphere is water vapor-dominated. As in \cite{Tian15} and \cite{Schaefer16}, we account for the increasing
mixing ratio of oxygen at the base of the hydrodynamic flow, which
slows the escape of hydrogen. \cite{Tian15} finds that, as oxygen
becomes the dominant species in the upper atmosphere, the
\cite{Hunten87} formalism predicts that an oxygen-dominated flow can
rapidly lead to the loss of all O$_2$ from planets around M
dwarfs. However,
hydrodynamic oxygen-dominated escape requires exospheric temperatures
$\sim m_\mathrm{O}/m_\mathrm{H} = 16$ times higher than that for a
hydrogen-dominated flow, which is probably unrealistic for most planets. Following the prescription of \cite{Schaefer16}, we therefore
shut off oxygen escape once its mixing ratio exceeds $X_\mathrm{O} =
0.6$ (corresponding to an equal number of O$_2$ and H$_2$O molecules at the base of the flow), switching to the diffusion-limited escape rate of
hydrogen. Finally, users can either choose a constant value for $\epsilon_\mathrm{XUV}$ or model it as a function of the incoming XUV flux as in \citet{Bolmont17}. In \citet{Bolmont17}, the authors
modeled atmospheric loss with a set of 1D radiation-hydrodynamic simulations that allowed
them to calculate the XUV escape efficiency,
$\epsilon_\mathrm{XUV}$. In Fig. \ref{fig:spinbody_cons}, we show a subset of
the range of $\epsilon_\mathrm{XUV}$ values
calculated in \citet{Bolmont17} for some values
of stellar XUV flux received by the planet. As
\stellar can calculate the star's changing XUV
flux over time, the XUV escape efficiency
will change over time if set by the
\citet{Bolmont17} model.

Currently, \atmesc does not model the absorption of oxygen by surface sinks, although users can run the code in two limiting cases: efficient surface sinks, corresponding to (say) a reducing magma ocean that immediately absorbs any photolytically produced oxygen; and inefficient surface sinks, corresponding to (say) a fully oxidized surface, leading to atmospheric buildup of O$_2$ over time. Upcoming modifications to \atmesc will couple it to the geochemical evolution of the planet's mantle in order to more realistically compute the rate of oxygen buildup in a hydrodynamically escaping atmosphere.

As a word of caution, it is important to reiterate that the energy-limited formalism we adopt in \atmesc is a very approximate description of the escape of an atmosphere to space. The heating of the upper atmosphere that drives hydrodynamic escape is strongly wavelength dependent and varies with both the composition and the temperature structure of the atmosphere, which we do not model. Moreover, line cooling mechanisms such as recombination radiation scale non-linearly with the incident flux. Non-thermal escape processes, such as those controlled by magnetic fields, flares, and/or coronal mass ejections, lead to further departures from the simple one-dimensional energy-limited escape rate. Nevertheless, as we argued above, several studies show that for small planets the escape rate does indeed scale with the stellar XUV flux and inversely with the gravitational potential energy of the gas \citep[\eg][]{Lopez12,Lammer2013,OwenWu13,OwenWu17} and that $\epsilon_\mathrm{xuv} \approx 0.1$ is a reasonable median value that predicts the correct escape fluxes within a factor of a few.
Since presently we have little information about the
atmospheric structure of exoplanets,
we choose to employ the energy-limited approximation and fold all of our uncertainty regarding the physics of the escape process into the XUV escape efficiency $\epsilon_\mathrm{xuv}$. Future versions of \vplanet can include diverse models, such as radiation-recombination limited escape \citep[\eg][]{MurrayClay09}, to more accurately track atmospheric evolution.

\section{The \texttt{BINARY} Module\label{app:binary}}

Here we describe the analytic theory for circumbinary orbits of test particles derived by \citet{Leung2013}. We adopt a cylindrical coordinate system centered on the binary center of mass, the system barycenter in this case, and consider the test particles to be massless circumbinary planets (CBPs). Assuming that the binary orbit lies in the $x-y$ plane, and that the orbit of the CBP is nearly coplanar, \citet{Leung2013} approximate the gravitational potential felt by the CBP due to the binary at the position $(R,\phi,z)$ as
\begin{equation} \label{eqn:binary_pot}
\begin{split}
    \Phi(R,\phi,z) &= \sum_{k=0}^{\infty} \Biggl[ \Phi_{0k0}(R) - \frac{1}{2}\left( \frac{z}{R} \right)^2 \Phi_{2k0}(R) + ... \Biggr] \cos k(\phi - M_b - \varpi_B) \\
    &+ e_{AB} \sum_{k=0}^{\infty} \Biggl( k \Biggl[\Phi_{0k0}(R) - \frac{1}{2}\left(\frac{z}{R}\right)^2 \Phi_{2k0}(R) + ... \Biggr]\\
    &- \frac{1}{2} \Biggl[ \Phi_{0k1}(R) - \frac{1}{2}\left( \frac{z}{R} \right)^2 \Phi_{2k1}(R) + ... \Biggr] \Biggr) \cos(k(\phi - \varpi_B) - (k+1)M_B) \\
    & + e_{AB} \sum_{k=0}^{\infty} \Biggl( -k \left[\Phi_{0k0}(R) - \frac{1}{2} \left( \frac{z}{R} \right)^2 \Phi_{2k0}(R) + ... \right] \\
    & - \frac{1}{2} \left[\Phi_{0k1}(R) - \frac{1}{2} \left( \frac{z}{R} \right)^2 \Phi_{2k1}(R) + ... \right] \Biggr) \cos(k(\phi - \varpi_B) - (k-1)M_B),
\end{split}
\end{equation}
for integer $k$, binary mean anomaly $M_b$, binary orbital eccentricity $e_{AB}$, binary longitude of the periapse $\varpi_B$, and R is the radial distance from the CBP to the binary center of mass. This expression, correct to first order in $e_{AB}$ as indicated by the ellipses, contains the two gravitational potential components from the stars that are in general not axisymmetric.  The two components, $\Phi_{jk0}(R)$ and $\Phi_{jk1}(R)$, are given by the following expressions
\begin{equation} \label{eqn:binary_potjk0}
    \Phi_{jk0}(R) = -\frac{2-\delta_{k0}}{2} \Biggl[ (-1)^k \frac{m_A}{(m_A+m_B)}b^k_{(j+1)/2}(\alpha_A) + \frac{m_B}{(m_A+m_B)}b^k_{(j+1)/2}(\alpha_B) \Biggr]\frac{G(m_A+m_B)}{R},
\end{equation}
and
\begin{equation} \label{eqn:binary_potjk1}
    \Phi_{jk1}(R) = -\frac{2-\delta_{k0}}{2} \Biggl[ (-1)^k \frac{m_A}{(m_A+m_B)}\alpha_A Db^k_{(j+1)/2}(\alpha_A) + \frac{m_B}{(m_A+m_B)}\alpha_B Db^k_{(j+1)/2}(\alpha_B) \Biggr]\frac{G(m_A+m_B)}{R},
\end{equation}
where $b^k_{(j+1)/2}$ is a Laplace coefficient, $D = \partial/\partial \alpha$,  $m_A$ and $m_B$ are the masses of the primary and secondary star, respectively, $G$ is the Universal Gravitational constant, and $\delta_{k0}$ is the Kroeneker delta function.  For a CBP located at cylindrical position $R$, $\alpha_A$ and $\alpha_B$ are the normalized semi-major axis of the CBP relative to each star, given by
\begin{equation} \label{eqn:binary_alpha}
    \alpha_i = \frac{a_{AB} m_i}{R(m_A + m_B)},
\end{equation}
where $a_{AB}$ is the binary orbital semi-major axis and the index $i$ is $A$ for the primary and $B$ for the secondary star, respectively.

Given the approximation for the binary gravitational potential in Eq.~(\ref{eqn:binary_pot}), $\Phi$, the equations that govern the motion of the CBP in cylindrical coordinates are given by
\begin{equation} \label{eqn:binary_EqnOfMotion}
    \ddot{R} - R\dot{\phi}^2 = -\frac{\partial \Phi}{\partial R}, R \ddot{\phi} + 2\dot{R}\dot{\phi} = -\frac{1}{R}\frac{\partial \Phi}{\partial \phi}, \ddot{z} = - \frac{\partial \Phi}{\partial z}.
\end{equation}

Following \citet{LeePeale2006}, \citet{Leung2013} approximate the orbit of the CBP as small epicyclic deviations from the circular motion of the guiding center via
\begin{equation} \label{eqn:binary_RPhiZ}
    R = R_0 + R_1(t), \phi = \phi_0(t) + \phi_1(t), z = z_1(t),
\end{equation}
where $R_0$ is the cylindrical radius of the CBP guiding center in the plane of the binary orbit, $0$ subscripts denote the position of the guiding center, $R_1$ is the small, time-dependent epicyclic radial deviation from the guiding center radius.

The Keplerian mean motion at the radius of the CBP's guiding center is
\begin{equation} \label{eqn:binary_nk}
    n_k = \sqrt{G(m_A + m_B)/R_0^3}
\end{equation}
and the phase angle of the circular motion of the CBP's guiding center is
\begin{equation} \label{eqn:binary_phi0}
    \phi_0(t) = n_0t + \psi_0
\end{equation}
for time $t$ and arbitrary constant phase offset, $\psi_0$.

The CBP mean motion is given by
\begin{equation} \label{eqn:binary_n0}
    n_0^2 = \left[ \frac{1}{R}\frac{d\Phi_{000}}{dR} \right]_{R_0},
\end{equation}
where this expression is evaluated at the radius of the guiding center, $R=R_0$.

Given these assumptions and definitions, \citet{Leung2013} solve the equations of motion given in Eq.~(\ref{eqn:binary_EqnOfMotion}) to obtain the radial position, $R$, of the CBP relative to the binary center of mass, the position of the CBP above or below the orbital plane of the binary, $z$, and the phase angle of the CBP, $\phi$, over the course of its orbit as an analytic function of time.  The CBP's radial position, $R$, is given by the following expression
\begin{equation} \label{eqn:binary_CBPR}
\begin{split}
    R &= R_0 \Biggl(1 - e_{\text{free}} \cos(\kappa_0 t + \psi) - C_0 \cos M_\text{B} - \sum_{k=1}^\infty \Biggl[ C_k^0 \cos k(\phi_0 - M_b - \varpi_B) \\
    &+ C_k^+ \cos(k(\phi_0 - \varpi_B) - (k+1)M_b) + C_k^- \cos(k(\phi_0 - \varpi_B) - (k-1)M_b) \Biggr] \Biggr),
\end{split}
\end{equation}
where $e_{free}$ is the CBP's free eccentricity, a free parameter of the model, and $\psi$ is an arbitrary phase offset.  Although the summations in Eq.~(\ref{eqn:binary_CBPR}) extend to $\infty$, we follow \citet{Leung2013} and truncate the summation at $k=3$ for computational speed with minimal loss of accuracy.

The variables $C_0$, $C_k^0$, and $C_k^{\pm}$ represent the approximate radial amplitudes for the forced oscillations due to the non-axisymmetric components of the binary gravitational potential and are given by the following
\begin{equation} \label{eqn:binary_CBPC0}
    C_0 = -e_{AB} \left[ \frac{d\Phi_{001}}{dR} \right]_{R_0} \Biggl/ [R_0 (k_0^2 - n_{AB}^2)],
\end{equation}

\begin{equation} \label{eqn:binary_CBPC0k}
    C_k^0 = \left[\frac{d\Phi_{0k0}}{dR} + \frac{2n\Phi_{0k0}}{R(n-n_{AB})} \right]_{R_0} \Biggl/ (R_0 [\kappa_0^2 - k^2(n_0 - n_{AB})^2]),
\end{equation}
and
\begin{equation} \label{eqn:binary_CBPCKPM}
    C_k^{\pm} = e_{AB} \left[ \pm k \frac{d\Phi_{0k0}}{dR} - \frac{d\Phi_{0k1}}{2dR} + \frac{kn(\pm 2k\Phi_{0k0} - \Phi_{0k1})}{R(kn - (k \pm 1)n_{AB}} \right]_{R_0} \Biggl/ (R_0 [\kappa_0^2 - (kn_0 - (k \pm 1)n_{AB})^2]).
\end{equation}
In these expressions, $k_0$ is the epicyclic frequency and is given by
\begin{equation} \label{eqn:binary_k0}
    \kappa_0^2 = \left[ R \frac{dn^2}{dR} + 4n^2 \right]_{R_0}
\end{equation}
for CBP mean motion, $n$ given by Eq.~(\ref{eqn:binary_n0}), and is evaluated at CBP radial position $R=R_0$.

The CBP's cylindrical phase angle relative to the binary's center of mass is
\begin{equation} \label{eqn:binary_CBPPhi}
\begin{split}
    \phi &= n_0 t + \phi_0 + \frac{2n_0}{\kappa_0}e_{\text{free}} \sin(\kappa_0 t + \psi) + \frac{n_0}{n_{AB}}D_0 \sin M_b \sum_{k=1}^{\infty} \Biggl[ \frac{n_0}{k(n_0 - n_{AB})}D_k^0 \sin k(\phi_0 - M_b - \varpi_B) \\
    &+ \frac{n_0}{kn_0 - (k+1)n_{AB}}D_k^+ \sin(k(\phi_0 - \varpi_B) - (k+1)M_b) + \frac{n_0}{kn_0 - (k-1)n_{AB}}D_k^- \sin(k(\phi_0 - \varpi) \\
    &- (k-1)M_b) \Biggr],
\end{split}
\end{equation}
where $\varphi$ is an arbitrary phase offset and the variables $D_0$, $D_k^0$, and $D_k^{\pm}$ are given by
\begin{equation} \label{eqn:binary_D0}
    D_0 = 2C_0,
\end{equation}

\begin{equation} \label{eqn:binary_Dk0}
   D_k^0 =  2C_k^0 - \left[ \frac{\Phi_{0k0}}{R^2n(n-n_{AB})} \right]_{R0},
\end{equation}
and
\begin{equation} \label{eqn:binary_Dkpm}
    D_k^{\pm} = 2C_k^{\pm} - e_{AB} \Biggl[\frac{k(\pm2k\Phi_{0k0}-\Phi_{0k1})}{2R^2n(kn-(k\pm1)n_{AB})} \Biggr]_{R_0}.
\end{equation}

The cylindrical position of the CBP above or below the plane of the binary, $z$, is decoupled from the epicyclic motion of the CBP orbital radius and phase angle and is simply
\begin{equation} \label{eqn:binary_CBPZ}
    z = R_0 i_{\text{free}} \cos(\nu_0 t + \zeta),
\end{equation}
where $\zeta$ is an arbitrary phase offset and $i_{\text{free}}$ is the free inclination, a free parameter of the model.  The vertical frequency $\nu_0$ is given by
\begin{equation} \label{eqn:binary_nu0}
    \nu_0^2 = \left[ - \frac{\Phi_{200}}{R^2} \right]_{R_0}.
\end{equation}

\section{The \texttt{DistOrb} Module\label{app:distorb}}

Our model for the orbital evolution, called \distorb (for ``Disturbing function
Orbits''), uses the literal, 4th order disturbing function developed in
\cite{MurrayDermott99} and \cite{Ellis2000}. We use only the secular terms, meaning
that the rapidly varying terms that depend on the mean longitudes of the planets are
ignored on the assumption that these terms will average to zero over long timescales.
This assumption is valid as long as no planets are in the proximity of mean-motion
resonances.

There are two solution methods in \distorb: the first is
a direct Runge-Kutta integration of the fourth-order equations of motion; the second is the Laplace-Lagrange eigenvalue solution, which
reduces the accuracy in the disturbing function to 2nd order, but returns a solution
that is explicit in time, and thus provides a solution in much less computation time and is stable by assumption.
The 4th order solution tends to produce results that match better with N-body; however,
the Laplace-Lagrange solution can be a powerful
predictive tool because it provides secular frequencies directly. One example of the use of
such frequencies is in the prediction of Cassini
states, as shown in $\S$~\ref{sec:cassini} \citep[see also][]{WardHamilton04,Brasser2014,Deitrick18a}.

The equations of motion are Lagrange's equations \citep[see][]{MurrayDermott99}. In
the secular approximation the equations for semi-major axis and mean longitude, and any
disturbing function derivative with respect to these variables, are ignored.
Additionally, to avoid singularities in the equations for the longitudes of pericenter
and ascending node, which occur at zero eccentricity and inclination, respectively, we
rewrite Lagrange's equations and the disturbing function in terms of the variables (a
form of Poincar\'{e} coordinates):
\begin{align}
h & = e \sin{\varpi} \\
k & = e \cos{\varpi} \\
p & = \sin{\frac{i}{2}} \sin{\Omega} \label{eqnp}\\
q & = \sin{\frac{i}{2}} \cos{\Omega} \label{eqnq},
\end{align}
where $e$ is the orbital eccentricity, $i$ is the inclination, $\Omega$ is the longitude
of ascending node, and $\varpi = \omega+\Omega$ is the longitude of periastron (see
Figure \ref{diagpA}).

Lagrange's equations for secular theory are then:
\begin{align}
&\begin{aligned}
\frac{dh}{dt} & = \frac{\sqrt{1-e^2}}{na^2} \frac{\partial\mathcal{R}}{\partial k} +
\frac{k p}{2na^2\sqrt{1-e^2}} \frac{\partial\mathcal{R}}{\partial p} \\
  &\qquad + \frac{k q}{2na^2\sqrt{1-e^2}} \frac{\partial\mathcal{R}}{\partial q},
\end{aligned}\label{eqnhdot}\\
&\begin{aligned}
\frac{dk}{dt} & = -\frac{\sqrt{1-e^2}}{na^2} \frac{\partial\mathcal{R}}{\partial h} -
\frac{h p}{2na^2\sqrt{1-e^2}} \frac{\partial\mathcal{R}}{\partial p} \\
&\qquad-\frac{h q}{2na^2\sqrt{1-e^2}} \frac{\partial\mathcal{R}}{\partial q},
\end{aligned}\label{eqnkdot}\\
&\begin{aligned}
\frac{dp}{dt} & = - \frac{k p}{2na^2\sqrt{1-e^2}} \frac{\partial\mathcal{R}}{\partial h}
+\frac{h p}{2na^2\sqrt{1-e^2}} \frac{\partial\mathcal{R}}{\partial k} \\
&\qquad + \frac{1}{4na^2\sqrt{1-e^2}} \frac{\partial\mathcal{R}}{\partial q},
\end{aligned}\label{eqnpdot}\\
&\begin{aligned}
\frac{dq}{dt} & = - \frac{k q}{2na^2\sqrt{1-e^2}} \frac{\partial\mathcal{R}}{\partial h}
+\frac{h q}{2na^2\sqrt{1-e^2}} \frac{\partial\mathcal{R}}{\partial k}\\
&\qquad - \frac{1}{4na^2\sqrt{1-e^2}} \frac{\partial\mathcal{R}}{\partial p},
\end{aligned} \label{eqnqdot}
\end{align}
where $\mathcal{R}$ is the disturbing function, and $a$, $n$, and $e$ are
the semi-major axis, mean motion, and eccentricity, respectively. See \cite{Berger1991}
for the complete set of Lagrange's equations in $h$, $k$, $p$, and $q$, including
mean-motion (\ie resonant) effects.

General relativity is known to affect the apsidal precession (associated with
eccentricity) of planetary orbits, so we include a correction to Equations
(\ref{eqnhdot}) and (\ref{eqnkdot}). Following \cite{Laskar1986}, the apsidal
corrections are:
\begin{align}
\frac{dh}{dt}\Bigm\lvert_{GR} &= \delta_R k \label{hGRcorr} \\
\frac{dk}{dt}\Bigm\lvert_{GR}& = -\delta_R h,\label{kGRcorr}
\end{align}
where
\begin{equation}
\delta_R = \frac{3 n^3 a^2}{c^2 (1-e^2)}, \label{deltaReqn}
\end{equation}
and $c$ is the speed of light.

For systems well away from mean-motion resonances, \distorb does a reasonable job at
modeling the orbital evolution, up to eccentricities of $\sim0.4$
or mutual inclinations of $\sim35^{\circ}$, largely capturing the frequencies and amplitudes of oscillations to within $5-10\%$. Note, however, that the accuracy is difficult to
quantify in general because of the highly non-linear nature of the orbital problem. See
\cite{Deitrick18a} for further comparisons. However, as noted in \cite{MurrayDermott99}, one of the series expansions used in deriving the disturbing
function diverges at $e=0.6627434$. If the eccentricity of a planet ever exceeds this value,
\distorb immediately halts. In this case, the user will need to switch to an N-body integrator
such as \spinbody, or another N-body code, which can still be
coupled to \distrot and \poise if desired.

Because planetary systems have a large number
of parameters (masses, number of planets, eccentricities, inclinations, etc.), the full limitations of \distorb are difficult to map. Thus we advise
users to always compare a small selection of
cases with an N-body model, to understand \distorb's applicability to the desired planetary
system.

Finally we present, for the sake of completeness, the disturbing function as used in \distorb, in the variables $h,k,p,$ and $q$ in Table \ref{distfxn}. These were originally derived by \cite{Ellis2000}; we have simply applied coordinate transformations and calculated derivatives with respect to the new coordinates. This disturbing function, in its original form, can also be seen in \cite{MurrayDermott99}. We will not restate the semi-major axis functions, $f_1, f_2, f_3$ and so on, in this paper, as they are taken directly from Table B.3 of \cite{MurrayDermott99}.

The secular disturbing function, for any pair of planets, is:
\begin{equation}
\mathcal{R} = \frac{\mu'}{a'}\mathcal{R}_D,
\end{equation}
for the inner body, and
\begin{equation}
\mathcal{R'} = \frac{\mu}{a'}\mathcal{R}_D,
\end{equation}
for the outer body. Here, $a'$ is the semi-major axis of the exterior planet and the mass factors are $\mu = \kappa^2 m$ and $\mu' = \kappa^2 m'$, where $m$ is the mass of the interior planet, $m'$ is the mass of the exterior planet, and $\kappa$ is Gauss' gravitational constant. Finally,
\begin{equation}
\mathcal{R}_D = \text{D}0.1 + \text{D}0.2 +\text{D}0.3 +\dots,
\end{equation}
where the terms D0.1, D0.2, and so on, are given in Table \ref{distfxn}.

\begin{table*}[ht]
\caption{Disturbing function}
\begin{tabular}{lp{0.8\textwidth}}
\hline\hline \\ [-1.5ex]
Term &   \\ [0.5ex]
\hline \\ [-1.5ex]
D0.1 & $f_1 + (h^2+k^2+h'^2+k'^2)f_2+(p^2+q^2+p'^2+q'^2)f_3$\\
	&~$+(h^2+k^2)^2 f_4+(h^2+k^2)(h'^2+k'^2)f_5+(h'^2+k'^2)^2 f_6$\\
	&~$+[(h^2+k^2)(p^2+q^2) + (h'^2+k'^2)(p^2+q^2)$\\
	&~~$ + (h^2+k^2)(p'^2+q'^2)+(h'^2+k'^2)(p'^2+q'^2)] f_7$\\
	&~$ +[(p^2+q^2)^2+(p'^2+q'^2)^2] f_8 +
(p^2+q^2)(p'^2+q'^2)f_9$   \\[0.5ex]
\hline \\[-1.5ex]
D0.2 & $(h h' + k k') [f_{10} + (h^2+k^2)f_{11}+(h'^2+k'^2)f_{12}$\\
  &~~$+(p^2+q^2+p'^2+q'^2)f_{13}]$\\[0.5ex]
\hline \\[-1.5ex]
D0.3 & $(p p' + q q') [f_{14} + (h^2+k^2+h'^2+k'^2)f_{15}$\\
  &~~$+(p^2+q^2+p'^2+q'^2)f_{16}]$\\[0.5ex]
\hline \\[-1.5ex]
D0.4 & $(h^2 h'^2 - k^2 h'^2 - h^2 k'^2 +k^2 k'^2+4 h h' k k')f_{17}$\\[0.5ex]
\hline \\[-1.5ex]
D0.5 & $(h^2 p^2 - h^2 q^2 - k^2 p^2 + k^2 q^2 +4 h k p q)f_{18}$\\[0.5ex]
\hline \\[-1.5ex]
D0.6 & $[h h' (p^2-q^2) - k k' (p^2-q^2) + 2 p q (h k'+k h')]f_{19}$\\[0.5ex]
\hline \\[-1.5ex]
D0.7 & $(h'^2 p^2 - h'^2 q^2 - k'^2 p^2 + k'^2 q^2 +4 h' k' p q)f_{20}$\\[0.5ex]
\hline \\[-1.5ex]
D0.8 & $(h^2 p p' - h^2 q q' - k^2 p p' + k^2 q q' +2 h k p' q + 2 h k p q')f_{21}$\\[0.5ex]
\hline \\[-1.5ex]
D0.9 & $[(h h' +k k') (p p' + q q') +(h k' -k h') (p q' - q p') ]f_{22}$\\[0.5ex]
\hline \\[-1.5ex]
D0.10 & $[(h h' +k k') (p p' + q q') +(h k' -k h') (q p' - p q') ]f_{23}$\\[0.5ex]
\hline \\[-1.5ex]
D0.11 & $[(h h' - k k') (p p' - q q') +(h k' +k h') (p q' +  q p') ]f_{24}$\\[0.5ex]
\hline \\[-1.5ex]
D0.12 & $(h'^2 p p' - h'^2 q q' - k'^2 p p' + k'^2 q q' +2 h' k' p' q + 2 h' k' p q')f_{25}$\\[0.5ex]
\hline \\[-1.5ex]
D0.13 & $(h^2 p'^2 - h^2 q'^2 - k^2 p'^2 + k^2 q'^2 +4 h k p' q')f_{18}$\\[0.5ex]
\hline \\[-1.5ex]
D0.14 & $[h h' (p'^2-q'^2) - k k' (p'^2-q'^2) + 2 p' q' (h k'+k h')]f_{19}$\\[0.5ex]
\hline \\[-1.5ex]
D0.15 & $(h'^2 p'^2 - h'^2 q'^2 - k'^2 p'^2 + k'^2 q'^2 +4 h' k' p' q')f_{20}$\\[0.5ex]
\hline \\[-1.5ex]
D0.16 & $(p^2 p'^2 - p^2 q'^2 - q^2 p'^2 + q^2 q'^2 +4 p q p' q')f_{26}$\\[0.5ex]
\hline \\[-1.5ex]
\end{tabular}
\label{distfxn}
\end{table*}

\section{The \texttt{DistRot} Module\label{app:distrot}}

\begin{figure}[th!]
\begin{center}
\includegraphics[width=\textwidth]{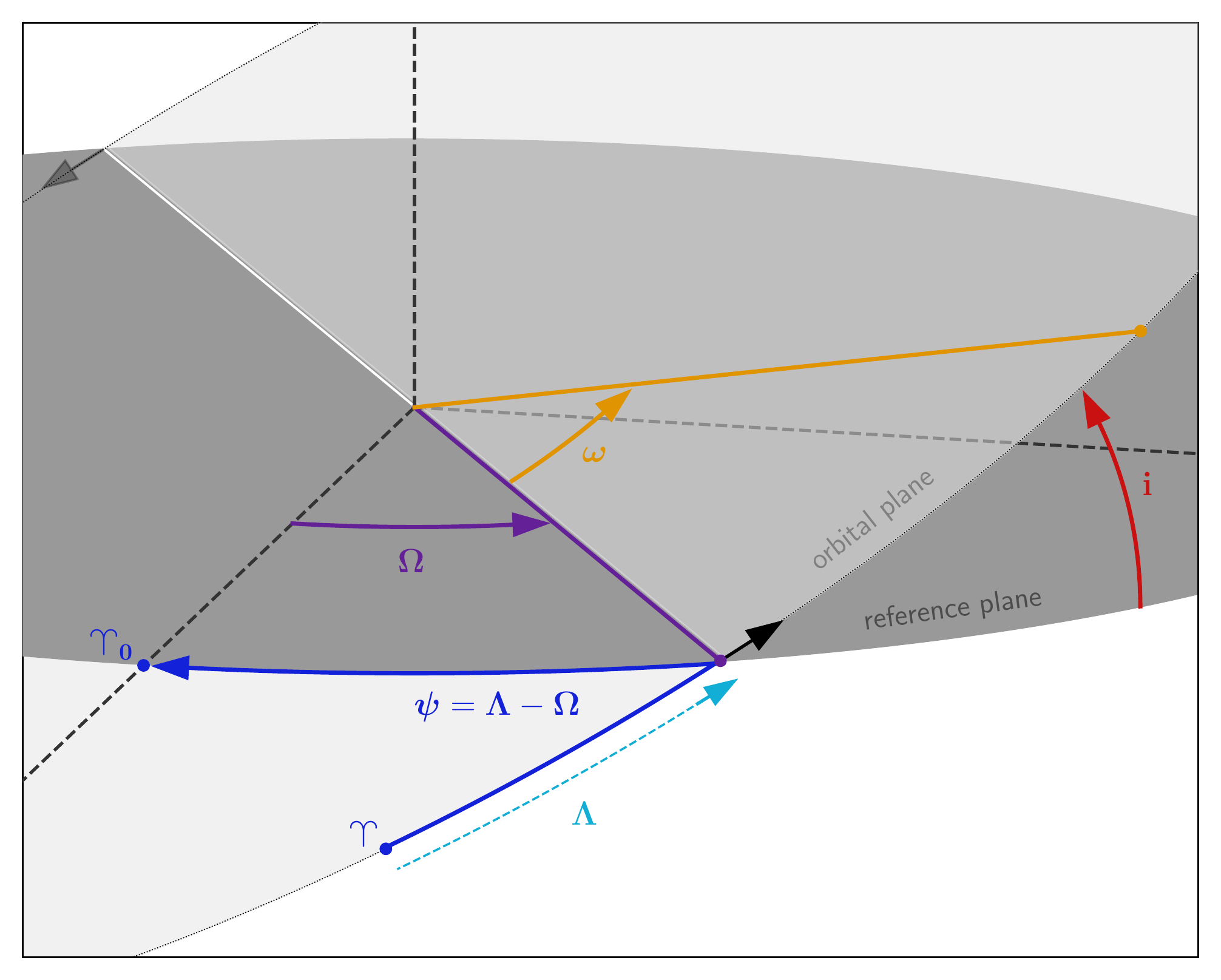}
\end{center}
\caption{\label{diagpA} Geometry used in the obliquity model, \distrot. The light
gray represents the planet's orbital plane, while the darker gray represents a plane of
reference. The important orbital angles are the inclination, $i$, the longitude of
ascending node, $\Omega$, and the argument of pericenter, $\omega$. The \emph{longitude}
of pericenter is a ``dog-leg'' angle, $\varpi = \Omega + \omega$. The angle $\Lambda$ is
measured from the vernal point $\vernal$ at time $t$, to the ascending node, $\Omega$.
The precession angle is defined as $\psi = \Lambda - \Omega$ (also a dog-leg angle). The
reference point for $\Omega$ is usually chosen as the vernal point at some known date for
solar system; however, there is probably a more sensible choice for exoplanetary
systems.}
\end{figure}

The rotational axis model, \distrot (for ``Disturbing function Rotation''), is derived
from the Hamiltonian for rigid body motion introduced by
\cite{Kinoshita1975, Kinoshita1977} and has been used extensively
\citep[\eg][]{Laskar1986, Laskar1993, Armstrong2004, Armstrong14}. In the
absence of large satellites (such as the Moon), the equations of motion for a rigid
planet are:
\begin{align}
&\begin{aligned}
\frac{d\psi}{dt} & = R(\varepsilon) - \cot(\varepsilon)~[A(p,q) \sin{\psi} \\
 &\qquad +  B(p,q) \cos{\psi}] - 2\Gamma(p,q) - p_g
\end{aligned}\label{eqnpA}\\
&\begin{aligned}
\frac{d\varepsilon}{dt} & = -B(p,q) \sin{\psi} + A(p,q) \cos{\psi} \label{eqnpsi},
\end{aligned}
\end{align}
where $\psi $ is the ``precession angle'' (see Figure \ref{diagpA} and the following
paragraph), $\varepsilon$ is the obliquity, and,
\begin{align}
R(\varepsilon) & = \frac{3 \kappa^2 M_{\star}}{a^3 \nu} \frac{J_2 M r^2}{C} S_0
\cos{\varepsilon} \label{eqnR}\\
S_0 & = \frac{1}{2} (1-e^2)^{-3/2}  \\
A(p,q) & = \frac{2}{\sqrt{1-p^2-q^2}} [\dot{q}+p \Gamma(p,q)] \label{eqnA}\\
B(p,q) & = \frac{2}{\sqrt{1-p^2-q^2}} [\dot{p}-q \Gamma(p,q)] \label{eqnB} \\
\Gamma(p,q) & = q\dot{p}-p\dot{q}. \label{eqnC}
\end{align}
Note the sign error in Equation (8) of \cite{Armstrong14}, corrected in our Equation
(\ref{eqnA}). Our Equation (\ref{eqnR}) does not contain the lunar constants present in
Equation~(24) of \cite{Laskar1986}. Here, $\varepsilon$ represents the obliquity, $p$ and
$q$ are the inclination variables from Eqs.~(\ref{eqnp}) and (\ref{eqnq}), $\dot{p}$ and $\dot{q}$ are their time
derivatives (Equations [\ref{eqnpdot}] and [\ref{eqnqdot}]), $\kappa$ is the Gaussian
gravitational constant, $M_{\star}$ is the mass of the host star in solar units, $\nu$ is
the rotation frequency of the planet in rad day$^{-1}$, $C M^{-1}r^{-2}$ is the specific
polar moment of inertia of the planet, and $J_2$ is the gravitational quadrupole of the
(oblate) planet. The final term in Equation (\ref{eqnpA}), $p_g$, accounts for precession due to general
relativity and is equal to $\delta_R/2$ \citep{Barker1970}, where $\delta_R$ is the apsidal precession rate
and is given by Equation (\ref{deltaReqn}).

The symbol $\psi$ refers to
the precession angle, defined as $\psi = \Lambda - \Omega$, where $\Lambda$ is the angle
between the vernal point $\vernal$, the position of the Sun/host star at the planet's
northern spring equinox, and the location of the ascending node, $\Omega$, measured from
some reference direction $\vernal_0$ (often taken to be the direction of the vernal point
at some reference date, hence the use of the symbol $\vernal$\footnote{The vernal
point occurred in the constellation Aries during Ptolemy's time; thus it is also called
the ``first point of Aries'' and the ``ram's horn'' symbol is used.}), see Figure
\ref{diagpA}. The convention of defining $\vernal$ as the location of the Sun at northern
spring equinox is sensible for the solar system since we observe from Earth's surface;
however, it is a confusing definition to use for exoplanets, for which the direction of
the rotation axis is unknown. We adhere to the convention for the sake of consistency
with prior literature. In the coming decades, it may become possible to determine the
obliquity and orientation of an exoplanet's spin axis; in that event, care should be
taken in determining the initial $\psi$ for obliquity modeling. One can equivalently
refer to $\vernal$ as the position of the planet at its northern spring equinox $\pm
180^{\circ}$. The relevant quantity for determining the instellation, however, is the angle
between periastron and the spring equinox, $\Delta^* = \omega + \Lambda +180^{\circ} = \varpi + \psi+180^{\circ}$ (See Section \ref{app:poise}).

An additional complication for obliquity evolution is, of course, the presence of a large
moon. We do not include the component of the \cite{Kinoshita1975} model that accounts for
the lunar torque because the coefficients used are specific to the Earth-Moon-Sun three
body problem and were calculated from the Moon's orbital evolution (and are therefore not
easily generalized). However, in \distrot we can approximate the effect of the Moon by \emph{forcing}
the precession rate, Eq.~(\ref{eqnR}), to be equal to the observed terrestrial
value. The effect of this is shown in Figure \ref{fig:obliquity}.

Equation (\ref{eqnpA}) for the precession angle contains a singularity at $\varepsilon =
0$. To avoid numerical instability, we instead recast Equations (\ref{eqnpA}) and
(\ref{eqnpsi}) in terms of the rectangular coordinates:
\begin{align}
\xi & = \sin{\varepsilon} \sin{\psi} \\
\zeta & = \sin{\varepsilon} \cos{\psi} \\
\chi & = \cos{\varepsilon}.
\end{align}
The third coordinate, $\chi$, is necessary to preserve sign information when the
obliquity crosses 90$^{\circ}$ \citep[see][]{Laskar1993}. The equations of motion for
these variables are then:
\begin{align}
\frac{d\xi}{dt} & = -B(p,q)\sqrt{1-\xi^2-\zeta^2} +
\zeta[R(\varepsilon)-2\Gamma(p,q)-p_g] \\
\frac{d\zeta}{dt} & = A(p,q)\sqrt{1-\xi^2-\zeta^2} - \xi[R(\varepsilon)-2\Gamma(p,q)-p_g] \\
\frac{d\chi}{dt} & = \xi B(p,q) - \zeta A(p,q).
\end{align}

The value of $J_2$ is a function of the planet's rotation rate and density structure. In
hydrostatic equilibrium the planet's
gravitational quadrupole moment scales with the rotation rate squared \citep{Cook1980,Hubbard1984}. This result simply comes
from force balance: equipotential surfaces will be those for which the combined gravitional and centrifugal force is constant. Then in hydrostatic equilibrium, the physical surface will be an equipotential surface. The rapid rotators of the Solar System (Earth, Mars, and the giants) appear to be close to hydrostatic equilibrium, in that
the $J_2$ is dominated by the rotational bulge, rather than topographic or other non-uniform features. For the slow rotators, such as Venus, the contribution to $J_2$ from rotation is small compared with
the contribution from topographic features \citep{Yoder95}, indicating a departure from hydrostatic balance.
In \distrot, we give the option to assume that planets are in hydrostatic
equilibrium, and scale $J_2$ according to:
\begin{equation}
J_2 = J_{2\oplus} \left (\frac{\nu}{\nu_{\oplus}} \right)^2 \left (\frac{r}{R_{\oplus}}
\right)^3 \left (\frac{M}{M_{\oplus}} \right)^{-1} \label{eqnJ2},
\end{equation}
where $\nu$ is the rotation rate, $r$ is the mean planetary radius, and $M$ is its mass. Earth's measured values of $J_{2\oplus} = 1.08265 \times 10^{-3}$,
$\nu_{\oplus}=7.292115 \times 10^{-5}$ radians s$^{-1}$, $R_{\oplus} =  6.3781\times
10^{6}$ m, and $M_{\oplus}= 5.972186\times 10^{24}$ kg are used for reference ($J_2$ from
\citealt{Cook1980}, $R_{\oplus}$ and $M_{\oplus}$ from \citealt{Prsa2016}). This approach is identical
to the method used by \cite{Brasser2014}. Like that study, we assume that $J_2$ cannot go
below the measured value of Venus from \cite{Yoder95}, which would ordinarily occur at
rotation periods $\gtrsim 13$ days. The assumption behind this scaling is then is that there is a limit to
hydrostatic equilibrium for a partially rigid body. Alternatively, the dynamical ellipticity,
$E_D = J_2 M^{-1} r^{-2}$, and the polar moment of
inertia (see below) can be independently set, allowing for departures from hydrostatic
equilibrium.

Note that the current version of \vplanet does not yet take into account the
additional contribution to $J_2$ from tidal distortion \citep[see][]{Hubbard1978}.

The \emph{specific} polar moment of inertia of a planet, $C M^{-1} r^{-2}$, is always
between 0.2 and 0.4.
The $C M^{-1} r^{-2}$ value of the Earth, for example, is $0.33$, \citep{Cook1980}.

Finally, we note that \distrot can also ingest output from previous orbital simulations, such as from N-body models, and compute the rotational axis through finite differencing. This functionality thus allows for coupling to external orbital models, if desired.

\section{The \texttt{EqTide} Module\label{app:eqtide}}

The tidal model we use is commonly called the equilibrium tide
model and was first conceived by George Darwin, grandson of Charles
\citep{Darwin1880}. This model assumes the gravitational potential of
the tide raiser on an unperturbed spherical surface can be expressed
as the sum of Legendre polynomials (\ie surface waves) and that the
elongated equilibrium shape of the perturbed body is slightly
misaligned with respect to the line that connects the two centers of
mass. This misalignment is due to dissipative processes within the
deformed body and leads to a secular evolution of the orbit, as well as
the spin angular momenta of the two bodies. Furthermore, the bodies
are assumed to respond to the time-varying tidal potential as though
they are damped, driven harmonic oscillators. As described below, this approach leads to a set of six coupled,
non-linear differential equations, but note that the model is linear
in the sense that there is no coupling between the surface waves which
sum to the equilibrium shape. A substantial body of research is
devoted to tidal theory
\cite[\eg][]{Darwin1880,GoldreichSoter66,Hut81,FerrazMello08,Wisdom08,EfroimskyWilliams09,Leconte10}, and the reader is
referred to these studies for a more complete description of the
derivations and nuances of equilibrium tide theory.

\subsection{Constant-Phase-Lag Model\label{app:eqtide:cpl}}

In the CPL model of tidal evolution, the angle between the line connecting the
centers of mass and the tidal bulge is constant. This approach is commonly
utilized in planetary studies \citep[\eg][]{GoldreichSoter66,Greenberg09} and the
evolution is described by the following equations:

\begin{equation}\label{eq:e_cpl}
  \frac{\mathrm{d}e}{\mathrm{d}t} \ = \ - \frac{ae}{8 G M_* M_p}
  \sum\limits_{i = 1}^2Z'_i \Bigg(2\varepsilon_{0,i} - \frac{49}{2}\varepsilon_{1,i} + \frac{1}{2}\varepsilon_{2,i} + 3\varepsilon_{5,i}\Bigg),
\end{equation}

\begin{equation}\label{eq:a_cpl}
  \frac{\mathrm{d}a}{\mathrm{d}t} \ = \ \frac{a^2}{4 G M_* M_p}
  \sum\limits_{i = 1}^2 Z'_i  \ {\Bigg(} 4\varepsilon_{0,i} + e^2{\Big [} -20\varepsilon_{0,i} + \frac{1
47}{2}\varepsilon_{1,i} + \nonumber \frac{1}{2}\varepsilon_{2,i} - 3\varepsilon_{5,i} {\Big ]} -4\sin^2(\psi_i){\Big [}\varepsilon_{0,i}-\varepsilon_{8,i}{\Big ]}{\Bigg )},
\end{equation}

\begin{equation}\label{eq:o_cpl}
  \frac{\mathrm{d}\Omega_i}{\mathrm{d}t} \ = \ - \frac{Z'_i}{8 M_i r_{\mathrm{g},i}
^2 R_i^2 n} {\Bigg (}4\varepsilon_{0,i} + e^2{\Big [} -20\varepsilon_{0,i} + 49\varepsilon_{1,i} + \varepsilon_{2,i} {\Big ]} + \nonumber \ 2\sin^2(\psi_i) {\Big [} -
2\varepsilon_{0,i} + \varepsilon_{8,i} + \varepsilon_{9,i} {\Big ]} {\Bigg )},
\end{equation}

\noindent and

\begin{equation}\label{eq:psi_cpl}
  \frac{\mathrm{d}\psi_i}{\mathrm{d}t} \ = \ \frac{Z'_i \sin(\psi_i)}{4 M_i r_{\mathrm{g},i}^2 R_i^2 n \Omega_i} {\Bigg (} {\Big [} 1-\xi_i {\Big ]}\varepsilon_{0,i}
+ {\Big [} 1+\xi_i {\Big ]}{\Big \{}\varepsilon_{8,i}-\varepsilon_{9,i}{\Big \}} {\Bigg)}.
\end{equation}

\noindent  The quantity $Z'_i$ is

\begin{equation}\label{eq:Zp}
Z'_i \equiv 3 G^2 k_{2,i} M_j^2 (M_i+M_j) \frac{R_i^5}{a^9} \ \frac{1}{n Q_i} \ ,
\end{equation}

\noindent where $k_{2,i}$ are the Love numbers of order 2, and $Q_i$ are the tidal quality factors. The parameter $\xi_i$ is

\begin{equation}\label{eq:chi}
\xi_i \equiv \frac{r_{\mathrm{g},i}^2 R_i^2 \Omega_i a n }{ G M_j},
\end{equation}

\noindent where $i$ and $j$ refer to the two bodies, and $r_g$ is the ``radius of gyration,'' \ie the moment of inertia is $M(r_gR)^2$. The signs of the phase lags are

\begin{equation}\label{eq:epsilon}
\begin{array}{l}
\varepsilon_{0,i} = \textrm{sgn}(2 \Omega_i - 2 n)\\
\varepsilon_{1,i} = \textrm{sgn}(2 \Omega_i - 3 n)\\
\varepsilon_{2,i} = \textrm{sgn}(2 \Omega_i - n)\\
\varepsilon_{5,i} = \textrm{sgn}(n)\\
\varepsilon_{8,i} = \textrm{sgn}(\Omega_i - 2 n)\\
\varepsilon_{9,i} = \textrm{sgn}(\Omega_i) \ ,\\
\end{array}
\end{equation}

\noindent with sgn($x$) the sign of any physical quantity $x$, \ie
sgn($x)~=~+1, -1$ or 0.

The tidal heating of the $i$th body is due to the transformation of
rotational and/or orbital energy into frictional heating. The heating
from the change in the orbital energy is

\begin{equation}\label{eq:E_orb_cpl}
\dot{E}_{\mathrm{orb},i} = \ \frac{Z'_i}{8} \ \times \ {\Big (} \ 4 \varepsilon_{0,i} + e^2 {\Big [-20 \varepsilon_{0,i} + \frac{147}{2} \varepsilon_{1,i} + \frac{1}{2} \varepsilon_{2,i} - 3 \varepsilon_{5,i} {\Big ]} - 4 \sin^2(\psi_i) \ {\Big [}\varepsilon_{0,i} - \varepsilon_{8,i}{\Big ]} \ {\Big )}},
\end{equation}

\noindent and that from the change in rotational energy is

\begin{equation}\label{eq:E_rot_cpl}
\dot{E}_{\mathrm{rot},i} = \ - \frac{Z'_i}{8} \frac{\omega_i}{n} \ \times \ {\Big (} \ 4 \varepsilon_{0,i} + e^2 {\Big [}-20 \varepsilon_{0,i} + 49 \varepsilon_{1,i}
+ \varepsilon_{2,i}{\Big ]} + 2 \sin^2(\psi_i) \ {\Big [}- 2 \varepsilon_{0,i} + \varepsilon_{8,i} + \varepsilon_{9,i}{\Big ]}
\ {\Big )}.
\end{equation}

\noindent The total heat in the $i$th body is therefore

\begin{equation}\label{eq:E_tide_cpl}
\dot{E}_{\mathrm{tide},i}^{\mathrm{CPL}} = - \ (\dot{E}_{\mathrm{orb},i} + \dot{E}_{\mathrm{rot},i}) > 0.
\end{equation}
The rate of evolution and amount of heating are set by three free
parameters: $Q, k _2$, and $r_g$. $\dot{E}_{\mathrm{tide},i}^{\mathrm{CPL}}$ can be plugged into Eq.~(\ref{eq:mantle_energy}) for worlds that are experiencing tidal heating.

\cite{Goldreich66} suggested that the equilibrium rotation period for both bodies is

\begin{equation}\label{eq:p_eq_cpl}
P_{eq}^{CPL} = \frac{P}{1 + 9.5e^2}.
\end{equation}
\cite{MurrayDermott99} present a derivation of this expression, which
assumes the rotation rate may take a continuum of values. However, the
CPL model described above only permits 4 ``tidal waves'', and hence does
not contain the resolution to permit this continuum. Instead the rotation rate is synchronous up to $e = \sqrt{1/19}~\approx~0.23$ and then jumps instantaneously to $\omega/n = 1.5$, \ie a 3:2 spin-orbit frequency ratio. The next phase
jump occurs at $e = \sqrt{2/19} \approx 0.32$ but is not present in the 2nd order CPL model. Therefore the evolution at larger $e$ predicted by the
CPL model may not be qualitatively correct. We urge caution when
interpreting CPL results above $e = 0.32$.

\subsection{The Constant-Time-Lag Model\label{app:eqtide:ctl}}

The constant-time-lag (CTL) model assumes that the time interval
between the passage of the perturber and the tidal bulge is
constant. This assumption allows the tidal response to be continuous
over a wide range of frequencies, unlike the CPL model. But, if
the phase lag is a function of the forcing frequency, then the system is no longer analogous to a
damped, driven harmonic oscillator. Therefore, this model should only
be used over a narrow range of frequencies, see
\cite{Greenberg09}. However, this model's use is widespread,
especially at high $e$, so we use it to evaluate tidal effects as
well. Compared to CPL, this model predicts larger tidal heating and evolution rates
at high $e$ due to the inclusion of higher order terms. Therefore,
the CPL and CTL models probably bracket the actual evolution.

The evolution is described by the following equations:

\begin{equation} \label{eq:e_ctl}
  \frac{\mathrm{d}e}{\mathrm{d}t} \ = \ \frac{11 ae}{2 G M_1 M_2}
  \sum\limits_{i = 1}^2Z_i \Bigg(\cos(\psi_i) \frac{f_4(e)}{\beta^{10}(e)}  \frac{\omega_i}{n} -\frac{18}{11} \frac{f_3(e)}{\beta^{13}(e)}\Bigg),
\end{equation}

\begin{equation}\label{eq:a_ctl}
  \frac{\mathrm{d}a}{\mathrm{d}t} \ = \  \frac{2 a^2}{G M_1 M_2}
  \sum\limits_{i = 1}^2 Z_i \Bigg(\cos(\psi_i) \frac{f_2(e)}{\beta^{12}(e)} \frac{\omega_i}{n} - \frac{f_1(e)}{\beta^{15}(e)}\Bigg),
\end{equation}

\begin{equation}\label{eq:o_ctl}
  \frac{\mathrm{d}\omega_i}{\mathrm{d}t} \ = \ \frac{Z_i}{2 M_i r_{\mathrm{g},i}^2
R_i^2 n} \Bigg( 2 \cos(\psi_i) \frac{f_2(e)}{\beta^{12}(e)} - \left[ 1+\cos^2(\psi)
 \right] \frac{f_5(e)}{\beta^9(e)},
\frac{\omega_i}{n} \Bigg)
\end{equation}

\noindent and

\begin{equation}\label{eq:psi_ctl}
  \frac{\mathrm{d}\psi_i}{\mathrm{d}t} \ = \ \frac{Z_i \sin(\psi_i)}{2 M_i r_{\mathrm{g},i}^2 R_i^2 n \omega_i}\left( \left[ \cos(\psi_i) - \frac{\xi_i}{ \beta} \right] \frac{f_5(e)}{\beta^9(e)} \frac{\omega_i}{n} - 2 \frac{f_2(e)}{\beta^{12}(e)} \right),
\end{equation}

\noindent where

\begin{equation}\label{eq:Z}
 Z_i \equiv 3 G^2 k_{2,i} M_j^2 (M_i+M_j) \frac{R_i^5}{a^9} \ \tau_i \ ,
\end{equation}

\noindent and

\begin{equation}\label{eq:f_e}
\begin{array}{l}
\beta(e) = \sqrt{1-e^2},\\
f_1(e) = 1 + \frac{31}{2} e^2 + \frac{255}{8} e^4 + \frac{185}{16} e^6 + \frac{25}{
64} e^8,\\
f_2(e) = 1 + \frac{15}{2} e^2 + \frac{45}{8} e^4 \ \ + \frac{5}{16} e^6,\\
f_3(e) = 1 + \frac{15}{4} e^2 + \frac{15}{8} e^4 \ \ + \frac{5}{64} e^6,\\
f_4(e) = 1 + \frac{3}{2} e^2 \ \ + \frac{1}{8} e^4,\\
f_5(e) = 1 + 3 e^2 \ \ \ + \frac{3}{8} e^4.
\end{array}
\end{equation}

\noindent The tidal heating of the $i$th body is therefore

\begin{equation}\label{eq:E_tide_ctl}
\dot{E}_{\mathrm{tide},i}^{\mathrm{CTL}} = \ Z_i {\Bigg (} \frac{f_1(e)}{\beta^{15}
(e)} - 2 \frac{f_2(e)}{\beta^{12}(e)} \cos(\psi_i) \frac{\omega_i}{n} + \ {\Big [}
\frac{1+\cos^2(\psi_i)}{2} {\Big ]} \frac{f_5(e)}{\beta^9(e)}\Big({\frac{\omega_i}{n}}\Big)^2 {\Bigg )} \ .
\end{equation}
$\dot{E}_{\mathrm{tide},i}^{\mathrm{CTL}}$ can be plugged into Eq.~(\ref{eq:mantle_energy}) for worlds that are experiencing tidal heating.

It can also be shown that the equilibrium rotation period for both bodies is
\begin{equation}\label{eq:p_eq_ctl_obl}
P_{eq}^{CTL}(e,\psi) = P\frac{\beta^3f_5(e)(1 + \cos^2\psi)}{2f_2(e)\cos\psi},
\end{equation}
which for low $e$ and $\psi = 0$ reduces to
\begin{equation}\label{eq:p_eq_ctl}
P_{eq}^{CTL} = \frac{P}{1 + 6e^2}.
\end{equation}

There is no general conversion between $ Q_\mathrm{p}$ and
$\tau_\mathrm{p}$. Only if $e~=~0$ and $\psi_\mathrm{p}~=~0$, when
merely a single tidal lag angle $\varepsilon_\mathrm{p}$ exists, then
\begin{equation}\label{eq:qtau}
Q_\mathrm{p}~\approx~1/(2|n-\omega_\mathrm{p}|\tau_\mathrm{p}),
\end{equation}
as long as $n-\omega_\mathrm{p}$ remains unchanged. Hence,
a dissipation value for an Earth-like planet of $Q_\mathrm{p}~=~12$ \citep{Williams78}
is not necessarily equivalent to a tidal time lag of 638\,s \citep{Lambeck77}, so the results for
the tidal evolution will intrinsically differ among the CPL and the
CTL model.

\subsection{The Orbit-Only Model\label{app:eqtide:orbit}}
The calculations of $\S$~\ref{sec:tidalheat} use the orbital evolution model of \cite{DriscollBarnes15}, which only considered the orbital effects and ignored dissipation in the star. We refer to this case as the ``orbit-only model.'' In this case the planet's semi-major axis $a$ and eccentricity $e$ evolution rates are \citep{Goldreich66,jackson2009,ferrazmello2008}
\begin{equation} \dot{e}=\frac{21}{2} Im(k_2) \frac{M_*}{M_p} \left(\frac{R_p}{a}\right)^5 n e,
\end{equation}

\noindent and

\begin{equation}
\dot{a}=2ea\dot{e}.
\label{eq:a_dot}
\end{equation}
The mean motion can be replaced with $n^2=GM_*/a^3$, and after rearrangement we obtain
\begin{equation}
\dot{e}=\frac{21}{2} Im(k_2) \frac{M_*^{3/2} G^{1/2} R_p^{5}}{M_p} \frac{e}{a^{13/2}}
\label{eq:ec_dot}
\end{equation}
The differential equations for thermal evolution (\ref{eq:dot_T_m}, \ref{eq:dot_T_c}) and orbital evolution (\ref{eq:ec_dot}, \ref{eq:a_dot}) are solved simultaneously to compute coupled thermal-orbital evolutions.  The tidally heated Earth model example uses the approximation $Im(k_2)=k_2/\mathcal{Q}$, where $\mathcal{Q}=\eta\omega/\mu$, in Eqs.~ (\ref{eq:a_dot}--\ref{eq:ec_dot}).

\section{The \texttt{GalHabit} Module \label{app:galhabit}}

The module \galhabit (``Galactic Habitability'') is designed to account for the effects of galactic scale
processes on binary stars. The main processes are the galactic tide, stellar migration,  and stellar
encounters, which, as shown by \cite{Kaib13}, can lead to the destabilization of
planetary systems.

\subsection{Galactic tides}
The galactic tide is a differential force on gravitationally-bound, widely separated
objects, such as wide binary star systems or Oort cloud comets orbiting the sun.
This force arises because of the variation in the gravitational force with distance from the
midplane and distance from the galactic center. At the sun's
galactocentric distance ($\sim$ 8 kpc), the galactic gravitational potential is nearly
axisymmetric (disk-like), where the density is highest in the mid-plane and gradually
decreases with some scale height in the $Z-$direction, \ie perpendicular to the plane of
the disk \citep{HeislerTremaine86}. However, the galactic bulge and halo do provide a
spheroidal
component to the galactic potential that becomes more important close to the galactic
center and far from the mid-plane \citep{Kordopatis15}. \galhabit does not at present account for
such departures from axisymmetry, and so should be used with caution when
modeling systems in such regions.

The effects of the axisymmetric galactic tide on a system may be described by the
time-averaged (or secular) Hamiltonian \citep{HeislerTremaine86}:
\begin{align}
    &\begin{aligned}
    H_{\text{av}} = & -\frac{\mu^2}{2L^2}+\frac{\pi G \rho_0}{\mu^2} \frac{L^2}{J^2} \cdot (J^2-J_z^2)[J^2 + 5(L^2-J^2) \sin^2{\omega}] \label{eqn:Havg},
    \end{aligned}
\end{align}
where $\mu = G(M_{\text{c}}+M)$, $L = \sqrt{\mu a}$, $J = \sqrt{\mu a (1-e^2)}$,
$J_z = J \cos{i}$, $\rho_0$ is the local galactic density, $M_{c}$ is the central
mass, $M$ the mass of the orbiter,
$a$ is its semi-major axis, $e$ is its eccentricity, $i$ is its inclination
of the orbit with respect to the galactic mid-plane, and $\omega$ is its argument of
periastron. The variables $L$, $J$, and $J_z$ are thus canonical momenta associated
with the energy of the orbit (in momentum units), the total angular momentum,
and the $Z-$component of the angular momentum, respectively. From Hamilton's equations
we know that, since the canonical angles associated with $L$ and $J_z$ do not appear
explicitly in $H_{\text{av}}$, $L$ and $J_z$ are perfectly conserved. Since
$\omega$, which is associated with $J$ does appear, however, $J$ is free to vary with
time. The eccentricity and inclination are thus coupled through the definitions of $J$
and $J_z$, and as one changes under the influence of the tide, the other must
compensate, resulting in ``Lidov-Kozai-like'' cycling of $e$ and $i$.

The Hamiltonian above results in four equations of motion (Hamilton's equations):
\begin{align}
    &\frac{dJ}{dt} = -\frac{\partial H_{\text{av}}}{\partial \omega} = -\frac{5\pi G \rho_0}{\mu^2}\frac{L^2}{J^2}(J^2-J_z^2)(L^2-J^2) \sin{2 \omega},\\
    &\frac{d\omega}{dt} = \frac{\partial H_{\text{av}}}{\partial J} = \frac{2 \pi G \rho_0}{\mu^2} \frac{L^2}{J^2}\left[J^3 + 5\left(\frac{L^2 J_z^2}{J}-J^3\right)\sin^2{\omega} \right],\\
    &\frac{d\Omega}{dt} = \frac{\partial H_{\text{av}}}{\partial J_z} = -\frac{2 \pi G \rho_0}{\mu^2} \frac{L^2}{J^2} J_z [J^2+5(L^2-J^2)\sin^2{\omega}], \\
    &\frac{dl}{dt} = \frac{\partial H_{\text{av}}}{\partial L}=\frac{\mu^2}{L^3}+\frac{2\pi G \rho_0}{\mu^2}\frac{L}{J^2}(J^2-J_z^2)[J^2+5(2L^2-J^2)\sin^2{\omega}], \label{eqn:galhmeana}
\end{align}
where $\Omega$ is the longitude of ascending node and $l$ is the mean anomaly. Since $L$ is
constant and we are modeling the orbit-averaged (secular) evolution of
the system, we can disregard Equation (\ref{eqn:galhmeana}) --- this choice ultimately affects
none of the other variables.

These equations are modeled in this form and the other orbital elements
($e$ and $i$) can be calculated from $J$ and $J_z$. Since in future versions of \vplanet we
intend to include the additional dynamics of triple star systems,
we desire a form of these equations that is more easily generalized to different
coordinate systems. Cartesian coordinate systems are ideal as rotational and translational
transformations are easily applied. In our case, we utilize the Cartesian components of
the angular momentum vector (per unit mass), $\vec{J} = (J_x,J_y,J_z)$, and the eccentricity
vector, $\vec{e} = (e_x,e_y,e_z)$. They are defined as:
\begin{align}
    &J_x = J \sin{\Omega} \sin{i} \label{eqn:Jx},\\
    &J_y = -J \cos{\Omega} \sin{i}, \\
    &J_z = J \cos{i} \label{eqn:Jz},\\
    &e_x = e [\cos{\Omega} \cos{\omega} - \sin{\Omega} \sin{\omega} \cos{i}],\\
    &e_y = e [\sin{\Omega} \cos{\omega} + \cos{\Omega} \sin{\omega} \cos{i}],\\
    &e_z = e \sin{\omega} \sin{i}.
\end{align}
The derivatives are calculated via chain-rule from Hamilton's equations above and
written in terms of orbital elements (when convenient), resulting in
\begin{align}
    &\frac{dJ_x}{dt} =  \frac{\sin{\Omega}}{\sin{i}} \frac{dJ}{dt} + J \sin{i} \cos{\Omega} \frac{d\Omega}{dt} \label{eqn:dJxdt},\\
    &\frac{dJ_y}{dt} = -\frac{\cos{\Omega}}{\sin{i}} \frac{dJ}{dt} + J \sin{i} \cos{\Omega} \frac{d\Omega}{dt} \label{eqn:dJydt},\\
    &\frac{dJ_z}{dt} = 0,
\end{align}
and
\begin{align}
    &\frac{de_x}{dt} = \frac{\partial e_x}{\partial J} \frac{dJ}{dt} +  \frac{\partial e_x}{\partial \Omega} \frac{d\Omega}{dt}+  \frac{\partial e_x}{\partial \omega} \frac{d\omega}{dt},\\
    &\frac{de_y}{dt} = \frac{\partial e_y}{\partial J} \frac{dJ}{dt} +  \frac{\partial e_y}{\partial \Omega} \frac{d\Omega}{dt}+  \frac{\partial e_y}{\partial \omega} \frac{d\omega}{dt},\\
    &\frac{de_z}{dt} = \frac{\partial e_z}{\partial J} \frac{dJ}{dt} +  \frac{\partial e_z}{\partial \omega} \frac{d\omega}{dt} \label{eqn:dezdt},
\end{align}
where, for the $x-$components,
\begin{align}
    &\frac{\partial e_x}{\partial J} = -\sqrt{\frac{1-e^2}{\mu a e^2}} \cos{\Omega} \cos{\omega} + \frac{\cos{i}}{e \sqrt{\mu a (1-e^2)}} \sin{\Omega} \sin{\omega} \label{eqn:dexdJ},\\
    &\frac{\partial e_x}{\partial \Omega} = -e \sin{\Omega} \cos{\omega} - e \cos{\Omega} \sin{\omega} \cos{i}, \\
    &\frac{\partial e_x}{\partial \omega} = -e \cos{\Omega} \sin{\omega} - e \sin{\Omega} \cos{\omega} \cos{i},
\end{align}
for the $y-$components,
\begin{align}
    &\frac{\partial e_y}{\partial J} = -\sqrt{\frac{1-e^2}{\mu a e^2}} \sin{\Omega} \cos{\omega} - \frac{\cos{i}}{e \sqrt{\mu a (1-e^2)}} \cos{\Omega} \sin{\omega} \label{eqn:deydJ}, \\
    &\frac{\partial e_y}{\partial \Omega} = e \cos{\Omega} \cos{\omega} - e \sin{\Omega} \sin{\omega} \cos{i}, \\
    &\frac{\partial e_y}{\partial \omega} = -e \sin{\Omega} \sin{\omega} + e \cos{\Omega} \cos{\omega} \cos{i},
\end{align}
and for the $z-$components,
\begin{align}
    &\frac{\partial e_z}{\partial J} = \frac{e^2-\sin^2{i}}{e \sin{i}\sqrt{\mu a (1-e^2)}} \sin{\omega} \label{eqn:dezdJ},\\
    &\frac{\partial e_z}{\partial \omega} = e \cos{\omega} \sin{i}.
\end{align}

Note that Equations (\ref{eqn:dJxdt}), (\ref{eqn:dJydt}), (\ref{eqn:dexdJ}), (\ref{eqn:deydJ}),
and (\ref{eqn:dezdJ}) contain singularities at either $e=0$ or $i=0$; however, the offending
terms cancel when the algebra in Equations (\ref{eqn:dJxdt} - \ref{eqn:dezdt}) is fully
carried out.
We perform this algebra as necessary to remove the singularities, which are
likely to trigger numerical instabilities or unphysical behavior. \vplanet contains the
resulting equations.

Our integration variables for the galactic tide have thus grown from three ($J$, $\Omega$, and
$\omega$) to five ($J_x$, $J_y$, $e_x$, $e_y$, and $e_z$); this is the trade-off of using this
more flexible coordinate system. In practice, the tidal model is still very fast (simulations of billions of years, with tides alone, take seconds on a CPU), and thus this
is an acceptable exchange for future flexibility.

For output, the orbital elements need to be calculated from the Cartesian vectors. The eccentricity
is simply $e = |\vec{e}|$, while $i$ and $\Omega$ can be calculated from $J = |\vec{J}|$, $J_x$,
$J_y$, and $J_z$ (Equations [\ref{eqn:Jx}]--- [\ref{eqn:Jz}]). For $\omega$, \cite{MardlingLin02}
give
\begin{align}
    &\cos{\omega} = \hat{e} \cdot \hat{n} \\
    &\sin{\omega} = \hat{e} \cdot (\hat{J} \times \hat{n}),
\end{align}
where $\vec{n} = \hat{Z} \times \vec{J}$ and $\hat{Z}$ is the unit vector perpendicular to the
galactic plane. Note that
$\vec{n}$ contains the full vector $\vec{J}$, not its unit vector $\hat{J}$, and that in the
equations above $\hat{n} = \vec{n}/|\vec{n}|$. The semi-major axis, $a$, is constant under
the galactic tide alone; however, additional forces (see next Section)
lead to changes in $a$. To
recalculate $a$, we can first calculate $e$ and then utilize the definition of $J$.

The local density around the host system is calculated using the mass model of
\cite{Kordopatis15}, which accounts for disk stars, gas, dark matter, and stars in the
galactic bulge. Technically, the latter two components have spheroidal distributions and
thus do not fit the axisymmetric assumption of the tidal model. The time-averaging of the
galactic Hamiltonian to produce Eq.~(\ref{eqn:Havg}) requires axisymmetry
\citep{HeislerTremaine86}.
Generally, within the disk of the Milky Way, these components are small and axisymmetry is a
decent approximation. However,
close to the galactic center or at high $Z$, this approximation may not hold. In that case, the
unaveraged Hamiltonian (or a reformulation of $H_{\text{av}}$) for the galactic tide may be
needed.

\subsection{Encounters with passing stars}
The second force supplied by the galactic environment is the gravitational perturbation
of wide orbits by passing stars. Close encounters between the solar system and nearby stars
are thought to affect the Oort cloud and thus to contribute to the appearance of long-period
comets \citep{Duncan87,Heisler87,KenyonBromley04,Rickman08,CollinsSari2010}. Similarly, such
close encounters will affect the orbits of widely separated
binary stars \citep{Kaib13}, potentially affecting orbiting planets. We use the techniques developed by \cite{Heisler87} and \cite{Rickman08} to account for these
perturbations. The process is described as follows.

\begin{table*}[ht]
    \centering
    \caption{Stellar properties used in encounter model \citep[reproduced from][]{GarciaSanchez2001}}
    \begin{tabular}{ccccc}
    \hline\hline \\ [-1.5ex]
Stellar type & $\Delta M_V$ & $v_{h,i}$ (km s$^{-1}$) & $\sigma_{\star i}$ (km s$^{-1}$) & $n_{\star i}$ (10$^{-3}$ pc$^{-3}$) \\ [0.5ex]
\hline \\ [-1.5ex]
        B0 & (-5.7, -0.2) & 18.6 & 14.7 & 0.06 \\
        A0 & (-0.2, 1.3) & 17.1 & 19.7 & 0.27 \\
        A5 & (1.3, 2.4) & 13.7 & 23.7 & 0.44 \\
        F0 & (2.4, 3.6) & 17.1 & 29.1 & 1.42 \\
        F5 & (3.6, 4.0) & 17.1 & 36.2 & 0.64 \\
        G0 & (4.0, 4.7) & 26.4 & 37.4 & 1.52 \\
        G5 & (4.7, 5.5) & 23.9 & 39.2 & 2.34 \\
        K0 & (5.5, 6.4) & 19.8 & 34.1 & 2.68 \\
        K5 & (6.4, 8.1) & 25.0 & 43.4 & 5.26 \\
        M0 & (8.1, 9.9) & 17.3 & 42.7 & 8.72 \\
        M5 & (9.9, 18) & 23.3 & 41.8 & 41.55 \\
        WD & - & 38.3 & 63.4 & 3.00\\
        Giants & - & 21.0 & 41.0 & 0.43
    \end{tabular}
    \label{tab:stellenc}
\end{table*}

At the start of a simulation, we calculate the total stellar encounter rate and the time until
the next encounter. The encounter rate is based on the local stellar properties determined
by \cite{GarciaSanchez2001}, see Table \ref{tab:stellenc}. Given an ``encounter radius''
(\ie the distance at which the system and a passing star are considered to encounter each other),
$R_{\text{enc}}$, the rate of encounters is given by
\begin{align}
    f_{\text{enc}} = \pi R^2_{\text{enc}} \sum_i v_i n_{\star i} \label{eqn:totalfenc},
\end{align}
where $n_{\star i}$ is the number density of each stellar type and
$v_i = \sqrt{v_{h,i}^2+\sigma_{\star i}^2}$ \citep{GarciaSanchez2001}. Note that $R_{\text{enc}}$ is a fixed distance at
which encounters are considered to begin (the size of the encounter box, say),
not the impact parameter (\emph{i.e.} the distance of closest approach). The encounter velocity, $v_i$,
is a function of both
the stellar type's velocity dispersion, $\sigma_{\star i}$, and the host system's peculiar (or
``apex'') velocity relative to each stellar type, $v_{h,i}$. Using the numbers in
\cite{GarciaSanchez2001} and $R_{\text{enc}} = 1$ pc yields a total encounter rate
(summed over 13 stellar types) of 10.5 Myr$^{-1}$ for the Sun in its current location.
The density of stars increases toward the galactic center, while the velocity dispersion
increases over time in galactic simulations \citep{Minchev12,Roskar12}. For systems placed
at different galactocentric distances, we scale the number density according to the mass
model of \cite{Kordopatis15}. The velocity dispersion is scaled as $\sqrt{t}$, where $t$ is
the simulation time. The next encounter time is then
\begin{align}
    t_{\text{next}} = t - \frac{\ln{\xi}}{f_{\text{enc}}},
\end{align}
where $\xi$ is a random number on the interval $(0,1]$ and is the time of
closest approach to the host system's primary star (see below).

When the simulation time exceeds the time of the next stellar encounter,
the initial position, mass,
and velocity of the passing star are selected randomly. The position is given by two random
numbers that give the angles $\theta_{\star}$ and $\phi_{\star}$---these are
the co-latitudinal and
longitudinal position of the passing star on a sphere of radius $R_{\text{enc}}$, centered
on the primary component of the host system. The passing star's initial position,
$\vec{R}_{\star} = (x_{\star},y_{\star},z_{\star})$, is then
\begin{align}
    x_{\star} = R_{\text{enc}} \sin{\theta_{\star}} \cos{\phi_{\star}},\\
    y_{\star} = R_{\text{enc}} \sin{\theta_{\star}} \cos{\phi_{\star}},\\
    z_{\star} = R_{\text{enc}} \cos{\theta_{\star}}.
\end{align}

To obtain the stellar mass, we need to build a distribution of the stellar encounter frequencies
for each stellar type and utilize rejection sampling. Similar to Eq.~(\ref{eqn:totalfenc}), the encounter frequencies are given for each stellar
type by
\begin{align}
    f_{\text{enc},i} = \pi R^2_{\text{enc}} v_i n_{\star i},
\end{align}
with values again given by \cite{GarciaSanchez2001}. The distribution of stellar types extends
across a magnitude range of
23.7 (the bin edges are given in Table \ref{tab:stellenc}) for the main sequence. Giant and
white dwarfs are given bin sizes of $\Delta M_{V,i} = 1$ each, resulting in a total magnitude
range of 25.7. We now draw a random magnitude, $\xi_{M_V}$, on the interval $[-7.7,18]$,
with
$-7.7<M_V<-6.7$ assigned to giants and $-6.7<M_V<-5.7$ assigned to white dwarfs (for
selection purposes only), and calculate the relative frequency of an encounter as
$f_{\text{rel}} = f_{\text{enc},i}/\Delta M_{V,i}$ for the corresponding stellar type.
We next draw another random number,
$\xi_{\text{samp}}$, on the interval
$[0,f_{\text{enc}}]$, where $f_{\text{enc}}$ is the total encounter rate and sets
the threshold for acceptance. If $f_{\text{rel}}<\xi_{\text{samp}}$, we then redraw
$\xi_{M_V}$ and $\xi_{\text{samp}}$ until $f_{\text{rel}}>\xi_{\text{samp}}$. Once a value for
the magnitude is accepted, the mass of the passing star is calculated using the formulae in
\cite{Reid02} for the main sequence. White dwarfs are assigned a mass of 0.9 $M_{\odot}$ and
giants a mass of 4 $M_{\odot}$. The white dwarf and giant masses are archetypal values from \emph{Allen's Astrophysical Quantities} \citep{Cox2000}. Following \cite{Rickman2005,Rickman08} and \cite{Kaib13}, we
have used these discrete values as a crude first approximation.
Future versions of \vplanet will sample their masses from a distribution.

The three components of the velocity of the star are selected from a normal distribution with
width $\sigma_{\star i}/\sqrt{3}$. This velocity is relative to the local standard of rest, and
thus we must account for the host system's apex velocity. The velocity between
the host and the passing star is then
\begin{align}
    \vec{v}_{h,\star} = \vec{v}_{\star} - \vec{v}_h,
\end{align}
where $\vec{v}_{\star}$ is the passing star's velocity as randomly selected and $\vec{v}_h$ is
the host system's apex velocity relative to the selected star's type. To account for the
additional effect of the host system's velocity on the encounter rate (and selection process),
we reject and redraw the velocity if a random number on $(0,1]$ is greater than
$|\vec{v}_{h,\star}|/v_{\text{max}}$, where $v_{\text{max}} = v_{h} + 3\sigma_{\star i}$
\citep[see][]{Rickman08}. We similarly reject and redraw $\vec{v}_{h,\star}$  if its radial
component
$(\vec{v}_{h,\star})_r$ is positive, in which case the passing star would be moving away from
the host system.

Our next step is to advance the mean anomaly, $l$, of the orbiting body in the host system
(whether it is a comet or secondary star). This value is calculated from
$l(t) = l(t-\Delta t) + n \Delta t$,
where $n = \sqrt{\mu/a^3}=\mu^2/L^3$. The mean motion, $n$, is merely the first term in Eq.~(\ref{eqn:galhmeana}). The model is not fully self-consistent in that the second term in
Eq.~(\ref{eqn:galhmeana}) (the effect of the tide on the mean anomaly) is not included
(indeed, this is what makes the model ``secular''),
however, this term is generally very small compared to $n$ \citep{HeislerTremaine86}.

Next, we calculate the impact parameters of the passing star. There are two that
interest us: the closest approach
to the primary (or central) star of the host system, $\vec{b}_1$, and the closest approach
to the orbiter, $\vec{b}_2$, which do not occur at the same instant in time. The
time of the encounter, $t_{\text{next}}$, is taken to be the time of closest approach
to the primary. The impact parameters are calculated using
simple kinematics assuming that the
position of each body in the host system is fixed at the time of the encounter,
$t_{\text{next}}$, and that the passing star's velocity is constant, \ie the encounter timescale is much shorter than the orbital timescale. This approach translates to
\begin{align}
    &\vec{b}_1 = \vec{v}_{h,\star} \Delta t_1 + \vec{R}_{\star} \\
    &\vec{b}_2 = \vec{v}_{h,\star} \Delta t_2 + \vec{R}_{\star} - \vec{R}_2,
\end{align}
where $\vec{R}_2$ is the position of orbiting body of the host system at time $t_{\text{next}}$
and
\begin{align}
    &\Delta t_1 = \frac{-\vec{R}_{\star}\cdot \vec{v}_{h,\star}}{v^2_{h,\star}} \\
    &\Delta t_2 = \frac{-(\vec{R}_{\star}-\vec{R}_2)\cdot \vec{v}_{h,\star}}{v^2_{h,\star}}.
\end{align}

Finally, we apply the perturbation using the impulse approximation \citep{RemyMignard85}. This
takes the form of an instantaneous change in the orbiter's velocity, given by
\begin{align}
    \Delta \vec{v}_2 = \frac{2 G m_{\star}}{v_{\star,2}b^2_2} \vec{b}_2 - \frac{2 G m_{\star}}{v_{h,\star} b^2_1}\vec{b}_1,
\end{align}
where $m_{\star}$ is the mass of the passing star and $v_{\star,2}$ is the relative speed
between the orbiter and the passing star. We then recalculate the osculating elements based on
the new velocity and finally recalculate $\vec{e}$ and $\vec{J}$. In summary, the path of the passing star
is as follows: the star begins at some random location a distance of $R_{\text{enc}}$ away from the host,
moves toward the host and companion until it reaches the distances $b_1$ and $b_2$, where it interacts gravitationally with
each body, then proceeds to leave the encounter area.

We then randomly select the time until the next encounter, $t_{\text{next}}$, using the
procedure previously described, and then continue integration of the tidal model, etc., until
$t_{\text{next}}$ is exceeded.

\subsection{Radial migration}
Radial migration of the host system is treated as a sudden change in the
galactocentric position,
since galactic simulations show the process is typically very rapid \citep{Roskar08}. In this
case, the encounter rate is recalculated at the time of migration, as is the local mass
density, $\rho_0$. Currently, a system is only
able to migrate once per simulation.

\section{The \texttt{POISE} Module\label{app:poise}}

The climate model, \texttt{POISE} (Planetary Orbit-Influenced
Simple EBM), is a one-dimensional EBM
based on \cite{NorthCoakley1979}, with a number of modifications,
foremost of which is the inclusion of a model of ice sheet
growth, melting, and flow. The model is one-dimensional in $x =
\sin{\phi}$, where $\phi$ is the latitude. In this fashion,
latitude cells of size $dx$ do not have equal width in
latitude, but are equal in area. The general energy balance
equation is
\begin{align}
&\begin{aligned}
C(x) \frac{\partial T}{\partial t}(x,t) -& D(x,t) \nabla^2 T(x,t) + I(x,T,t) = S(x,t) (1-\alpha(x,T,t)), \label{eqn:ebmeq} \\
\end{aligned}
\end{align}
where $C(x)$ is the heat capacity of the surface at location $x$,
$T$ is the surface temperature, $t$ is time, $D$ is the
coefficient of heat diffusion between latitudes (due to
atmospheric circulation), $I(x,t)$ is the outgoing long-wave
radiation (OLR) to space (\ie the thermal infrared
flux), $S(x,t)$ is the incident instellation (stellar flux), and
$\alpha$ is the planetary albedo and represents the
fraction of the instellation that is reflected back into space. Note that \poise assumes a constant atmospheric mass, and so should not be coupled to \atmesc.

Though the model lacks a true longitudinal dimension, each
latitude is divided into a land portion and a water portion. The
land and water have distinct heat capacities and albedos, and
heat is allowed to flow between the two regions. The energy
balance equation can then be separated into two equations \citep{NorthCoakley1979,Deitrick18b}:
\begin{align}
&\begin{aligned}
C_L \frac{\partial T_L}{\partial t} - D \frac{\partial}{\partial x} (1-x^2) \frac{\partial T_L}{\partial x} +& \frac{\nu}{f_L} (T_L-T_W) + I(x,T_L,t) \\& = S(x,t) (1-\alpha(x,T_L,t)),\label{eqn:ebland}\\
\end{aligned}\\
&\begin{aligned}
C_W^{eff} \frac{\partial T_W}{\partial t} - D \frac{\partial}{\partial x} (1-x^2) \frac{\partial T_W}{\partial x} + &\frac{\nu}{f_W} (T_W-T_L) + I(x,T_W,t)\\& = S(x,t) (1-\alpha(x,T_W,t)),\label{eqn:ebwater}\\
\end{aligned}
\end{align}
where we have employed the co-latitudinal component of the
spherical Laplacian, $\nabla^2$ (the radial and
longitudinal/azimuthal components vanish). The effective heat
capacity of the ocean is $C_W^{eff} = m_d C_W$, where $m_d$ is
an adjustable parameter representing the
mixing depth of the ocean.
The parameter $\nu$ is used to adjust the land-ocean heat
transfer to reasonable values, and $f_L$ and $f_W$ are the
fractions of each latitude cell that are land and ocean,
respectively.

The instellation received as a function of
latitude, $\phi$, and declination of the host star, $\delta$, is
calculated using the formulae of \cite{Berger1978}. Declination,
$\delta$, varies over the course of the planet's orbit for
nonzero obliquity. For Earth, for example, $\delta \approx
23.5^{\circ}$ at the northern summer solstice, $\delta =
0^{\circ}$ at the equinoxes, and $\delta \approx -23.5^{\circ}$
at the northern winter solstice. Because $\delta$ is a function
of time (or, equivalently, orbital position), the instellation varies, and gives
rise to the seasons (again, assuming the obliquity is nonzero).
For latitudes and times where there is no sunrise (\eg polar
darkness during winter):
\begin{equation}
S(\phi,\delta) = 0,
\end{equation}
while for latitudes and times where there is no sunset:
\begin{equation}
S(\phi,\delta) = \frac{S_{\star}}{\rho^2} \sin{\phi} \sin{\delta},
\end{equation}
and for latitudes with a normal day/night cycle:
\begin{equation}
S(\phi,\delta) = \frac{S_{\star}}{\pi \rho^2} (H_0 \sin{\phi} \sin{\delta} + \cos{\phi} \cos{\delta} \sin{H_0}).
\end{equation}
Here, $S_{\star}$ is the solar/stellar constant,
$\rho$ is the distance between the planet and host star
normalized by the semi-major axis (\ie $\rho = r/a$), and
$H_0$ is the hour angle of the star at sunrise and sunset,
and is defined as:
\begin{equation}
\cos{H_0} = - \tan{\phi}\tan{\delta}.
\end{equation}
The declination of the host star with respect to the planet's
celestial equator is a simple function of its obliquity $\varepsilon$
and its true longitude $\theta$:
\begin{equation}
\sin{\delta} = \sin{\varepsilon} \sin{\theta}.
\label{eqn:decl}
\end{equation}
See also \cite{Laskar1993} for a comprehensive derivation. For
these formulae to apply, the true longitude should be defined as
$\theta = f + \Delta^*$, where $f$ is the true anomaly (the
angular position of the planet with respect to its periastron) and
$\Delta^*$ is the angle between periastron and the planet's position at
its northern spring equinox, given by
\begin{equation}
    \Delta^* = \varpi + \psi + 180^{\circ}.
\end{equation}
Above, $\varpi$ is the longitude of periastron, and $\psi$ is
the precession angle.
Note that we add $180^{\circ}$ because of the convention of
defining $\psi$ based on the vernal point, $\vernal$, which is
the position of the \emph{sun} at the time of the northern
spring equinox.

A point of clarification is in order: EBMs (at least, the models
employed in \vplanet) can be either \emph{seasonal} or
\emph{annual}. The EBM component of \poise is a seasonal
model---the variations in the instellation throughout the
year/orbit are resolved and the temperature of the surface at
each latitude varies in response, according to the leading terms
in Equations (\ref{eqn:ebland}) and (\ref{eqn:ebwater}). In an annual
model, the instellation at
each latitude is averaged over the year, and the energy balance
equation, Eq. (\ref{eqn:ebmeq}), is forced into ``steady state'' by
setting $\partial T/\partial t$ equal to zero (either
numerically or analytically). By ``steady state'', we mean that
the surface conditions (temperature and albedo) come to final
values and remain there. Seasonal EBMs, on the other
hand, can be in a stable balance, in that the orbit-averaged
surface conditions remain the same from year to year, but the
surface conditions vary \emph{throughout} the year.

The planetary albedo is a function of surface type (land or
water), temperature, and zenith angle. For land grid cells, the
albedo is:
\begin{equation}
\alpha = \left\{ \begin{array}{cc}
			\alpha_L + 0.08 P_2(\sin{Z}) & \begin{array}{c}\hspace{1mm} \text{if } M_{\text{ice}} = 0 \text{ and } T > -2^{\circ} \text{ C} \end{array}\\
			\alpha_i & \begin{array}{c} \hspace{1mm}  \text{if }M_{\text{ice}} > 0\text{ or } T <= -2^{\circ} \text{ C},
			\end{array}\\
		     \end{array} \right.
\label{eqn:albland}
\end{equation}
while for water grid cells it is:
\begin{equation}
\alpha = \left\{ \begin{array}{cc}
			\alpha_W + 0.08 P_2(\sin{Z}) & \hspace{1mm} \text{if } T > -2^{\circ} \text{ C}\\
			\alpha_i &\hspace{4mm}  \text{if } T <= -2^{\circ} \text{ C},\\
		     \end{array} \right.
		     \label{eqn:albwater}
\end{equation}
where $Z$ is the zenith angle of the sun at noon and $P_2(x) =
1/2 (3 x^2-1)$ (the second Legendre polynomial). This last
quantity is used to approximate the additional reflectivity seen
at shallow incidence angles, \emph{\eg} at high latitudes on
Earth. The zenith angle at each latitude is given by
\begin{equation}
Z = | \phi - \delta |.
\end{equation}

The albedos, $\alpha_L$, $\alpha_W$ (see Table
\ref{tab:ebm_tab}), not accounting for zenith angle effects, are
chosen to match Earth data \citep{NorthCoakley1979} and account,
over the large scale, for clouds, various surface types, and
water waves. Additionally, the factor of $0.08$ in Equations
(\ref{eqn:albland}) and (\ref{eqn:albwater}) is chosen to reproduce the
albedo distribution in \cite{NorthCoakley1979}.
The ice albedo, $\alpha_i$, is a single value that
does not depend on zenith angle due to the fact that ice
tends to occur at high zenith angle, so that the zenith angle
is essentially already accounted for in the choice of $\alpha_i$.
Equation (\ref{eqn:albland}) indicates that when there is
ice on land ($M_{\text{ice}}>0$), or the temperature is below freezing,
the land takes on the albedo of ice. Though there are multiple
conditionals governing the albedo of the land, in practice the
temperature condition is only used when ice sheets are turned off
in the model, since ice begins to accumulate at $T = 0^{\circ}$
C, so that ice is always present when $T < -2^{\circ}$ C. Equation
(\ref{eqn:albwater}) indicates a simpler relationship for the albedo
over the oceans: when it is above freezing, the albedo is that of
water (accounting also for zenith angle effects); when it is
below freezing, the albedo is that of ice, $\alpha_i$.

The land fraction and water fraction are constant
across all latitudes, which is roughly equivalent to a single
continent that extends from pole to pole. The current version of \poise does not contain other geographies because we have
not yet developed a consistent method to
handle the flow of ice when land fraction varies.

\cite{NorthCoakley1979} utilized a linearization of the OLR with
temperature:
\begin{equation}
I = A + BT, \label{eqn:olrnc}\\
\end{equation}
where, for Earth, $A = 203.3$ W m$^{-2}$ and $B = 2.09$ W
m$^{-2}$ $^{\circ}$C$^{-1}$, and $T$ is the surface temperature
in $^{\circ}$C. This linearization is a good fit to the observations of Earth \citep{WarrenSchneider79}. A side benefit is that it
allows the coupled set of equations to be formulated as a matrix
problem that can be solved using an implicit Euler scheme \citep{Press1987}
with
the following form:
\begin{equation}
\mathscr{M}\cdot T_{n+1} = \frac{C T_n}{\Delta t} - A + S (1-\alpha),\label{eqn:matrixform}\\
\end{equation}
where $T_n$ is a vector containing the current surface
temperatures, $T_{n+1}$ is a vector representing the temperatures
to be calculated, and $C$, $A$, $S$, and $\alpha$ are vectors
containing the heat capacities, OLR offsets (Eq.~[\ref{eqn:olrnc}]), instellation at each latitude, and albedos,
respectively. The matrix $\mathscr{M}$ contains all of the
information on the left-hand sides of Eqs.~(\ref{eqn:ebland}) and
(\ref{eqn:ebwater}) related to temperature. The time-step, $\Delta t$,
is chosen so that conditions do not change significantly between
steps, resulting in typically 60 to 80 time-steps per orbit. The
new temperature values can then be calculated by taking the
dot-product of $\mathscr{M}^{-1}$ with the right-hand side of
Eq.~(\ref{eqn:matrixform}). The large time step allowed by this
integration scheme greatly speeds the climate model, permitting simulations for millions of years in hours of computation time.

We model ice accumulation and ablation in a similar fashion to
\cite{Armstrong14}. Ice accumulates on land at a constant rate,
$r_{\text{snow}}$, when temperatures are below 0$^{\circ}$ C.
Melting/ablation occurs when ice is present and temperatures are
above 0$^{\circ}$ C, according to the formula:
\begin{equation}
\frac{dM_{\text{ice}}}{dt} = \frac{\xi \sigma (T_{\text{freeze}}^4 - (T+T_{\text{freeze}})^4)}{L_h},
\label{eqn:ablation}
\end{equation}
where $M_{\text{ice}}$ is the surface mass density of ice, $\sigma =
5.67 \times 10^{-8}$ W m$^{-2}$ K$^{-4}$ is the Stefan-Boltzmann
constant, $L_h$ is latent heat of fusion of ice, $3.34 \times
10^5$ J kg$^{-1}$ and $T_{\text{freeze}} = 273.15$ K. The factor $\xi$ is used to scale the ice
ablation. This can be done to reproduce values similar to Earth ($\sim$3 mm
$^{\circ}$C$^{-1}$ day$^{-1}$
\citep[see][]{Braithwaitezhang2000,Lefebre2002,Huybers2008} or to experiment with other melting rates.

The ice sheets flow across the surface via
deformation and sliding at the base. We use the formulation from
\cite{Huybers2008} to model the changes in ice height due to
these effects. Bedrock depression is important in this model
(despite the fact that we have only one atmospheric layer and
thus do not resolve elevation-based effects), because the flow
rate is affected. The ice flow \citep[via][]{Huybers2008} is:
\begin{align}
&\begin{aligned}
\frac{\partial h}{\partial t} = \frac{\partial}{\partial y} &\left[ \frac{2A_{\text{ice}}(\rho_i g)^n}{n+2} \left | \left ( \frac{\partial (h+H)}{\partial y} \right )^{n-1} \right | \right. \left. \cdot \frac{\partial (h+H)}{\partial y}~ (h+H)^{n+2} + u_b h \right], \label{eqn:iceflow1}\\
\end{aligned}
\end{align}
where $h$ is the height of the ice, $H$ is the height of the
bedrock (always negative or zero, in this case), $A_{\text{ice}}$
represents the deformability of the ice, $\rho_i$ is the density
of ice, $g$ is the acceleration due to gravity, and $n$ is the
exponent in Glen's flow law \citep{Glen1958}, where $n=3$. The ice height and ice
surface mass density, $M_{\text{ice}}$ are simply related via $M_{\text{ice}} =
\rho_i h$. The first term inside the derivative represents the
ice deformation; the second term is the sliding of the ice at the base.
The latitudinal coordinate, $y$, is related to the radius of the
planet and the latitude, $y = R \phi$, thus $\Delta y = R \Delta
x (1-x^2)^{-1/2}$. Finally, $u_b$, the ice velocity across the
sediment, is:
\begin{align}
&\begin{aligned}
u_b = &\frac{2 D_0 a_{\text{sed}}}{(m+1)b_{\text{sed}}} \left ( \frac{ |a_{\text{sed}}|}{2D_0 \mu_0} \right )^m \cdot  \left ( 1 - \left [ 1- \frac{b_{\text{sed}}}{|a_{\text{sed}}|} \min \left ( h_s,\frac{|a_{\text{sed}}|}{b_{\text{sed}}} \right ) \right ]^{m+1} \right ), \label{eqn:basalvel}
\end{aligned}
\end{align}
as described by \cite{Jenson1996}. The constant $D_0$ represents
a reference deformation rate for the sediment, $\mu_0$ is the
reference viscosity of the sediment, $h_s$ is the depth of the
sediment, and $m=1.25$. The shear stress from the ice on the
sediment is:
\begin{equation}
a_{\text{sed}} = \rho_i g h \frac{\partial (h+H)}{\partial y}. \label{eqn:shear}
\end{equation}
The shear strength increases linearly with depth, with a constant rate of increase, $b_{\text{sed}}$, given by the material properties of the soil:
\begin{equation}
b_{\text{sed}} = (\rho_s-\rho_w)g \tan{\phi_s},
\end{equation}
where $\rho_s$ and $\rho_w$ are the density of the sediment and
water, respectively, and $\phi_s$ is the internal deformation
angle of the sediment. We adopt the same numerical values as
\cite{Huybers2008} for all parameters related to ice and sediment
(see Table \ref{tab:ice_tab}), with a few exceptions. We use a value
of $A_{\text{ice}}$ (ice deformability) that is consistent with ice at
270 K \citep{Paterson1994}, and a value of $r_{\text{snow}}$ (the
precipitation rate) that best reproduces Milankovitch
cycles on Earth (see Section \ref{sec:milankovitch}). Note also that the
value of $D_0$ in Table A2 of \cite{Huybers2008} appears to be
improperly converted for the units listed (the correct value,
from \cite{Jenson1996}, is listed in the text, however). With
Eqs. (\ref{eqn:basalvel}) and (\ref{eqn:shear}), Eq. (\ref{eqn:iceflow1})
can be treated numerically as a diffusion equation, with the
form:
\begin{equation}
\frac{\partial h}{\partial t} = D_{\text{ice}} \frac{\partial^2 (h+H)}{\partial y^2},\\
\end{equation}
where,
\begin{align}
&\begin{aligned}
D_{\text{ice}} & = \frac{2A_{\text{ice}}(\rho_i g)^n}{n+2} \left | \left ( \frac{\partial (h+H)}{\partial y} \right )^{n-1} \right | ~ (h+H)^{n+2}\\
&+\frac{2 D_0 \rho_i g h^2}{(m+1)b_{\text{sed}}} \left ( \frac{ |a_{\text{sed}}|}{2D_0 \mu_0} \right )^m \cdot  \left ( 1 - \left [ 1- \frac{b_{\text{sed}}}{|a_{\text{sed}}|} \min \left ( h_s,\frac{|a_{\text{sed}}|}{b_{\text{sed}}} \right ) \right ]^{m+1} \right ),
\end{aligned}
\end{align}
and $D_{\text{ice}}$ is evaluated at each time-step, at every boundary
to provide mass continuity. We solve the diffusion equation numerically using
a Crank-Nicolson scheme \citep{Crank1947}.

The bedrock depresses and rebounds locally in response to the changing weight of ice above, always seeking isostatic equilibrium. The equation governing the bedrock height, $H$, is
\begin{equation}
    \frac{\partial H}{\partial t} = \frac{1}{T_b}\left( H_{eq} - H - \frac{\rho_i h}{\rho_b} \right),
    \label{eqn:brock}
\end{equation}
where $T_b$ is a characteristic relaxation time scale, $H_{eq} = 0$ is the ice-free equilibrium height, and $\rho_b$ is the bedrock density \citep{Clark1998,Huybers2008}. We again adopt the values used by \cite{Huybers2008} (see Table \ref{tab:ice_tab}).

Because of the longer time-scales (years) associated with the ice
sheets, the growth/melting and ice-flow equations are run asynchronously in \poise. First, the
EBM, Eq.~(\ref{eqn:ebmeq}), is run for a user set number of
orbital periods (for example, 3-5), and ice accumulation and ablation is tracked
over this time frame, but ice-flow (Equation \ref{eqn:iceflow1}) is
ignored. Integrating the EBM without ice flow for a few orbits is generally acceptable as long as the total time
is lower than the required time step for the ice flow, which is typically around 5-10 years. The annually-averaged ice accumulation/ablation is then
calculated from this time-frame and passed to the ice-flow
time-step, which can be much longer (years). The EBM is then
re-run periodically to update accumulation and ablation and ensure that conditions vary smoothly and
continuously.

To clarify, the hierarchy of models and their time-steps (when \poise is part of a \vplanet calculation) is as follows:
\begin{enumerate}
\item The EBM (shortest time-step): run for a duration of several
orbital periods with time-steps on the order of days. The model
is then rerun at the end of every orbital/obliquity time-step and
at user-set intervals throughout the ice-flow model.
\item The ice-flow model (middle time-step): run at the end of
every orbital time-step (with time-steps of a few orbital
periods), immediately after the EBM finishes. The duration of the
model will follow one of two scenarios:
\begin{enumerate}[label=(\alph*)]
\item If the orbital/obliquity time-step is sufficiently long,
the EBM is rerun at user-set intervals, then the ice-flow model
continues. The ice-flow model and the EBM thus alternate
back-and-forth until the end of the orbit/obliquity time-step.
\item If the orbital/obliquity time-step is shorter than the
user-set interval, the ice-flow model simply runs until the end
of the orbital time-step.
\end{enumerate}
\item The rest of \vplanet (longest time-step). The
time-steps are set by the fastest changing variable amongst those parameters, see $\S$~\ref{sec:vplanet}.
\end{enumerate}

\begin{figure*}[ht!]
\includegraphics[width=\textwidth]{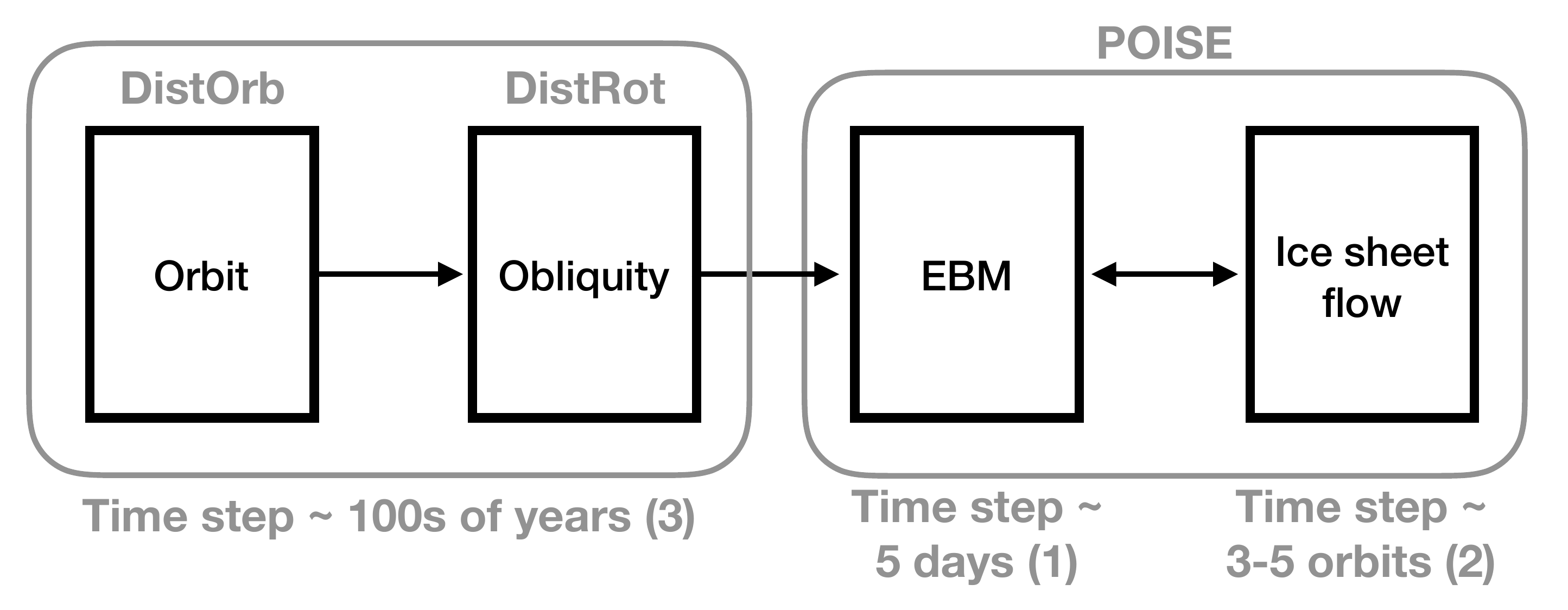}
\caption{\label{fig:poisestruct} Hierarchy of \texttt{POISE} and
the orbit and obliquity models. The orbit and obliquity models
(\distorb and \distrot) are run for $\sim$
hundreds of years (with an adaptive time step determined by the
rates of change of the orbital/obliquity parameters).
\texttt{POISE} is run at the end of each orbit/obliquity time
step. First, the EBM is run for several orbits, with time steps
of $\sim$ 5 days. Then the ice flow model is run with time steps
of $\sim 3-5$ orbits. The ice flow model runs until the next
orbit/obliquity time step, or until a user-set time, at which
point the EBM is rerun for several orbits.}
\end{figure*}

This approach is shown schematically in
Figure \ref{fig:poisestruct}. The user-set interval
discussed above must be considered carefully. The assumption is
that annually-averaged climate conditions like surface
temperature and albedo do not change much during the time span
over which the ice-flow model runs. For the results in Section \ref{sec:milankovitch}, we choose a value that
ensures that the ice-flow does not run so long that it
changes the albedo over more than one grid point without updating the temperature
and ice balance (growth/ablation) via the EBM.

The initial conditions for the EBM are as follows: The first time the
EBM is run, the planet has zero ice mass on land, the temperature
on both land and water is set by the function
\begin{equation}
T_0 = 7.5^{\circ}\text{C} + (20^{\circ}\text{C})(1-2\sin^2{\phi}),\label{eqn:inittemp}
\end{equation}
where $\phi$ is the latitude. This choice gives the planet a mean
temperature of $\sim14^{\circ}$ C, ranging from
$\sim28^{\circ}$ C in the tropics to $\sim-13^{\circ}$ at the
poles, or a ``warm start'' condition. The initial albedo
of the surface is calculated from the initial temperatures. In order to adjust the
initial conditions to better suit non-Earth-like initial conditions (\emph{i.e.}, different insolation parameters), we
then perform a ``spin-up'' phase, running the EBM iteratively
until the mean temperature between iterations changes by
$<0.1^{\circ}$ C, \emph{without} running the orbit, obliquity, or
ice-flow models, to bring the seasonal EBM close to equilibrium at
the actual stellar flux the planet receives and its actual
initial obliquity. This spin up phase generally takes a few seconds of computation time. The threshold value of $0.1^{\circ}$
was chosen so that the EBM is close to equilibrium with the stellar forcing without the spin up phase running
longer than 10-20 seconds, however, a different value may be more appropriate for different scenarios. In that case, users
should adjust this value and/or the initial temperature distribution to suit the nature of the planet being modeled. There does exist the possibility that, for certain values of this error threshold, the solution will oscillate indefinitely between two solutions and the spin up phase will never end. We advise users to be mindful of  whether  the combination of initial conditions and spin up process produces sensible results.
Then, every time the EBM is rerun (at the
user-set interval or the end of the orbit/obliquity time-step),
the initial conditions are taken from the previous EBM run
(temperature distribution) and the end of the ice-flow run
(albedo, ice mass).

\begin{table*}[ht]
\caption{Parameters used in the EBM}
\begin{center}
\begin{tabular}{lllp{0.5\linewidth}}
\hline\hline \\ [-1.5ex]
Variable & Value & Units & Physical description \\ [0.5ex]
\hline \\ [-1.5ex]
$C_L$ & $1.55 \times 10^7$ & J m$^{-2}$ K$^{-1}$ & land heat capacity \\
$C_W$ & $4.428 \times 10^6$ & J m$^{-2}$ K$^{-1}$ m$^{-1}$ & ocean heat capacity per meter of depth \\
$m_d$ & 70 & m & ocean mixing depth \\
$D$ & 0.58 & W m$^{-2}$ K$^{-1}$ & meridional heat diffusion coefficient\\
$\nu$ & 0.8 &  & coefficient of land-ocean heat flux\\
$A$ & 203.3 & W m$^{-2}$ & OLR parameter\\
$B$ & 2.09 & W m$^{-2}$ K$^{-1}$& OLR parameter\\
$\alpha_L$ & 0.363 &  & albedo of land \\
$\alpha_W$& 0.263 &  & albedo of water \\
$\alpha_i$& 0.6  & & albedo of ice\\
$f_L$ & 0.34 & & fraction of latitude cell occupied by land\\
$f_W$ & 0.66 & & fraction of latitude cell occupied by water\\
\hline
\end{tabular}
\end{center}
\label{tab:ebm_tab}
\end{table*}
\begin{table*}[ht]
\caption{Parameters used in the ice sheet model}
\begin{center}
\begin{tabular}{lllp{0.5\linewidth}}
\hline\hline \\ [-1.5ex]
Variable & Value & Units & Physical description \\ [0.5ex]
\hline \\ [-1.5ex]
$T_{freeze}$ & 273.15 & K & freezing point of water \\
$L_h$ & $3.34 \times 10^5$ & J kg$^{-1}$ & latent heat of fusion of water\\
$r_{\text{snow}}$ & $2.25 \times 10^{-5}$ & kg m$^{-2}$ s$^{-1}$ & snow/ice deposition rate \\
$A_{\text{ice}}$ & $2.3 \times 10^{-24}$ & Pa$^{-3}$ s$^{-1}$ & deformability of ice \\
$n$ & 3 & & exponent of Glen's flow law \\
$\rho_i$ & 916.7 & kg m$^{-3}$ & density of ice \\
$\rho_s$ & 2390 & kg m$^{-3}$ & density of saturated sediment \\
$\rho_w$ & 1000 & kg m$^{-3}$ & density of liquid water\\
$D_0$ & $7.9 \times 10^{-7}$ & s$^{-1}$ & reference sediment deformation rate\\
$\mu_0$ & $3 \times 10^9$ & Pa s & reference sediment viscosity \\
$m$ & 1.25 & & exponent in sediment stress-strain relation\\
$h_s$ & 10 & m & sediment depth \\
$\phi_s$ & 22 & degrees & internal deformation angle of sediment \\
$T_b$ & 5000 & years & bedrock depression/ rebound timescale\\
$\rho_b$ & 3370 & kg m$^{-3}$ & bedrock density\\
\hline
\end{tabular}
\label{tab:ice_tab}
\end{center}
\end{table*}

\section{The \texttt{RadHeat} Module\label{app:radheat}}

Radiogenic heat production in the Earth is generated primarily by the decay of $^{238}U$, $^{235}U$, $^{232}Th$, and $^{40}K$ in the crust, mantle, and core. Other species, such as $^{26}$Al, are also potential heat sources on exoplanets. The radiogenic power produced by species $i$ in reservoir $j$ is
\be
Q_{i,j}=Q_{i,j}(0)\exp(-\lambda_{i,1/2}t)
\label{radiogenicheat}
\ee
where $\lambda_{i,1/2}=\ln 2/\tau_{i,1/2}$, $\tau_{i,1/2}$ is the  half-life, $t$ is time, $Q_{i.j}(0)$ is the initial heat production at $t=0$, and $j = $ core, mantle, crust.
The initial radiogenic heat can be input as a power, number of atoms  (typically $\sim 10^{42}$), or mass within each reservoir. Table ~\ref{tab:radheat} shows our default Earth values.

\begin{table}[ht]
\begin{center}
\caption{Initial Radiogenic Properties of Earth}
\begin{tabular}{cccccc}
\hline\hline
Isotope & Half-life & Energy/decay & $Q_{rad,core}$(0) & $Q_{rad,man}$(0) & $Q_{rad,crust}$(0)\\
 & (Gyr) &  ($10^{-12}$ J) & (TW) & (TW) &  (TW)\\
\hline
$^{26}$Al & 0.717 & 0.642 & 0 & 0 & 0\\
$^{40}$K & 1.8178 & 0.213 & 33.86 & 36.16 & 13.89\\
$^{232}$Th & 20.202 & 6.834 & 0.145 & 6.52 & 3.83\\
$^{235}$U & 1.015 & 6.555 & 0.50 & 20.25 & 10.39\\
$^{238}$U & 6.452 & 8.283 & 0.12 & 11.67 & 5.16\\
\hline
\end{tabular}
\label{tab:radheat}
\end{center}
\end{table}

\section{The \texttt{SpiNBody} Module\label{app:spinbody}}

In addition to the \distorb method described in Section \ref{app:distorb}, orbital evolution can also be modelled using a classic N-body simulation. The \vplanet module \spinbody simulates the orbital evolution of a system by calculating the gravitational forces on each body using Newtonian gravity. The gravitational force on an orbital body $i$ by $j$ additional bodies is given by:

\be F = \sum_j G \frac{m_i m_j}{r_{i,j}^2},\ee

\noindent where $G$ is Newton's gravitational constant, $m_i$ is the mass of body $i$, and $r_{i,j}$ is the distance between body $i$ and body $j$. Thus, the barycentric position and velocity derivatives are given by:

\be \dot{x_i} = v_i \ee
\be \dot{v_i} = \sum_j G \frac{m_j}{r_{i,j}^2} \ee

\begin{figure}[ht!]
\begin{center}
\includegraphics[width=0.9\textwidth]{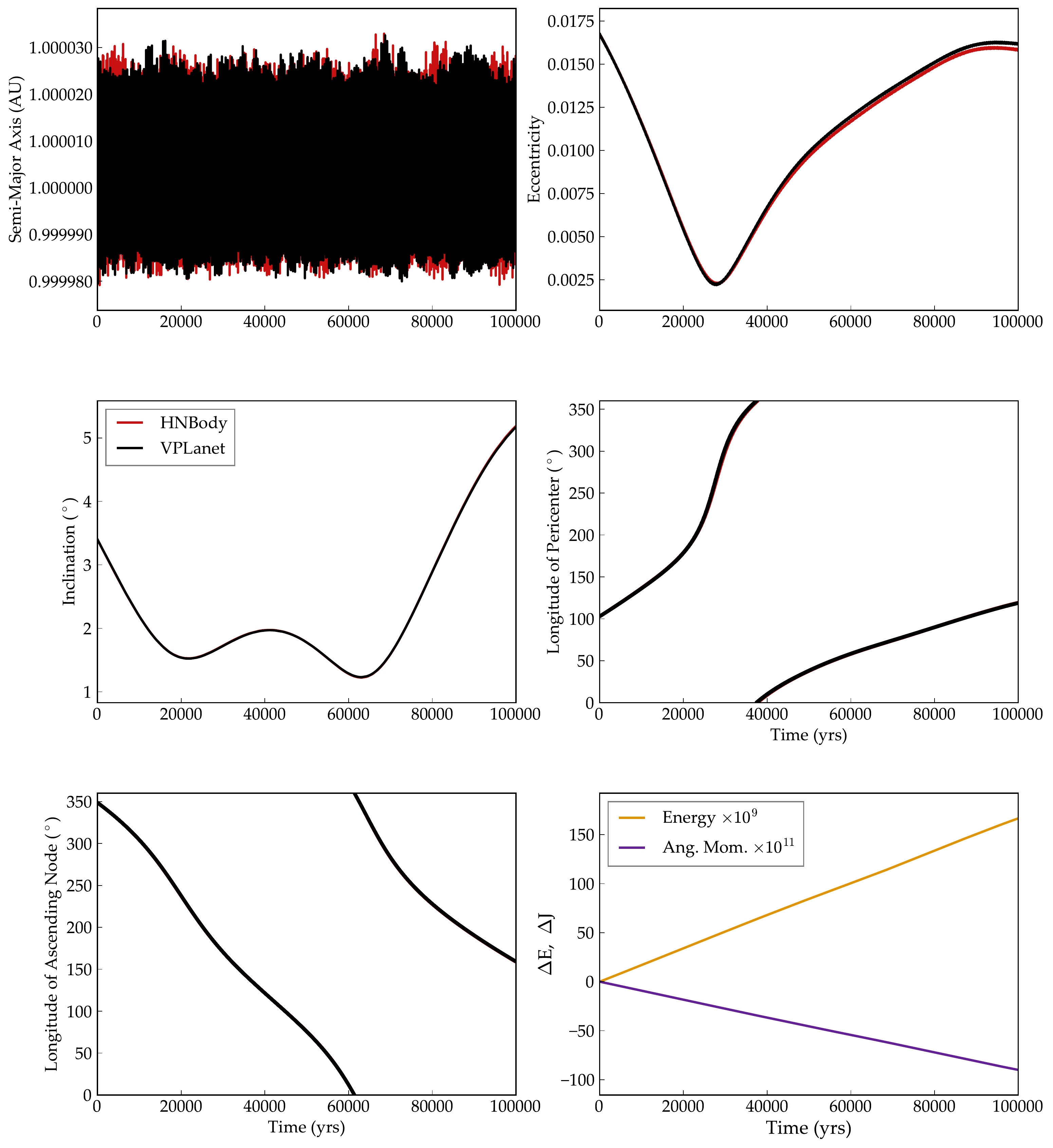}
\end{center}
\caption{\label{fig:SS_Nbody}Evolution of Earth's orbit according to \spinbody (black) and \hnbody (red). The two trajectories are nearly identical. The bottom right panel shows the conservation of energy ($\Delta E$) and angular momentum ($\Delta J$) for \spinbody only. \href{https://github.com/VirtualPlanetaryLaboratory/vplanet/tree/master/examples/SS_NBody}{\link{examples/SS\_NBody}}.}
\end{figure}

Because this is an N-body model, it is several orders of magnitude slower than \distorb. Unlike \distorb, however, \spinbody can simulate systems of large eccentricity and mutual inclination, as well as systems with orbital resonances. Note, because \spinbody uses Cartesian coordinates and derivatives, it cannot currently couple to modules (\eg \distorb and \eqtide) that calculate orbital evolution based on the derivatives of the osculating elements. Fig.~\ref{fig:SS_Nbody} shows the evolution of Earth's orbit due to the Sun and other 7 planets for both \vplanet in black, and \hnbody's 4th order Runge-Kutta method (\ie the same method as \vplanet) in red. At the end of the simulations, the difference in eccentricity is about 2\%, but is a function of timestep.

\section{The \texttt{STELLAR} Module\label{app:stellar}}

The \stellar module simulates the evolution of the stellar radius, radius of gyration ($r_g$), effective temperature, luminosity, XUV luminosity, and rotation rate over time. \stellar evolves the first four quantities using a bicubic spline interpolation over mass and time of the evolutionary tracks of \citet{Baraffe15} for solar-metallicity stars. \stellar tracks the XUV luminosity of stars using the empirical broken power law model of \cite{Ribas05},
\begin{align}
\label{eq:lxuv}
\frac{L_\mathrm{XUV}}{L_\mathrm{bol}} = \left\{
				\begin{array}{lcr}
					f_\mathrm{sat} &\ & t \leq t_\mathrm{sat} \\
					f_\mathrm{sat}\left(\frac{t}{t_\mathrm{sat}}\right)^{-\beta_\mathrm{XUV}} &\ & t > t_\mathrm{sat},
				\end{array}
				\right.
\end{align}
where $L_\mathrm{bol}$ is the total (bolometric) stellar luminosity, \ie from the \citet{Baraffe15} grids, $\beta_\mathrm{XUV}$ is the power law exponent \citep[equal to $-1.23$ for Sun-like stars; ][]{Ribas05}, and $f_\mathrm{sat}$ is the initial ratio of XUV to bolometric luminosity. Prior to $t = t_\mathrm{sat}$, the XUV luminosity is said to be ``saturated,'' as observations show that the ratio $L_\mathrm{XUV}/L_\mathrm{bol}$ remains relatively constant at early times \citep[\eg][]{Wright11}.
For G and K dwarfs, $t_\mathrm{sat} \approx 100 \mathrm{Myr}$ and $f_\mathrm{sat} \approx 10^{-3}-10^{-4}$ \citep{Jackson12}.
Because of poor constraints on the ages of field M dwarfs, the saturation timescale for these stars is not known, although it is likely longer than that for K dwarfs. Following \citet{LugerBarnes15}, we adopt $t_\mathrm{sat} = 1 \,\mathrm{Gyr}$ and $f_\mathrm{sat} = 10^{-3}$ as default values in \stellar.

In addition to tracking the evolution of fundamental stellar properties, \stellar models the angular momentum evolution of stars undergoing stellar evolution, \eg radius contraction, and angular momentum loss due to magnetic braking. \vplanet includes three different prescriptions from \citet{Reiners2012}, \citet{Repetto2014}, and \citet{Matt2015}, with the latter including the corrections of \citet{Matt2019}.

The \citet{Reiners2012} magnetic braking model, which links angular momentum loss with stellar magnetic field strength, is given by
\begin{equation} \label{eqn:stellar:reiners}
\begin{split}
\frac{dJ_{\star}}{dt} & = -C \left[ \omega \left(\frac{R^{16}}{M^2} \right)^{1/3} \right] \text{for $\omega \geq \omega_{crit}$} \\
\frac{dJ_{\star}}{dt} & = -C \left[ \left( \frac{\omega}{\omega_{crit}} \right)^4 \omega \left(\frac{R^{16}}{M^2} \right)^{1/3} \right] \text{for $\omega < \omega_{crit}$},
\end{split}
\end{equation}
where $C = 2.66 \times 10^3 \text{ (gm$^5$ cm$^{-10}$ s$^3$)$^{1/3}$}$, $\omega_{crit} = 8.56 \times 10^{-6}\text{ s$^{-1}$}$ for $M > 0.35$ M$_{\odot}$, and $\omega_{crit} = 1.82 \times 10^{-6} \text{ s$^{-1}$}$ for $M \leq 0.35 M_{\odot}$. In this model, the dichotomy between unsaturated ($\omega < \omega_{crit}$) and saturated ($\omega \geq \omega_{crit}$) assumes that for fast enough rotation rates, \ie the stars in the ``saturated" regime, the stellar magnetic field strength is constant, whereas in the unsaturated regime, the stellar magnetic field strength depends on the rotation rate. The \citet{Repetto2014} model is simply based off of the observed spin-down of Sun-like stars by \citet{Skumanich1972} and is given by
\begin{equation} \label{eqn:stellar:repetto}
\frac{dJ_{\star}}{dt} = - \gamma M r_g^2 R^4 \omega^3,
\end{equation}
where $\gamma = 5 \times 10^{-25} \text{ s m}^{-2}$. \stellar also allows the user to select the \citet{Matt2015} magnetic braking model that successfully reproduced observed features in the rotation period distribution of \textit{Kepler} field stars.  The \citet{Matt2015} model is based on scaling relations with the stellar mass, radius, and Rossby number, the ratio of the convective turnover timescale to the rotation period, and is given by the following equations:
\begin{equation} \label{eqn:stellar:matt}
\begin{split}
\frac{dJ_{\star}}{dt} & = -T_0 \left(\frac{\tau_{cz}}{\tau_{cz\odot}}\right)^2 \left(\frac{\omega_{\star}}{\omega_{\odot}}\right)^3, \text{(unsaturated)} \\
\frac{dJ_{\star}}{dt} & = -T_0 \chi^2 \left(\frac{\omega_{\star}}{\omega_{\odot}}\right), \text{(saturated)}
\end{split}
\end{equation}
where $\tau_{cz}$ is the convective turnover timescale, $\tau_{cz\odot}$ is the solar convective turnover timescale, 12.9 days, and
\begin{equation} \label{eqn:stellar:mattchi}
\chi = \frac{\text{Ro}_{\odot}}{\text{Ro}_{sat}} = \frac{\omega_{sat} \tau_{cz}}{\omega_{\odot} \tau_{cz\odot}}
\end{equation}
is the inverse critical Rossby number for saturation in solar units. \citet{Matt2015} finds that saturation occurs when $Ro \leq Ro_{\odot}/\chi$, where $\chi = 10$.  The normalized torque is given by
\begin{equation} \label{eqn:stellar:mattT0}
T_0 = 6.3 \times 10^{23} \text{ Joules } \left(\frac{R}{R_{\odot}}\right)^{3.1} \left(\frac{M}{M_{\odot}}\right)^{0.5}.
\end{equation}
We compute $\tau_{cz}$ using the fit presented in \citet{Cranmer2011}, given by
\begin{equation} \label{eqn:stellar:mattcz}
\tau_{cz} = 314.24\exp\left[-\left(\frac{T_{eff}}{1952.5\text{ K}}\right)-\left(\frac{T_{eff}}{6250\text{ K}}\right)^{18}\right] + 0.002
\end{equation}
for stellar effective temperature $T_{eff}$.

\begin{figure}[t!]
\begin{center}
\includegraphics[width=\textwidth]{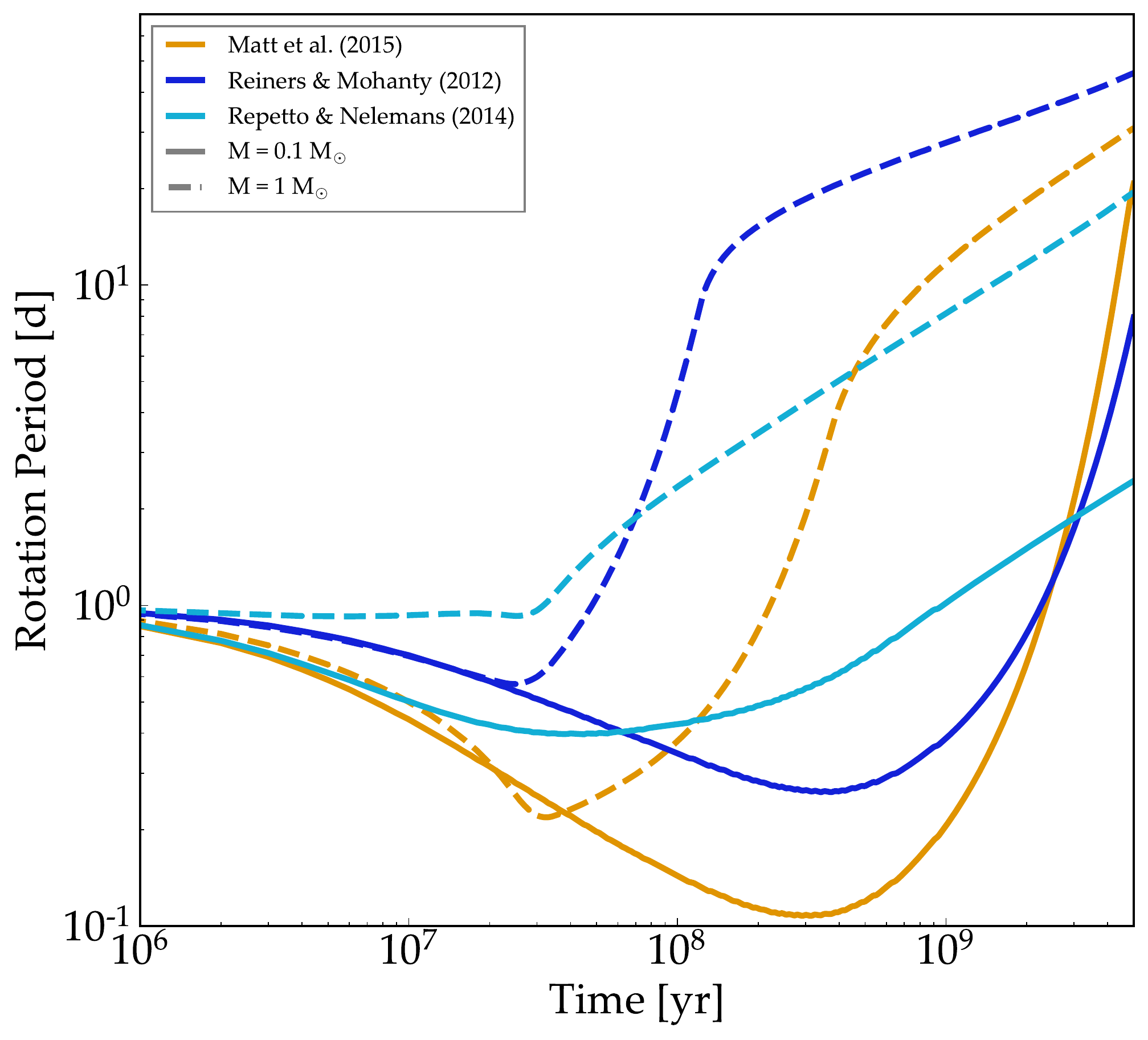}
\end{center}
\caption{ \label{fig:magneticBraking} Rotation period evolution for $0.1 \msun$ (late M dwarf, solid lines) and $1 \msun$ (Sun-like star, dashed lines) mass stars evolved using \stellar according to Eq.~(\ref{eqn:stellar_rot_rate_dt}) for the \citet{Matt2015} (blue), \citet{Reiners2012} (orange), and \citet{Repetto2014} magnetic braking models. Initially as the stars contract along the pre-main sequence, the rotation periods decrease via conservation of angular momentum. Once the stars reach the main sequence, stellar radii and $r_g$ remain approximately constant allowing magnetic braking to remove angular momentum from the stars, increasing rotation periods in the long terms. Approximate runtime: 2 minutes. \href{https://github.com/VirtualPlanetaryLaboratory/vplanet/tree/master/examples/MagneticBraking}{\link{examples/MagneticBraking}}}
\end{figure}

We model the total change in a star's rotation rate, $\omega$ due to radius and $r_g$ evolution, assuming conservation of angular momentum, and magnetic braking according to
\begin{equation} \label{eqn:stellar_rot_rate_dt}
\dot{\omega} = \frac{\dot{J}_{\star}}{I} - \frac{2 \dot{R} \omega}{R} - \frac{2 \dot{r_g} \omega}{r_g}
\end{equation}
for the stellar moment of inertia, $I = m r_g^2 R^2$. We compute numerical time derivatives of the stellar radius and $r_g$ using our interpolation of the \citet{Baraffe15} grids and $\dot{J}_{\star}$ is the angular momentum loss due to magnetic braking given by one of the above models. In Fig.~\ref{fig:magneticBraking}, we plot the rotation period evolution for late M-dwarfs and Sun-like stars subject to Eq.~(\ref{eqn:stellar_rot_rate_dt}) for the three magnetic braking models implemented in \vplanet.

\begin{figure}[t!]
\begin{center}
\includegraphics[width=\textwidth]{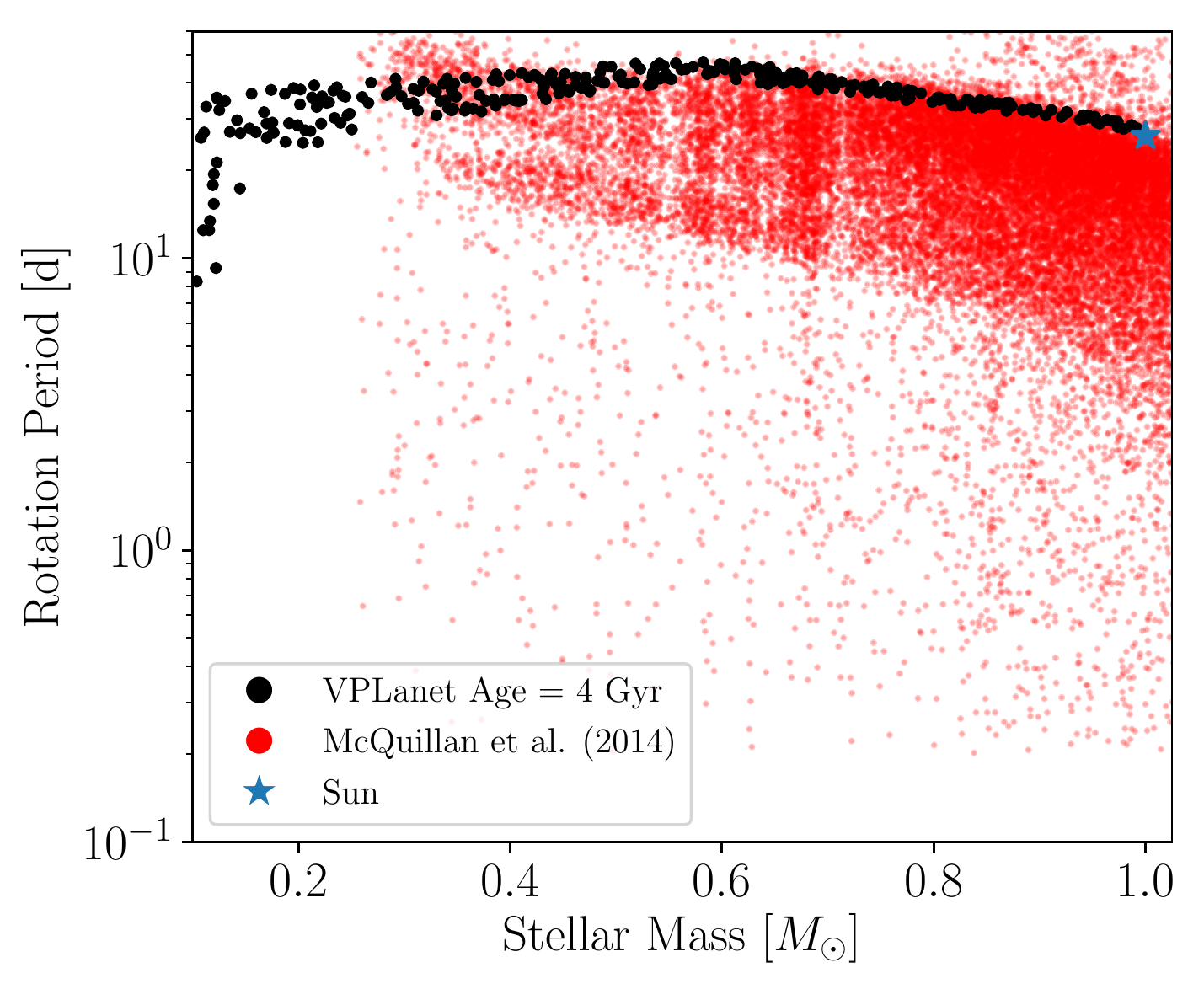}
\end{center}
\caption{Rotation period distribution of a ${\sim}4$ Gyr-old synthetic cluster of stars simulated using \stellar with the \citet{Matt2015} magnetic braking model (black, adapting \vplanet simulations from \citet{Fleming19}). Following Fig. (3) in \citet{Matt2015}, we compare the \citet{Fleming19} simulated distribution to the rotation distribution of \kepler field stars (red) measured by \citet{McQuillan2014}. For reference, we plot the modern solar rotation period as a blue star. Using \stellar, \citet{Fleming19} recover the \citet{Matt2015} result that the upper envelope of the \kepler stellar rotation period distribution is well-matched by a 4 Gyr-old synthetic cluster, validating the \stellar implementation of the \citet{Matt2015} magnetic braking model. \label{fig:kepler} \href{https://github.com/VirtualPlanetaryLaboratory/vplanet/tree/master/examples/MagneticBraking}{\link{examples/MagneticBraking}}}
\end{figure}

In Fig.~\ref{fig:kepler}, we compare our \stellar magnetic braking implementation to the \citet{Matt2015} model by reproducing their Fig. 3. We plot the rotation period distribution of a synthetic cluster of single stars with ages ${\sim} 4$ Gyr from the publicly-available \vplanet \stellar simulations of \citet{Fleming19}\footnote{Simulation data are available at \href{https://github.com/dflemin3/sync/tree/master/Data}{https://github.com/dflemin3/sync/tree/master/Data} and is given in the MagneticBraking \vplanet example directory (see the Fig.~\ref{fig:kepler} caption).} in Fig.~\ref{fig:kepler} and overplot the \kepler stellar rotation period distribution measured by \citet{McQuillan2014}. We find that our 4 Gyr-old synthetic cluster matches the upper envelope of the \kepler distrubtion, in good agreement with \citet{Matt2015} and validating our implementation.

\section{The \texttt{ThermInt} Module\label{app:thermint}}

The thermal history of the interior is determined by the time evolution of the average mantle $T_m$ and average core $T_c$ temperatures.  The energetic state of the core governs inner core crystallization and the dynamo.

\subsection{Governing Differential Equations}

The conservation of energy in the mantle is
\be Q_{surf}=Q_{conv}+Q_{melt} =Q_{i,man}+Q_{cmb}+Q_{man}+Q_{tidal}+Q_{L,man},
\label{eq:mantle_energy} \ee
where $Q_{surf}$ is the total mantle surface heat flow, $Q_{conv}$ is heat conducted through the lithospheric thermal boundary layer that is supplied by mantle convection, $Q_{melt}$ is heat loss due to the eruption of upwelling mantle melt at the surface, $Q_{i,man}$ is heat generated by radioactive decay of species $i$ in the mantle, $Q_{cmb}$ is heat lost from the core across the core-mantle boundary (CMB), $Q_{man}$ is the secular heat lost from the mantle, $Q_{tidal}$ is heat generated in the mantle by tidal dissipation, and $Q_{L,man}$ is latent heat released by the solidification of the mantle.  Crustal heat sources have been excluded because they do not contribute to the mantle heat budget.  Note that heat can be released from the mantle in two ways: via conduction through the upper mantle thermal boundary layer ($Q_{conv}$) and by melt eruption ($Q_{melt}$).  Detailed expressions for heat flows and temperature profiles as functions of mantle and core properties are given below.

Similarly, the thermal evolution of the core is governed by the conservation of energy in the core,
\be Q_{cmb}=Q_{core}+Q_{icb}+Q_{i,core}   \label{eq:core_energy} \ee
where $Q_{core}$ is core secular cooling, $Q_{i,core}$ is radiogenic heat production of species $i$ in the core, and heat released by the solidification of the inner core is $Q_{icb}=\dot{M}_{ic}(L_{icb}+E_{icb})$, where $\dot{M}_{ic}$ is the change in inner core mass $M_{ic}$, and $L_{icb}$ and $E_{icb}$ are the latent and gravitational energy released per unit mass at the inner-core boundary (ICB).

Thermal evolution equations for the average mantle $T_m$ and core $T_c$ temperatures are derived by using the secular cooling equation $Q_{j}=-c_j M_i \dot{T}_j$, where $c$ is specific heat and $j$ refers to either mantle or core, in Eqs. (\ref{eq:mantle_energy}) and (\ref{eq:core_energy}).  Solving for $\dot{T}_m$ and $\dot{T}_c$ gives the mantle and core thermal evolution equations,
\begin{equation}
 \dot{T}_m=\left( Q_{cmb} + Q_{i,man} +Q_{tidal}+Q_{L,man} -Q_{conv}-Q_{melt} \right)  / M_m c_m
\label{eq:dot_T_m} \end{equation}
\begin{equation}
\dot{T}_c= -\frac{ (Q_{cmb}-Q_{i,core})} {M_c c_c - A_{ic} \rho_{ic} \eta_c \frac{d R_{ic}}{d T_{cmb}} (L_{Fe}+E_G) }
\label{eq:dot_T_c} \end{equation}
where the denominator of (\ref{eq:dot_T_c}) is the sum of core specific heat and heat released by the inner core growth, $A_{ic}$ is inner core surface area, $\rho_{ic}$ is inner core density, $\eta_c$ is a constant that relates average core temperature to CMB temperature, $dR_{ic}/dT_{cmb}$ is the rate of inner core growth as a function of CMB temperature, and $L_{Fe}$ and $E_G$ are the latent and gravitational energy released at the ICB per unit mass. Table \ref{tab:thermint} lists the values for these parameters, which are taken from \cite{mckenzie1984}, \cite{mckenzie1988}, \cite{hirschmann2000}, \cite{elkinstanton2008b}, \cite{Barnes13}, and \cite{jaupart2015}.

An additional conservation of energy can be added for the cooling of the lithosphere for a stagnant lid planet.  Alternatively, the stagnant lid heat flow can be approximated by
\be Q_{conv,stag} = \epsilon_{stag} Q_{conv}
\label{eq:Q_conv_stag}\ee
where $\epsilon_{stag}\approx 1/25$ \citep{DriscollBercovici14}.

\subsection{Mantle Rheology}
Effective mantle viscosity follows an Arrhenius Law form,
\be
\nu=\nu_{ref}\mbox{exp}\left(\frac{E_\nu}{R_g T}\right)/\epsilon_{phase}
\label{eq:nu} \ee
where $\nu=\eta/\rho_m$ is kinematic viscosity, $\rho_m$ is mantle density, $\nu_{ref}$ is a reference viscosity, $E_\nu$ is the viscosity activation energy, $R_g$ is the gas constant, $T$ is either upper or lower mantle temperature, and $\epsilon_{phase}$ accounts for the weakening effect of a solid to liquid phase change (see Table \ref{tab:thermint} for a list of constants).  Shear modulus, which is required to calculate tidal heating (see App.~\ref{app:thermint}.3), is similarly described,
\be
\mu=\mu_{ref}\mbox{exp}\left(\frac{E_\mu}{R_g T_m}\right)/\epsilon_{phase}~.
\label{eq:mu}\ee
This model predicts the rapid drop in shear modulus with melt fraction demonstrated experimentally by \citet{jackson2004}.
The reference shear modulus $\mu_{ref}=10^5$ Pa and effective stiffness $\beta_{st}=1.71\times10^4$ GPa are calibrated by $k_2=0.3$ and \tidalq = 100 for the present-day mantle.

The influence of melt fraction $\phi$ on viscosity is modeled following the parameterization of \citet{costa2009},
\be
\epsilon_{phase}(\phi) = \frac{1+\Phi^{\delta_{ph}}}{ \left[ 1- F \right]^{B \phi^*} },
\label{eq:epsilon_phase}
\ee
and
\be
F =  (1-\xi) \mbox{erf} \left[ \frac{\sqrt{\pi}}{2(1-\xi)} \Phi(1+\Phi^{\gamma_{ph}}) \right],
\label{eq:epsilon_phase2}
\ee
where $\Phi=\phi/\phi^*$, ``erf'' is the error function, and $\phi^*$, $\xi$, $\gamma_{ph}$, and $\delta_{ph}$ are empirical constants (Table \ref{tab:thermint}). The nominal value of $A_\mu=2\times10^5$ J mol$^{-1}$ produces a dissipation peak when the melt fraction is about $50\%$.

The mid-mantle and lower mantle viscosities can be related to the upper mantle viscosity by
\be \nu_{man}=f_{\nu,man} \nu_{UM} ~~,~~ \nu_{LM}=f_{\nu,LM} \nu_{UM},
\label{eq:nu_man} \ee
where $f_{\nu,man}$ and $f_{\nu,LM}$ are constant coefficients.  The mantle Rayleigh number is
\be Ra_{man} = \frac{ \alpha g \Delta T_{man} D_{man}^3} {\kappa \nu_{man}},
\label{eq:ra_man} \ee
where $\alpha$ is thermal expansivity, $D_{man}=R-R_{cmb}$ is mantle shell thickness, $\kappa$ is thermal diffusivity, and mantle convective temperature jump is,
\be \Delta T_{man}=\Delta T_{UM} + \Delta T_{LM} = (T_{UM}-T_g) + (T_{cmb}-T_{LM}),
\label{eq:delta_T_man} \ee
where $T_{UM}$ and $\Delta T_{UM}$ are the temperatures at the bottom of and across the upper mantle thermal boundary layer; $T_{LM}$ and $\Delta T_{LM}$ are the temperatures at the top of and across the lower mantle thermal boundary layer; $T_{cmb}$ is CMB temperature.

\subsection{Tidal Heating}
Tidal heating associated with gravitational tides is assumed to occur by visco-elastic dissipation in the mantle.
The power dissipated by tidal strain in a planet in synchronous rotation is \citep{segatz1988,DriscollBarnes15},
\be
Q_{tidal}=-\frac{21}{2} \mbox{Im}(k_2) \frac{G M_*^2 R_p \omega e^2}{a^6},
\ee
where $G$ is the gravitational constant, $M_*$ is the central mass, $R_p$ is planet radius, $\omega$ is orbital frequency, $e$ is orbital eccentricity, $a$ is orbital semi-major axis, and $\mbox{Im}(k_2)$ is the imaginary part of the complex second order love number $k_2$.
If planetary rotation is synchronous, then the tidal frequency is equal to the mean motion $\omega=n=\sqrt{GM_*/a^3}$, and the tidal power reduces to
\be
Q_{tidal}= - \frac{21}{2}  \mbox{Im}(k_2)G^{3/2} M_*^{5/2} R_p \frac{e^2}{a^{15/2}}. \label{eq:q_tidal}
\ee
This expression for tidal dissipation is the product of three physical components: (1) tidal efficiency ($-$Im$(k_2)$), (2) star-planet size ($M_*^{5/2}R_p$), and (3) orbit ($e^2/a^{15/2}$).

The one-dimensional dissipation model in Eq.~(\ref{eq:q_tidal}) assumes a homogeneous body with uniform stiffness and viscosity.  We assume that dissipation is dominated by the material properties at the base of the upper mantle thermal boundary layer where viscosity is expected to be at a minimum in the mantle.
To derive the dissipation efficiency ($-$Im$(k_2)$) we first define the Love number,
\be k_2=\frac{3}{2}\frac{1}{1+\frac{19}{2}\frac{\mu}{\beta_{st}} },  \label{eq:k2}  \ee
where $\mu$ is shear modulus and $\beta_{st}$ is effective stiffness.  Writing shear modulus as a complex number and using the constitutive relation for a Maxwell body, one can derive the dissipation efficiency in Eq.~(\ref{eq:q_tidal}) as
\be -\mbox{Im}(k_2)=\frac{57 \eta \omega}{4 \beta_{st}\left(1+\left[\left(1+\frac{19\mu}{2\beta_{st}}\right) \frac{\eta\omega}{\mu}\right]^2 \right)} \label{eq:im_k2}, \ee
where $\eta=\nu\rho$ is dynamic viscosity \citep{henning2009}.  We note that this model does not involve a tidal $\mathcal{Q}$ factor, rather the rheological response of the mantle is described entirely by $\mbox{Im}(k_2)$.
For comparison with other models, one can compute the standard tidal $\mathcal{Q}$ factor of the Maxwell model as
\be  \mathcal{Q} =\frac{\eta \omega}{\mu}. \label{eq:tidal_q} \ee
The common approximation is then -Im$(k_2)\approx k_2/Q$.

\subsection{Geotherm}
The mantle temperature profile is assumed to be adiabatic everywhere except in the thermal boundary layers where it is conductive.  The adiabatic temperature profile in the well mixed region of the mantle is approximated to be linear in radius, which is a good approximation considering that mantle thickness $D=2891$ km is significantly less than the adiabatic scale height $H=c_p/\alpha g\approx12650$ km,
\be
T_{ad}=T_{UM} + \gamma_{ad}(R-r-\delta_{UM})~,
\label{eq:T_ad} \ee
where the solid adiabatic gradient is $\gamma_{ad}\approx0.5~\mbox{K/km}$.  In the thermal boundary layers the conductive temperature solutions,
\begin{eqnarray}
\Delta T_{UM} \mbox{erf}\left[ \frac{R-r}{\delta_{UM}}\right] + T_g ~~,~~ \mbox{Upper mantle} \\
\Delta T_{LM} \mbox{erf}\left[ \frac{R_c-r}{\delta_{LM}}\right] + T_{cmb} ~~,~~ \mbox{Lower mantle}
\label{eq:T_erf} \end{eqnarray}
replace the adiabat.  Thermal boundary layer temperature jumps are $\Delta T_{UM}=T_{UM}-T_g$ and $\Delta T_{LM}=T_{cmb}-T_{LM}$, where $T_g$ is ground temperature, and thermal boundary layer depth is $\delta$  \citep{DriscollBercovici14}.

The core temperature profile is assumed to be adiabatic throughout the entire core, and the thermal boundary layers within the core are ignored.  This approximation is valid because the low viscosity and high thermal conductivity of liquid iron produce very small thermal boundary layers that are insignificant.  The core adiabatic profile is approximated by
\begin{equation}
T_c(r)=T_{cmb}\exp \left( \frac{R_c^2-r^2}{D_N^2} \right),
\label{eq:T_c} \end{equation}
where $D_N\approx 6340$ km is an adiabatic length scale \citep{labrosse2001}.  The iron solidus is approximated by Lindemann's Law,
\begin{equation}
T_{Fe}(r)=T_{Fe,0} \exp \left[  -2 \left( 1- \frac{1}{3\gamma_c} \right) \frac{r^2}{D_{Fe}^2} \right] - \Delta T_\chi,
\label{eq:lindemann} \end{equation}
where $T_{Fe,0}=5600$ K, $\gamma_c$ is the core Gruneisen parameter, $D_{Fe}=7000$ km is a constant length scale \citep{labrosse2001}, and $\Delta T_\chi$ is a liquidus depression due to the presence of a light element, \eg potassium. If $\Delta T_\chi=0$, then the inner core radius can be written as a function of CMB temperature by equating Eqs.~(\ref{eq:T_c}) and (\ref{eq:lindemann}) and solving for $r=R_{ic}$, giving
\begin{equation}
R_{ic}=R_c \sqrt{ \frac{ (D_N/R_c)^2 \ln(T_{cmb}/T_{Fe,cen})+1}  {(D_N/R_c)^2 \ln(T_{Fe,cmb}/T_{Fe,cen}) +1} },
\label{eq:r_i} \end{equation}
where $T_{cmb}$ is CMB temperature, $T_{Fe,cen}$ is the liquidus at the center of the core, and $T_{Fe,CMB}$ is the liquidus at the CMB.
The time derivative of Eq.~(\ref{eq:r_i}) gives the rate of inner core growth as a function of core cooling rate $\dot{T}_c$ as,
\be
\dot{R}_{ic} = \frac{D_N^2/(2R_{ic})} { (D_N/R_c)^2\ln{( T_{Fe,cmb}/T_{Fe,cen} )} +1} \frac{\dot{T}_c}{T_c},
\label{eq:dot_r_i}
\ee
Assuming a constant inner core density $\rho_{ic}$, inner core mass growth rate is approximated by
\begin{equation}
\dot{M}_{ic}= A_{ic} \rho_{ic} \dot{T}_{cmb} \frac{d R_{ic}}{d T_{cmb}},
\label{eq:dot_m_i}\end{equation}
where $A_{ic}$ is the area of the inner core, and the change in inner core radius with CMB temperature can be computed by the temperature derivative of Eq.~(\ref{eq:r_i}), giving
\begin{equation}
\frac{d R_{ic}}{d T_{cmb}} = \frac{\dot{R}_{ic}}{\dot{T}_c}
 = \frac{D_N^2/(2R_{ic})} { T_{cmb}[ (D_N/R_c)^2\ln{( T_{Fe,cmb}/T_{Fe,cen} )} +1]}.
\label{eq:dri_dtcmb} \end{equation}

\subsection{Mantle and Core Heat Flows}
In this subsection we define the remaining heat flows that appear in the mantle and core energy balance, Eqs.~(\ref{eq:mantle_energy})---(\ref{eq:core_energy}). The convective cooling of the mantle $Q_{conv}$ is proportional to the temperature gradient in the upper mantle thermal boundary layer,
\begin{equation}
Q_{conv}=A k_{UM} \frac{\Delta T_{UM}}{\delta_{UM}},
\label{eq:q_conv}\end{equation}
where $A$ is surface area and $k_{UM}$ is upper mantle thermal conductivity.  $Q_{conv}$ is written in terms of $T_m$ and the thermal boundary layer thickness $\delta_{UM}$ by requiring that the Rayleigh number of the boundary layer $Ra_{UM}$ be equal to the critical Rayleigh number for thermal convection $Ra_c\approx660$ \citep{howard1966,solomatov1995,sotin1999,DriscollBercovici14}.
Solving for the critical thermal boundary layer thickness from $Ra_{UM}=Ra_c$ gives,
\begin{equation}
\delta_{UM}=D \left(Ra_c \frac{\nu_{UM}\kappa}{\alpha g \Delta T_{UM} D^3} \right)^{\beta}.
\label{eq:delta_1}\end{equation}
Using this relationship in Eq.~(\ref{eq:q_conv}) gives,
\begin{equation}
Q_{conv}=A k_{UM}  \left( \frac{\alpha g}{Ra_c \kappa }\right)^{\beta} \frac{(\epsilon_{UM}\Delta T_m)^{\beta+1} } {\nu_{UM}^{\beta}},
\label{eq:q_conv_m} \end{equation}
where the thermal boundary layer temperature jump $\Delta T_{UM}$ has been replaced by $\Delta T_{UM}\approx\epsilon_{UM}\Delta T_m$,
$\epsilon_{UM}=\exp(-(R_{UM}-R_m)\alpha g/c_p)\approx0.7$, which is the adiabatic temperature decrease from the average mantle temperature to the bottom of the upper mantle thermal boundary layer, $\Delta T_m=T_m-T_g$, and the mantle cooling exponent is $\beta=1/3$  \citep{DriscollBercovici14}.

Similar to the mantle convective heat flow, the CMB heat flow is,
\begin{equation}
Q_{cmb}=A_c k_{LM} \frac{\Delta T_{LM}}{\delta_{LM}},
\label{eq:q_cmb} \end{equation}
where $A_c$ is core surface area and $k_{LM}$ is lower mantle thermal conductivity. The lower mantle and CMB temperatures, $T_{LM}$ and $T_{cmb}$, are extrapolations along the mantle and core adiabats: $T_{LM}=\epsilon_{LM}T_m$ and $T_{cmb}=\epsilon_c T_c$, where  $\epsilon_{LM}=\exp(-(R_{LM}-R_{m})\alpha g /c_p) \approx 1.3$
and $\epsilon_c\approx0.8$.
Analogous to Eq.~(\ref{eq:delta_1}), the lower mantle thermal boundary layer thickness $\delta_{LM}$ is  derived by assuming the boundary layer Rayleigh number is critical, giving
\begin{equation}
\delta_{LM}=\left( \frac{\kappa \nu_{LM}} {\alpha g \Delta T_{LM}} Ra_c \right)^{1/3},
\label{eq:delta_2} \end{equation}

Core secular cooling is
\begin{equation}
Q_{core}=-M_c c_c \dot{T}_c,
\label{eq:q_core} \end{equation}
where $M_c$ is core mass, $c_c$ is core specific heat, and $\dot{T}_c$ is the rate of change of the average core temperature $T_c$.

\subsection{Mantle Melting}  \label{appendix_solidus}
The mantle solidus is approximated by a third-order polynomial \citep{elkinstanton2008b},
\be
T_{sol}(r)=A_{sol}r^3+B_{sol}r^2+C_{sol}r+D_{sol},
\label{eq:solidus} \ee
where the coefficients are constants (see Table \ref{tab:thermint}).  This solidus is calibrated to fit the following constraints: solidus temperature of $1450$ K at the surface, solidus temperature of $4150$ K at the CMB \citep{andrault2011}, and present-day upwelling melt fraction of $f_{melt}=8\%$.  The liquidus is assumed to be hotter by a constant offset $\Delta T_{liq}=500$ K, so $T_{liq}(r)=T_{sol}(r)+\Delta T_{liq}$.

Mantle melt heat loss (or advective heat flow) is modeled as,
\begin{equation}
Q_{melt}=\epsilon_{erupt}\dot{M}_{melt} \left( L_{melt} + c_{m} \Delta T_{melt} \right),
\label{eq:q_melt} \end{equation}
where $\epsilon_{erupt}=0.2$ is the assumed efficiency of magma eruption to Earth's surface, $\dot{M}_{melt}$ is melt mass flux (see below),
$L_{melt}$ is latent heat of the melt, $c_{m}$ is specific heat of the melt, and $\Delta T_{melt}$ is the excess temperature of the melt at the surface (see below).  The latent heat released by mantle melting does not contribute to the cooling of the mantle.  This formulation of heat loss is similar to the ``heat pipe'' mechanism invoked for Io \citep{oreilly1981,moore2003}, where melt eruption is a significant source of heat loss.   We note that this mechanism is more important for stagnant lid planets, where the normal conductive heat flow is lower \citep{DriscollBercovici14}.

The melt mass flux $\dot{M}_{melt}$ is the product of the upwelling solid mass flux times the melt mass fraction $f_{melt}$,
\begin{equation}
\dot{M}_{melt}=\dot{V}_{up} \rho_{solid} f_{melt}(z_{UM}),
\label{eq:m_melt}\end{equation}
where solid density is $\rho_{solid}$, volumetric upwelling rate is $\dot{V}_{up}=1.16\kappa A_p/ \delta_{UM}$, $z_{UM}=R-\delta_{UM}$, and melt fraction is
\be f_{melt}(z)=\frac{T_m(z)-T_{sol}}{T_{liq}-T_{sol}} ~. \label{f_melt} \ee
This model predicts a ridge melt production of $\dot{M}_{melt}=2.4\times10^6~\mbox{kg~s}^{-1}$ for $\delta_{UM}=80$ km and $f_{melt}=0.1$, similar to present-day global melt production estimates \citep{cogne2004}.

We define the magma ocean as the region of the mantle with temperature exceeding the liquidus.  Given the geotherm in Eqs.~(\ref{eq:T_ad}) and (\ref{eq:T_erf}) and the liquidus $T_{liq}(r)$ similar to Eq.~(\ref{eq:solidus}), the mantle will mainly freeze from the bottom of the convecting mantle up because the liquidus gradient is steeper than the adiabat \citep[\eg][]{elkinstanton2012}.  However, if the core is hot enough, a second melt region exists in the lower mantle boundary layer, where the temperature gradient exceeds the liquidus and the mantle freezes towards the CMB.

Latent heat released from the solidification of the mantle is
\be
Q_{L,man}=\dot{M}_{sol} L_{melt},
\label{eq:Q_latent_man} \ee
where $L_{melt}$ is the latent heat released per kg and $\dot{M}_{sol}$ is the solid mantle growth rate.  The growth rate is calculated assuming a uniform mantle density $\rho_m$ so that $\dot{M}_{sol}=\rho_m \dot{V}_{sol}$, where $\dot{V}_{sol}=-\dot{V}_{liq}$.  The rate of change of the liquid volume of the mantle is approximated by
\be \dot{V}_{liq} =\frac{dV_{liq}}{dT_m} \dot{T}_m,
\label{eq:dot_v_liq} \ee
where $\dot{T}_m$ is the mantle secular cooling rate and $dV_{liq}/dT_m$ is linearly approximated by $8\times10^{17}~\mbox{m}^3\mbox{K}^{-1}$, which is the change in liquid volume from a $90\%$ liquid to a completely solid mantle.
This approximation implies that the latent heat released due to mantle solidification is linearly proportional to the mantle secular cooling rate, and the ratio of the latent heat flow to the mantle secular cooling heat flow is $Q_{L,man}/Q_{sec,m}\approx0.24$.  For example, a mantle solidification time of 100 Myr corresponds to an average latent heat release of $Q_{L,man}\approx400$ TW over that time.

\subsection{Core Dynamo}
Given the thermal cooling rate of the core, the magnetic dipole moment $\mathcal{M}$ is estimated from the empirical scaling law,
\be \mathcal{M} = 4\pi R_c^3 \gamma_d \sqrt{\rho/2\mu_0} \left( F_c D_c \right)^{1/3}, \label{eq:magmom} \ee
where $\gamma_d$ is the saturation constant for fast rotating dipolar dynamos, $\mu_0=4\pi \times 10^{-7} \mbox{H~m}^{-1}$ is magnetic permeability, $D_c=R_c-R_{ic}$ is the dynamo region shell thickness, $R_c$ and $R_{ic}$ are outer and inner core radii, respectively, and $F_c$ is the core buoyancy flux \citep{olson2006}.
We assume that the field is dipolar, ignoring the complicating influences of shell thickness and heterogeneous boundary conditions.  In this formulation a positive buoyancy flux implies dynamo action, which is a reasonable approximation when the net buoyancy flux is large, but may overestimate the field strength at low flux.
The total core buoyancy flux $F_c$ is the sum of thermal and compositional buoyancy fluxes,
\be F_c=F_{th}+F_\chi, \label{eq:f_c}\ee
where the thermal and compositional buoyancy fluxes are
\begin{eqnarray} F_{th}&=&\frac{\alpha_c g_c}{\rho_c c_c} q_{c,conv}, \label{f_th}  \\
	F_\chi &=& g_i\frac{\Delta \rho_\chi}{\rho_c} \left(\frac{R_{ic}}{R_c}\right)^2 \dot{R}_{ic}, \label{eq:f_chi}
\end{eqnarray}
where the subscript $c$ refers to bulk core properties, core convective heat flux is $q_{c,conv}=q_{cmb}-q_{c,ad}$, gravity at the ICB is approximated by $g_{ic}=g_c R_{ic}/R_c$, and the outer core compositional density difference is $\Delta \rho_\chi=\rho_c-\rho_\chi$ with $\rho_\chi$ the light element density.  For simplicity, the expression for light element buoyancy Eq.~(\ref{eq:f_chi}) ignores buoyancy due to latent heat release at the ICB because it is a factor of $3.8$ less than buoyancy of the light elements.

The isentropic core heat flux at the CMB, proportional to the gradient of Eq.~(\ref{eq:T_c}), is
\be q_{c,ad}=k_c T_{cmb} R_c/D_N^2,\label{q_cad} \ee
where core thermal conductivity is approximated by the Wiedemann-Franz law,
\be
 k_c=\sigma_c L_c T_{cmb},\label{eq:core_conductivity}
 \ee
and electrical conductivity is $\sigma_c$ and $L_c$ is the Lorentz number.  For typical values of high pressure-temperature iron, $\sigma_c=10\times10^5$ $\Omega^{-1} m^{-1}$ \citep{pozzo2012,gomi2013}, $L_c=2.5\times10^{-8}~\mbox{W}\Omega\mbox{K}^{-1}$, and $T_{cmb}=4000$ K, the core thermal conductivity is $k_c=100~\mbox{Wm}^{-1}\mbox{K}^{-1}$.

\begin{longtable}{l | l | l | l}
\caption{\thermint constants.}
\label{tab:thermint}
\endfirsthead
\endhead
\hline
Symbol & Value & Units & Description\\
\hline
   $A_\nu$      &           $3\times10^5$              &  J mol$^{-1}$ &	 Viscosity activation energy in (\ref{eq:nu})	\\
   $A_\mu$	&	$2\times10^5$		&	J mol$^{-1}$ &	 Nominal shear modulus activation energy in (\ref{eq:mu})	\\
$A_{sol}$		&    $-1.160\times10^{-16}$	& K$/$m$^{3}$	&	Solidus coefficient in (\ref{eq:solidus}) \\
      $\alpha$ &       $3\times10^{-5}$    & K$^{-1}$ &	  Thermal expansivity of mantle	\\
   $\alpha_c$ &       $1\times10^{-5}$    & K$^{-1}$ &	  Thermal expansivity of core	\\
	$B$		&	$2.5$			& nd		& 	Melt fraction coefficient in (\ref{eq:epsilon_phase}) \\
   $B_{sol}$		&    $1.708\times10^{-9}$	& K$/$m$^{2}$	&	Solidus coefficient in (\ref{eq:solidus}), calibrated \\
   $\beta$ 	&	$1/3$			& nd		&	Convective cooling exponent in (\ref{eq:q_conv_m})\\
   $\beta_{st}$	&	$1.71\times10^{4}$	&	GPa		& Effective mantle stiffness \\
      $c_m$          &           1265    & J kg$^{-1}$ K$^{-1}$ &	  Specific heat of mantle	\\
   $c_c$         &           840    & J kg$^{-1}$ K$^{-1}$ &	  Specific heat of core	\\
      $C_{sol}$		&    $-9.074\times10^{-3}$	& K$/$m	&	Solidus coefficient in (\ref{eq:solidus}), calibrated \\
      $D$           &       2891    & km &	  Mantle depth	\\
     $D_{Fe}$              &       7000     & km &	  Iron solidus length scale	\\
  $D_N$              &       6340    & km &	  Core adiabatic length scale	\\
     $D_{sol}$		&    $1.993\times10^{4}$	& K	&	Solidus coefficient in (\ref{eq:solidus}), calibrated \\
     $\delta_{ph}$		&	$6$	& nd &	Rheology phase coefficient in (\ref{eq:epsilon_phase}, \ref{eq:epsilon_phase2}) \\
         $E_G$     &           $3\times10^5$    &  J kg$^{-1}$ &	  Gravitational energy density release at ICB 	\\
   $\epsilon_{UM}$        &          0.7   & nd &	  Upper mantle adiabatic temperature drop	\\
   $\epsilon_{LM}$        &           1.3    & nd &	 Lower mantle adiabatic temperature jump	\\
   $\epsilon_c$        &          0.8    & nd &	  Average core to CMB adiabatic temperature drop	\\
     $\phi^*$		&	$0.8$	& nd &	Rheology phase coefficient in (\ref{eq:epsilon_phase}, \ref{eq:epsilon_phase2}) \\
   $g_{UM}$          &           $9.8$  & $\mbox{m~s}^{-2}$ 	&	Upper mantle gravity	\\
   $g_{LM}$          &           $10.5$ & $\mbox{m~s}^{-2}$ 	 &	  Lower mantle gravity	\\
   $g_c$          &           $10.5$     & $\mbox{m~s}^{-2}$ 	 &	  CMB gravity	\\
     $\gamma_{c}$      &        $1.3$            & nd  &	  Core Gruneisen parameter	\\
      $\gamma_{dip}$       &          0.2    & nd &	  Magnetic dipole intensity coefficient in (\ref{eq:magmom})	\\
     $\gamma_{ph}$		&	$6$	&	 nd &  Rheology phase coefficient in (\ref{eq:epsilon_phase}, \ref{eq:epsilon_phase2}) \\
   $k_{UM}$          &           $4.2$    & W m$^{-1}$ K$^{-1}$ &	  Upper mantle thermal conductivity	\\
   $k_{LM}$         &           $10$    &  W m$^{-1}$ K$^{-1}$ &	  Lower mantle thermal conductivity	\\
   $\kappa$	&		$10^{-6}$	&	m$^2$ s$^{-1}$	&	Mantle thermal diffusivity \\
      $L_{Fe}$          &           750    & kJ kg$^{-1}$ &	  Latent heat of inner core crystallization	\\
         $L_{melt}$     &           320    & kJ kg$^{-1}$  &	  Latent heat of mantle melting	\\
            $L_e$         &       $2.5\times10^{-8}$                            &  W $\Omega$ K$^{-1}$&	  Lorentz number	\\
   $M_m$           &       $4.06\times10^{24}$    &   kg  &	  Mantle mass	\\
   	$M_c$	&	$1.95\times10^{24}$		& kg		&	Core mass \\
	$\mu_{ref}$	&	$10^5$	&	Pa	&	Reference shear modulus in (\ref{eq:mu}) \\
      $\mu_0$            &       $4\pi\times10^{-7}$     & H m$^{-1}$ &	  Magnetic permeability	\\
      $\nu_{ref}$        &       $6\times10^7$    &  m$^2$s$^{-1}$ &	  Reference viscosity	\\
   $\nu_{LM}/\nu_{UM}$   &    $2$          & nd &	  Viscosity jump from upper to lower mantle	\\
   $Q_{rad,0}$           &       $60$    & TW &	 Initial mantle radiogenic heat flow (J07) \\
   $R$         &       6371    &  km &	  Surface radius 	\\
   $R_c$          &       3480    & km  &	  Core radius	\\
   $R_m$		&	4925		& km		& Radius to average mantle temperature $T_m$\\  
   $Ra_c$             &           660    & nd  &	  Critical Rayleigh number	\\
   $\rho_c$        &           11900    & kg m$^{-3}$ &	  Core density	\\
   $\rho_{ic}$          &           13000    & kg m$^{-3}$ &	  Inner core density	\\
   $\rho_m$         &           4800    & kg m$^{-3}$ &	  Mantle density	\\
      $\rho_{melt}$         &           2700    & kg m$^{-3}$ &	  Mantle melt density	\\
      $\rho_{solid}$       &           3300                   & kg m$^{-3}$ &	  Mantle upwelling solid density 	\\
   $\Delta \rho_{\chi}$   &            700    & kg m$^{-3}$  &	  Outer core compositional density difference	\\
  $\sigma_c$      &           $10\times10^5$    & S m$^{-1}$ &	  Core electrical conductivity	\\
   $T_{Fe,0}$           &           5600    & K  &	  Iron solidus coefficient in (\ref{eq:lindemann})	\\
            $\xi$		&	$5\times10^{-4}$	& nd &	Rheology phase coefficient in (\ref{eq:epsilon_phase}, \ref{eq:epsilon_phase2}) \\
\hline
\end{longtable}

\section{List of Symbols\label{app:symbols}} 
In this appendix we present the definitions of all symbols used in this manuscript. Those that are general are listed first, followed by those that are specific to individual models.

\medskip

\twocolumngrid

\subsection{General Symbols}
\noindent $a$ = semi-major axis\\
$e$ = eccentricity\\
$G$ = Newton's gravitational constant\\
$j$ = index\\
$L$ = luminosity\\
$M$ = mass\\
$M_*$ = stellar mass\\
$M_p$ = planetary mass\\
$M_j$ = mass of body $j$\\
$n$ = mean motion (orbital frequency)\\
$\Omega$ = rotational frequency\\
$\Omega_j$ = rotational frequency of body $j$\\
$P$ = period\\
$\psi$ = obliquity\\
$\psi_j$ = obliquity of body $j$\\
$R$ = body radius\\
$R_p$ = planetary radius\\
$r_g$ = radius of gyration\\
$T_{eff}$ = effective temperature\\
$t$ = time\\

\subsection{\texttt{AtmEsc} Symbols}
\noindent XUV = X-ray + UV luminosity\\
$b_\mathrm{diff}$ = binary diffusion coefficient \\
$\epsilon_\mathrm{XUV}$ = XUV absorption efficiency \\
$F_\mathrm{diff}$ = diffusion-limited particle escape flux \\
$F_\mathrm{EL}$ = energy-limited particle escape flux \\
$F_\mathrm{H}$ = hydrogen particle escape flux \\
$\mathcal{F}_\mathrm{XUV}$ = XUV flux \\
$k_\mathrm{boltz}$ = Boltzmann constant \\
$K_\mathrm{tide}$ = tidal enhancement factor \\
$m_\mathrm{c}$ = crossover mass \\
$m_\mathrm{H}$ = hydrogen atom mass \\
$m_\mathrm{O}$ = oxygen atom mass \\
$T_\mathrm{flow}$ = temperature of hydrodynamic flow \\
$X_\mathrm{O}$ = oxygen mixing ratio \\

\subsection{\texttt{BINARY} Symbols}
\noindent CBP = circumbinary planet\\
$m_i$ = mass of $i^{th}$ star \\
$a_{AB}$ = binary semi-major axis \\
$e_{AB}$ = binary eccentricity \\
$M_{b}$ = binary mean anomaly \\
$\varpi_{B}$ = binary longitude of the periapse \\
$\alpha_i$ = normalized CBP orbital distance from the $i^{th}$ star \\
$\delta_{k0}$ = Kroenker delta function \\
$b^k_{(j+1)/2}$ = Laplace coefficient \\
$\Phi_{jk0}$ = 0th component of binary gravitational potential \\
$\Phi_{jk1}$ = 1st component of binary gravitational potential \\
$R$ = radial distance from binary center of mass \\
$R_0$ = CBP radial guiding center \\
$R_1$ = time-dependent epicyclic radial deviation from the guiding center radius \\
$\psi_0$ = arbitrary phase offset in the $R$ direction \\
$\varphi$ = arbitrary phase offset in the $\phi$ direction \\
$\xi$ = arbitrary phase offset in the $z$ direction \\
$n_k$ = CBP Keplerian mean motion \\
$n_0$ = CBP mean motion \\
$n_{AB}$ = binary mean motion \\
$e_{\textrm{free}}$ = binary free eccentricity \\
$i_{\textrm{free}}$ = binary free inclination \\
$\kappa_0$ = Radial CBP epicyclic frequency \\
$\nu_0$ = Vertical CBP epicyclic frequency \\

\subsection{\texttt{DistOrb} Symbols}
\noindent $e$ = orbital eccentricity \\
$\varpi$ = longitude of pericenter \\
$i$ = orbital inclination \\
$\Omega$ = longitude of ascending node \\
$h$ = first Poincar\'{e} coordinate for eccentricity \\
$k$ = second Poincar\'{e} coordinate for eccentricity \\
$p$ = first Poincar\'{e} coordinate for inclination \\
$q$ = second Poincar\'{e} coordinate for inclination \\
$\mathcal{R}$ = disturbing function \\
$n$ = mean motion \\
$\delta_R$ = post-Newtonian correction factor \\
$c$ = speed of light \\
$\mu$ = mass factor of inner planet in a pair \\
$\mu'$ = mass factor of outer planet in a pair \\
$\kappa$ = Gaussian gravitational constant \\
$\mathcal{R_D}$ = direct terms disturbing function \\
$D0.i$ = term $i$ of direct disturbing function \\
$f_i$ = semi-major axis functions in disturbing function\\

\subsection{\texttt{DistRot} Symbols}
\noindent $\psi$ =  precession angle \\
$\varepsilon$ =  obliquity of planet's spin axis\\
$R(\varepsilon)$ = precession rate due to stellar torque \\
$p$ = first Poincar\'{e} coordinate for inclination \\
$q$ = second Poincar\'{e} coordinate for inclination \\
$A(p,q)$ = precession/obliquity term from orbital plane motion \\
$B(p,q)$ = precession/obliquity term from orbital plane motion \\
$\Gamma(p,q)$ = precession/obliquity term from orbital plane motion \\
$p_g$ = geodetic (relativistic) precession rate \\
$\kappa$ = Gaussian gravitational constant \\
$M_{\star}$ = mass of host star\\
$M$ = mass of planet\\
$J_2$ = gravitational quadrupole moment due to planetary oblateness \\
$a$ = semi-major axis \\
$\nu$ = rotation rate of planet \\
$C$ = polar moment of inertia of planet\\
$r$ = equatorial radius of planet\\
$S_0$ = correction term for precession due to eccentric orbit \\
$\Lambda$ = angle between vernal point and ascending node\\
$\Omega$ = longitude of ascending node\\
$\vernal$ = vernal point (location of sun on northern spring equinox) \\
$\xi$ = first rectangular coordinate for obliquity and precession \\
$\zeta$ = second rectangular coordinate for obliquity and precession \\
$\chi$ = third rectangular coordinate for obliquity and precession \\
$J_{2\oplus}$ = gravitational quadrupole moment of Earth \\
$\nu_{\oplus}$ = rotation rate of Earth \\
$R_{\oplus}$ = equatorial radius of Earth\\
$M_{\oplus}$ = mass of Earth\\

\subsection{\texttt{EqTide} Symbols}
\noindent $\varepsilon$ = sign of the phase lag\\
$k_2$ = Love number of degree 2\\
$P_{eq}$ = equilibrium spin period\\
$Q$ = tidal quality factor \\
$\xi$ = constant in \eqtide~calculations\\
$Z$ = constant in \eqtide~calculations\\

\subsection{\texttt{GalHabit} Symbols}
\noindent $Z$ = height above galactic midplane \\
$H_{\text{av}}$ = time-averaged Hamiltonian for galactic tide \\
$\mu$ = mass factor \\
$G$ = Newtonian gravitational constant \\
$\rho_0$ = density of stars \\
$M_c$ = mass of central star \\
$M$ = mass of orbiter \\
$a$ = semi-major axis \\
$L$ = canonical momentum associated with orbital energy\\
$J$ = canonical momentum associated with total orbital angular momentum\\
$J_z$ = canonical momentum associated with $Z$ angular momentum \\
$\omega$ = argument of periastron \\
$i$ = inclination of system with respect to galactic plane\\
$\Omega$ = longitude of ascending node \\
$l$ = mean anomaly \\
$\vec{J}$ = angular momentum vector \\
$J_x$ = x component of angular momentum \\
$J_y$ = y component of angular momentum \\
$\vec{e}$ = eccentricity vector \\
$e_x$ = x component of eccentricity vector \\
$e_y$ = y component of eccentricity vector \\
$e_z$ = z component of eccentricity vector \\
$\hat{e}$ = unit vector with direction of eccentricity vector \\
$\hat{J}$ = unit vector with direction of angular momentum vector \\
$\hat{n}$ = unit vector used in calculation of $\omega$ \\
$\vec{n}$ = vector used in calculation of $\omega$ \\
$\hat{Z}$ = unit vector perpendicular to galactic plane \\
$f_{\text{enc}}$ = frequency of stellar encounters \\
$R_{\text{enc}}$ = radius at which stellar encounters begin \\
$v_i$ = velocity of typical encounter with stellar type $i$ \\
$n_{\star,i}$ = number density of stellar type $i$ \\
$v_{h,i}$ = average velocity of system relative to stellar type $i$ \\
$\sigma_{\star i}$ = velocity dispersion of stellar type $i$\\
$t_{\text{next}}$ = time of next encounter \\
$\xi$ = random number used in $t_{\text{next}}$ calculation \\
$\vec{R}_{\star}$ = starting position of passing star \\
$x_{\star}$ = x component of passing star's starting position \\
$y_{\star}$ = y component of passing star's starting position \\
$z_{\star}$ = z component of passing star's starting position \\
$x_{\star}$ = x component of passing star's starting position \\
$\theta_{\star}$ = co-latitudinal angle of passing star's starting position \\
$\phi_{\star}$ = longitudinal angle of passing star's starting position \\
$f_{\text{enc},i}$ = frequency of stellar encounters of stellar type $i$\\
$M_{V}$ = magnitude of passing star \\
$\Delta M_{V,i}$ = magnitude bin size of stellar type $i$ \\
$f_{\text{rel}}$ = frequency of stellar encounters of stellar type $i$ per magnitude\\
$\xi_{M_V}$ = randomly drawn magnitude of passing star \\
$\xi_{\text{samp}}$ = random number used in rejection sampling for magnitude\\
$\vec{v}_{h,\star}$ = relative velocity vector between system and passing star\\
$\vec{v}_{\star}$ = passing star's velocity vector \\
$\vec{v}_{h}$ = velocity of system relative to local standard of rest for stellar type $i$\\
$v_{\text{max}}$ = threshold velocity used for passing star velocity selection \\
$n$ = mean anomaly of orbiter \\
$\vec{b}_1$ = impact parameter of passing star relative to host system's primary star \\
$\vec{b}_2$ = impact parameter of passing star relative to host system's orbiter \\
$\vec{R}_2$ = position of orbiter relative to primary \\
$\Delta t_1$ = time difference until closest approach of passing star to primary \\
$\Delta t_2$ = time difference until closest approach of passing star to orbiter \\
$\Delta \vec{v}_2$ = instantaneous change in velocity of orbiter due to passing star \\
$m_{\star}$ = mass of passing star \\

\subsection{\texttt{POISE} Symbols}
\noindent OLR = outgoing longwave radiation\\
$\phi$ = latitude\\
$x$ = sine of latitude \\
$C$ = heat capacity of atmosphere\\
$T$ = temperature at/near surface \\
$D$ = coefficient of heat diffusion \\
$I$ = OLR \\
$S$ = instellation\\
$\alpha$ = albedo \\
$C_L$ = heat capacity over land \\
$C_W^{eff}$ = heat capacity over water \\
$T_L$ = temperature over land \\
$T_W$ = temperature over water \\
$f_L$ = land fraction of grid cell \\
$f_W$ = water fraction of grid cell \\
$\nu$ = land-ocean heat transport parameter \\
$C_W$ = head capacity over water per meter of depth \\
$m_d$ = depth of ocean mixing layer \\
$\delta$ = declination of sun/host star\\
$S_{\star}$ = flux at planet's distance from host star \\
$\rho$ = planet-star distance, in semi-major axis units \\
$H_0$ = hour angle of host star at sunrise/sunset \\
$\varepsilon$ = obliquity \\
$\theta$ = true longitude of planet \\
$f$ = true anomaly of planet \\
$\Delta^*$ = angle between periastron and planet's position at northern spring equinox \\
$\varpi$ = longitude of periastron\\
$\psi$ = precession angle \\
$\vernal$ = vernal point \\
$\alpha_L$ = albedo over land \\
$\alpha_W$ = albedo over water \\
$P_2$ = second Legendre polynomial \\
$M_{\text{ice}}$ = mass of ice on land \\
$Z$ = zenith angle of sun/star at noon \\
$A$ = OLR coefficient \\
$B$ = OLR coefficient \\
$\mathscr{M}$ = matrix used in energy balance model integration \\
$\xi$ = ice ablation tuning parameter \\
$\sigma$ = Stefan-Boltzmann constant \\
$L_h$ = latent heat of fusion for water\\
$T_{\text{freeze}}$ = freezing temperature of water \\
$h$ = height of ice sheet above ground \\
$H$ = height of bedrock \\
$y$ = latitudinal coordinate \\
$R$ = radius of planet \\
$A_{\text{ice}} $ = deformability of ice \\
$\rho_i$ = density of ice \\
$g$ = gravitational acceleration at the surface \\
$n$ = Glen's flow law exponent \\
$u_b$ = flow speed of ice at the base \\
$D_0$ = reference deformation rate of sediment \\
$a_{\text{sed}}$ = shear stress from ice on sediment \\
$b_{\text{sed}}$ = rate of increase of shear strength with depth of sediment \\
$\mu_0$ = reference viscosity of sediment \\
$m$ = coefficient used in basal flow calculation \\
$h_s$ = depth of sediment \\
$\rho_s$ = density of sediment \\
$\rho_w$ = density of water \\
$\phi_s$ = internal deformation angle of sediment \\
$D_{\text{ice}}$ = diffusion coefficient for ice flow \\
$r_{\text{snow}}$ = rate of snow accumulation \\
$T_b$ = relaxation time scale of bedrock \\
$H_{eq} $ = ice-free equilibrium height of bedrock \\
$\rho_b$ = bedrock density\\
$T_0$ = initial temperature distribution of surface \\

\subsection{\texttt{RadHeat} Symbols}
\noindent $Q_{i,j}$ = radiogenic heat production of species $i$ in reservoir $j$ \\
$\lambda_{i,1/2}$ = radiogenic decay constant of species $i$ = $\ln 2/\tau_{i,1/2}$ \\
$\tau_{i,1/2}$ = radiogenic halflife of species $i$ \\

\subsection{\texttt{SpiNBody} Symbols}
\noindent $F_i$ = Newtontian gravitational force on orbital body $i$ \\
$m_i$ = Mass of orbital body $i$\\
$r_{i,j}$ = Distance between $i$ and $j$ \\
$\dot{v_i}$ = Net gravitational acceleration of body $i$ \\
$\dot{x_i}$ = Net velocity of body $i$ \\

\subsection{\texttt{STELLAR} Symbols}
\noindent $\beta_\mathrm{sat}$ = XUV power law index \\
$L_\mathrm{bol}$ = Stellar bolometric (total) luminosity \\
$L_\mathrm{XUV}$ = Stellar XUV luminosity \\
$f_\mathrm{sat}$ = Saturation ratio of $L_\mathrm{XUV}$ to $L_\mathrm{bol}$\\
$t_\mathrm{sat}$ = XUV saturation timescale \\
$r_g$ = Stellar radius of gyration \\

\subsection{\texttt{ThermInt} Symbols}
\noindent CMB = core mantle boundary\\
ICB = inner core boundary\\
$Q_{surf}$ = total mantle surface heat flow \\
$Q_{conv}$ = heat conducted through the lithospheric thermal boundary layer that is supplied by mantle convection \\
$Q_{melt}$ = heat loss due to the eruption of upwelling mantle melt at the surface \\
$Q_{cmb}$ = heat lost from the core conducted across the CMB \\
$Q_{man}$ = secular (sensible) heat lost from the mantle \\
$Q_{tidal}$ = heat generated in the mantle by tidal dissipation \\
$Q_{L,man}$ = latent heat released by the solidification of the mantle \\
$Q_{core}$ = core secular cooling \\
$Q_{i,core}$ = radiogenic heat production of species $i$ in the core \\
$Q_{icb}$ = heat released by the solidification of the inner core \\
$M_{ic}$ = inner core mass \\
$L_{icb}$ = latent energy released per unit mass at the ICB \\
$E_{icb}$ = gravitational energy released per unit mass at the ICB \\
$T_g$ = ground temperature \\
$T_{UM}$ = upper mantle temperature \\
$T_m$ = average mantle temperature \\
$T_{LM}$ = lower mantle temperature \\
$T_{CMB}$ = core-mantle boundary temperature \\
$T_c$ = average core temperature \\
$\delta_{UM}$ = thermal boundary layer thickness of upper mantle \\
$\delta_{LM}$ = thermal boundary layer thickness of lower mantle \\
$\epsilon_{erupt}$ = mantle melt mass extrusive eruption fraction \\
$A$ = Surface area of planet \\
$A_{CMB}$ = CMB area \\
$A_{ICB}$ = ICB area \\
$A_\nu$ = Viscosity activation energy in (\ref{eq:nu})	\\
$A_\mu$ = Nominal shear modulus activation energy in (\ref{eq:mu})	\\
$A_{sol}$ = Solidus coefficient in (\ref{eq:solidus}) \\
$\alpha$ = Thermal expansivity of mantle \\
$\alpha_c$ = Thermal expansivity of core \\
$B$ = Melt fraction coefficient in (\ref{eq:epsilon_phase}) \\
$B_{sol}$ = Solidus coefficient in (\ref{eq:solidus}) \\
$\beta$ = Convective cooling exponent in (\ref{eq:q_conv_m}) \\
$\beta_{st}$ = Effective mantle stiffness \\
$c_m=$ = Specific heat of mantle \\
$c_c$ = Specific heat of core \\
$C_{sol}$ = Solidus coefficient in (\ref{eq:solidus}), calibrated \\
$D$ = Mantle depth \\
$D_{Fe}$ = Iron solidus length scale\\
$D_N$ = Core adiabatic length scale \\
$D_{sol}$ = Solidus coefficient in (\ref{eq:solidus}), calibrated \\
$\delta_{ph}$ = Rheology phase coefficient in (\ref{eq:epsilon_phase}, \ref{eq:epsilon_phase2}) \\
$E_G$ = Gravitational energy density release at ICB \\
$\epsilon_{UM}$ = Upper mantle adiabatic temperature drop \\
$\epsilon_{LM}$ = Lower mantle adiabatic temperature jump \\
$\epsilon_c$ = Average core to CMB adiabatic temperature drop \\
$\epsilon_{phase}$ = mantle rheology melt reduction factor \\
$F_c$ = total core buoyancy flux \\
$F_{th}$ = core thermal buoyancy flux \\
$F_{\chi}$ = core compositional buoyancy flux \\
$\phi^*$ = Rheology phase coefficient in (\ref{eq:epsilon_phase}, \ref{eq:epsilon_phase2}) \\
$g_{UM}$ = Upper mantle gravity \\
$g_{LM}$ = Lower mantle gravity \\
$g_c$ = CMB gravity	\\
$\gamma_{c}$ = Core Gruneisen parameter	\\
$\gamma_{dip}$ = Magnetic dipole intensity coefficient in (\ref{eq:magmom})	\\
$\gamma_{ph}$	= Rheology phase coefficient in (\ref{eq:epsilon_phase}, \ref{eq:epsilon_phase2}) \\
$\gamma_{ad}$ = solid mantle adiabatic temperature gradient \\
$k_2$ = real second order Love number in (\ref{eq:k2})\\
$Im(k_2)$ = imaginary second order Love number in (\ref{eq:im_k2}) \\
$k_{UM}$ = Upper mantle thermal conductivity \\
$k_{LM}$ = Lower mantle thermal conductivity \\
$\kappa$ = Mantle thermal diffusivity \\
$k_c$ = core thermal conductivitiy \\
$L_{Fe}$ = Latent heat of inner core crystallization \\
$L_{melt}$ = Latent heat of mantle melting \\
$L_e$ = Lorentz number \\
$M_m$ = Mantle mass \\
$M_{sol}$ = mass of solid mantle \\
$V_{liq}$ = volume of liquid mantle \\
$M_c$= Core mass \\
$M_{ic}$= Inner core mass \\
$\mathcal{M}$ = core dynamo dipolar magnetic moment \\
$\mu_{ref}$ = Reference shear modulus in (\ref{eq:mu}) \\
$\mu_0$ = Magnetic permeability \\
$\nu_{ref}$ = Reference viscosity \\
$f_{\nu,UM,LM}$ = Viscosity jump from upper to lower mantle	\\
$q_{c,conv}$ = core convective heat flux \\
$q_{c,ad}$ = core adiabatic heat flux \\
$\mathcal{Q}$ = tidal quality factor \\
$R$ = Surface radius \\
$R_c$ = Core radius \\
$R_{ic}$ = Inner core radius \\
$R_m$ = Radius to average mantle temperature $T_m$ \\
$Ra_c$ = Critical Rayleigh number \\
$\rho_c$ = Core density \\
$\rho_{ic}$ = Inner core density \\
$\rho_m$ = Mantle density \\
$\rho_{melt}$ = Mantle melt density	\\
$\rho_{solid}$ = Mantle upwelling solid density \\
$\Delta \rho_{\chi}$ = Outer core compositional density difference \\
$\sigma_c$ = Core electrical conductivity \\
$T_{Fe,0}$ = Iron solidus coefficient in (\ref{eq:lindemann}) \\
$T_{Fe,cmb}$ = Iron solidus temperature at CMB \\
$T_{Fe,cen}$ = Iron solidus temperature at center of core \\
$\xi$ = Rheology phase coefficient in (\ref{eq:epsilon_phase}, \ref{eq:epsilon_phase2}) \\

\onecolumngrid

\section{\texttt{VPLanet} Accessibility and Support Software\label{app:support}}

In this final appendix, we describe our approaches for accessibility and reproducibility. In general, we adhere to the recommendations of the National Acadamies 2018 report {\it Open Science by Design}\footnote{\url{https://www.nap.edu/catalog/25116/open-science-by-design-realizing-a-vision-for-21st-century}}. This study identified numerous challenges in science as software and datasets grow very large. \vplanet brings together numerous models from disparate fields of science and therefore risks becoming inscrutable to many professional scientists. Moreover, \vplanet can produce large amounts of high-dimensional data that can be difficult for the user to digest. More philosophically, given that the challenge of detecting life beyond the Solar System is daunting and the potential societal impact of such a discovery is huge, it is critical that the scientific process that culminates in such an announcement be as transparent as possible. To address these issues, \vplanet has been designed and released to maximize usability and clarity, including code testing and data management. The next subsections describe \texttt{Python} scripts (requiring \texttt{Python} 3.x) to explore parameter space, visualize \vplanet output, efficiently store that output, and the \texttt{git} repository that hosts \vplanet.

\subsection{Parameter Sweeps: \texttt{VSPACE}}
\vsp is a \texttt{Python} code that conveniently generates initial conditions for \vplanet simulations but does not run \vplanet itself. This option is ideal when the user wishes to run a large number of simulations, perhaps on a supercomputer. Initial conditions can be generated in a random (Monte-Carlo) fashion or over a regular grid. The user creates a template plain text file that contains a list of all desired input files, with a list of input options to be changed, iterated over, or randomly drawn. \vsp operates on this text file, determines which files are to be copied to new directories, which input parameters are to be varied, and then creates a new directory that contains (as individual directories) all the desired simulations. When used in random mode, the user sets the total number of simulations and the parameters to be sampled along with details about the distributions. Current distribution types are uniform, Gaussian/normal, uniform in sine, or uniform in cosine. When used in grid mode, the user does not need to specify the total number of simulations. One simply sets all of the parameters to be varied, along with the spacing type (linear or logarithmic), and spacing size, or the total range and number of grid points.  \vsp automatically determines the number of simulations and identifies every permutation of the desired set of input parameters using the \texttt{itertools} library. There is no limit to the number of parameters that can be iterated over in a single set of simulations, thus the user should be cautious about generating large numbers of files. The \link{examples/IoHeat} directory in the \vplanet repository contains a \vsp example.

\subsection{Plotting: \texttt{VPLot}}
The \vplot package is recommended for interacting with and visualizing the output of \vplanet simulations. \vplot is coded in
\texttt{Python} and is installable via the \texttt{pip} command or from source from its own  \texttt{GitHub} repository\footnote{\url{https://github.com/VirtualPlanetaryLaboratory/vplot}}. \vplot reads \texttt{.log} and
\texttt{.forward} files generated by a \vplanet run in the current working directory and provides an object-oriented interface to the simulation output, with convenient plotting functions. \vplot can also be run from the command line, providing a quick graphical view of the simulation output. For more information on \vplot, please see its online documentation. Note that all plots in this document that display \vplanet output have utilized \vplot.

\subsection{Data Management: \texttt{BigPlanet}}

\bigpl is an accompanying Python package for storing the output of a large number of \vplanet simulations.  For large numbers of \vplanet simulations, \ie a suite over a large grid of initial conditions, the total size of the output for all simulations can exceed what can be stored in memory, necessitating new solutions. \bigpl solves this issue by using the HDF5 data storage format\footnote{\url{https://support.hdfgroup.org/HDF5/}} for efficiently storing full time series simulation outputs for easy access. \bigpl leverages the hierarchical structure of the HD5F format by storing each simulation as a group, each simulation body as a subgroup of the simulation, and each variable as a subgroup of its parent body.  \bigpl wraps the HDF5 file interface in easy-to-use python code, using the \texttt{H5PY}\footnote{\url{https://www.h5py.org/}} python package, to allow users to easily read and manipulate \vplanet simulation outputs from storage for in-memory analysis.  \bigpl is extensively documented, and includes several example scripts and Jupyter notebooks that demonstrate its typical use cases.

\subsection{The \texttt{VPLanet} Repository}

The \vplanet repository contains the source code, example scripts, and documentation to facilitate its use.  The \vplanet team continuously checks that all new updates reproduce past results through continuous integration\footnote{\url{https://travis-ci.org/}}, and periodically searches for memory issues using \texttt{valgrind}\footnote{\url{http://valgrind.org/}} and \texttt{addresssanitizer}\footnote{\url{https://github.com/google/sanitizers/wiki/AddressSanitizer}}. Thus, the master branch in this \texttt{git} repository is stable, and the results presented in this manuscript can all be obtained with it. For those who prefer not use \texttt{git}, gzipped tarballs are also available. The figures in this document were generated with version 1.0.

Community input and additions are welcome. Pull requests should be issued to the ``dev'' branch and will be reviewed before being merged with the master branch. As new features and modules are added to the software suite, this document will be updated and available in the  \href{https://github.com/VirtualPlanetaryLaboratory/vplanet/tree/master/Manual}{\link{Manual}} sub-directory of the \vplanet repository.

\bibliography{bib}


\end{document}